\documentclass{aa}  

\usepackage{graphicx}
\usepackage{epstopdf}
\usepackage{txfonts}
\usepackage{bm}
\usepackage{amsmath}
\usepackage{amssymb}
\usepackage{wasysym}

\graphicspath{ {Figures/} }
\usepackage[font=large]{subfig}
\usepackage{array}
\def\cX{\mathcal{X}}
\def\cY{\mathcal{Y}}
\def\cZ{\mathcal{Z}}
\def\E{\mathrm{E}}
\newcolumntype{R}{ >{${}}r<{{}$} }
\newcolumntype{C}{ >{${}}c<{{}$}}
\newcolumntype{L}{ >{${}}l<{{}$}}

\newcommand{\Poincare}{Poincar\'{e}}
\newcommand{\arcsecyr}{$\arcsec\,{\textrm{yr}}^{-1}$ }
\newcommand{\arcsecyrNOSPACE}{$\arcsec\,{\textrm{yr}}^{-1}$}

\newcommand{\LaX}{LaX13b}

\let\vec\boldsymbol

\newcommand{\calX}{\mathcal{X}}
\newcommand{\calY}{\mathcal{Y}}

\newcommand{\subscript}[2]{#1_{\textnormal{\scriptsize #2}}}

\newcommand{\Hsec}{\widehat{H}}
\newcommand{\Hiss}{\mathcal{H}}
\newcommand{\HLL}{\subscript{\Hsec}{LL}}

\newcommand{\HissLL}[1]{\Hiss_{\textnormal{LL}(\vec{#1})}}
\newcommand{\matrOD}[2]{\vec{#1}_{\text{#2}}}

\newcommand{\inn}[1]{\subscript{#1}{i}}
\newcommand{\out}[1]{\subscript{#1}{o}}
\newcommand{\ii}[1]{\subscript{#1}{ii}}
\newcommand{\oo}[1]{\subscript{#1}{oo}}

\newcommand{\gLL}{\subscript{g}{LL}}
\newcommand{\sLL}{\subscript{s}{LL}}

\newcommand{\innxprime}{\inn{\vec{x}}^\prime}

\newcommand{\free}[1]{\subscript{#1}{i,F}}
\newcommand{\forc}[1]{\subscript{#1}{i,f}}

\newcommand{\vecu}{\vec{u}}
\newcommand{\vecv}{\vec{v}}

\newcommand{\Huv}{\mathbb{H}}
\newcommand{\HuvLL}[1]{\Huv_{\textnormal{LL}(\vec{#1})}}

\newcommand{\Chi}{X}
\newcommand{\vecChi}{\vec{\Chi}}
\newcommand{\vecchi}{\vec{\chi}}
\newcommand{\vecPsi}{\vec{\Psi}}
\newcommand{\vecpsi}{\vec{\psi}}

\newcommand{\vecI}{\vec{I}}
\newcommand{\vectheta}{\vec{\theta}}

\newcommand{\eccanomaly}{E}

\newcommand{\tcoll}{\tau_c}

\newcommand{\bpm}{\begin{pmatrix}}  
\newcommand{\epm}{\end{pmatrix}}
\newcommand\Rotx[1]{\bpm
             1& 0& 0\cr 0& \cos#1& -\sin#1\cr
             0& \sin#1& \cos#1\cr
             \epm}

\newcommand\Rotz[1]{\bpm
            \cos#1&-\sin#1&0\cr
            \sin#1&\cos#1&0\cr
             0& 0& 1\cr
             \epm}
\newcommand\abs[1]{\left\vert#1\right\vert}

\def\rms{\mathrm{rms}}

\begin{document} 

\title{Long-term dynamics of the solar system inner planets}
\titlerunning{Long-term dynamics of the inner solar system}

\author{Federico Mogavero \and Jacques Laskar}
\authorrunning{F. Mogavero \& J. Laskar}

\institute{Astronomie et Syst\`{e}mes Dynamiques, Institut de M\'{e}canique C\'{e}leste et de Calcul des \'{E}ph\'{e}m\'{e}rides \\ 
CNRS UMR 8028, Observatoire de Paris, Universit\'{e} PSL, Sorbonne Universit\'{e}, 77 Avenue Denfert-Rochereau, 75014 Paris, France \\
\email{federico.mogavero@obspm.fr}
}

\date{Received ; accepted }


\abstract{Although the discovery of the chaotic motion of the inner planets in the solar system (Mercury to Mars) 
  dates back to more than thirty years ago, the secular chaos of their orbits still dares more analytical analyses. 
  Apart from the high-dimensional structure of the motion, this is probably related to the lack of an adequately simple 
  dynamical model. 
  In this paper, we consider a new secular dynamics for the inner planets, with the aim of retaining a fundamental set of 
  interactions responsible for their chaotic behaviour, while being consistent with the predictions of the most precise orbital 
  solutions currently available. 
  We exploit the regularity in the secular motion of the outer planets (Jupiter to Neptune), to predetermine 
  a quasi-periodic solution for their orbits. This reduces the secular phase space to the degrees of freedom dominated by the 
  inner planets. On top of that, the smallness of the inner planet masses and the absence of strong mean-motion resonances 
  permits to restrict ourselves to first-order secular averaging. 
  The resulting dynamics can be integrated numerically in a very efficient way through Gauss's method, while computer algebra 
  allows for analytical inspection of planet interactions, once the Hamiltonian is truncated at a given total degree in eccentricities 
  and inclinations. 
  The new model matches very satisfactorily reference orbital solutions of the solar system over timescales shorter than or 
  comparable to the Lyapunov time. It correctly reproduces the maximum Lyapunov exponent of the inner system and the statistics 
  of the high eccentricities of Mercury over the next five billion years. The destabilizing role of the $g_1-g_5$ secular resonance 
  also arises. A numerical experiment, consisting of a thousand orbital solutions over one hundred billion years, 
  reveals the essential properties of the stochastic process driving the destabilization of the inner solar system and clarifies its 
  current metastable state.}

\keywords{Solar system -- Chaos -- Celestial mechanics -- Planets and satellites: dynamical evolution and stability -- Methods: analytical -- Methods: numerical}

\maketitle
%

\section{Introduction}
\label{sect:intro}
The inner solar system, with the secular chaos of its planetary orbits \citep{Laskar1989,Laskar1990,Sussman1992}, 
holds a special status among the dynamical systems of celestial mechanics. 
Even if its chaotic behaviour has been demonstrated numerically three decades ago, still no analytical study 
corroborates or rules out the role in chaos generation of the secular resonances $2(g_3-g_4)-(s_3-s_4)$ 
and $(g_3-g_4)-(s_3-s_4)$, between the Earth and Mars fundamental precession frequencies, as proposed 
in \citep{Laskar1990,Laskar1992} and supported by \citep{Laskar2004}. Such a fact probably relates to the complex network of dynamical interactions 
among the inner planets: the fundamental precession frequencies of their orbits vary in an intricate way over a 
0.1\arcsecyr scale in a few tens of million years (Myr), with the exception of the Venus-dominated eccentricity mode $g_2$ 
which has somewhat smaller variations \citep{Laskar1990,Laskar2004}. This behaviour reveals the essential high-dimensional 
structure of the inner solar system, which limits the possibility of faithfully modelling its dynamics with 
few effective degrees of freedom \citep[e.g.,][]{Lithwick2011,Batygin2015}. 

Getting analytical insight into the motion of the inner planets requires an appropriate dynamical modelling of the long-term evolution 
of their orbits. On the one hand, such a model must be consistent with the predictions of the reference numerical integrations available in 
literature \citep{Laskar1990,Laskar2004,Laskar2008,Laskar2009}, to ensure it reproduces the dynamical features of the inner 
system with sufficient precision. On the other hand, the corresponding Hamiltonian should be put in a form suitable for 
the systematic application of canonical perturbation techniques \citep{Hori1966,Deprit1969}, which is 
essential to an unbiased analysis of such a high-dimensional dynamics. Moreover, 
the possibility to numerically integrate the equations of motion in an efficient way is fundamental to study the 
chaotic evolution of the orbital solutions in a statistical way. Unfortunately, the construction of such a 
model turns out to be a delicate task. In principle, one could just consider the full N-body Hamiltonian of the Newtonian 
gravitational interactions among the solar system planets, with the addition of the leading corrections coming from 
general relativity and the Earth-Moon interaction. This already reproduces the precession frequencies of the inner orbits 
with a precision better than 0.01\arcsecyr \citep{Laskar1999}. However, such a Hamiltonian is unnecessarily complicated, as 
it includes short-time harmonics, with periods of less than 5\,000 yr \citep[e.g.,][]{Carpino1987}, which are known to generate 
small quasi-periodic oscillations in the inner orbits, without being implied in chaos generation. Indeed, the inner planets 
are not involved in any relevant mean-motion resonance. At the same time, the long-term numerical integration of the corresponding 
equations of motion is very time consuming, the resulting solutions needing to be filtered to extract the secular 
trend of the orbits \citep{Carpino1987,Nobili1989}. These facts suggest to consider a secular Hamiltonian to directly describe the slow 
movement of the planet perihelia and nodes, after proper averaging over the short-time orbital motion \citep{Laskar1984,Laskar1985}. 
Secular dynamics includes the essential planet interactions responsible for chaos in the inner solar system and allows to perform the 
fastest long-term numerical integrations \citep{Laskar1988,Laskar1989,Laskar1994,Laskar2008}. Indeed, this made the 
discovery of chaos in the inner system possible before the use of symplectic integration schemes \citep{Sussman1992}. 
Unfortunately, an effective secular model for the entire solar system has to be of high order in planet masses, 
principally because of the 5:2 near mean-motion resonance between Jupiter and Saturn, the so-called Great Inequality 
\citep{Laplace1785,Laskar1996}. A simple averaging of the N-body Hamiltonian over the planet mean longitudes, resulting in 
a first-order secular dynamics in planet masses, would reproduce very poorly the fundamental frequencies $g_5$ and $g_6$, which 
dominate the perihelion precession of Jupiter and Saturn, respectively \citep{Laskar1988}\footnote{Even the second-order secular dynamics 
in \citep{Laskar1985} needed an ad-hoc correction of 0.27\arcsecyr for the Saturn-dominated eccentricity mode $g_6$ \citep{Laskar1988}.}. 
The construction of such higher-order models requires the manipulation of large Poisson series and the use of sophisticated 
computer algebra systems. This is probably the reason why they are still not widely used, at least as a basis of extensive research. 

This paper introduces a new secular dynamics for the orbits of the inner planets. It is based on the 
practical long-term regularity of the outer planet trajectories \citep{Laskar1990,Laskar2004,Hoang2021}, the smallness of the inner 
planet masses and the absence of relevant mean-motion resonances in the inner system (Sect. \ref{sect:dyn_model}). 
We show that the present model can be numerically integrated via the so-called 
\citet{Gauss1818}'s method (Sect. \ref{sect:gauss_method}), while its Hamiltonian is suitable for a systematic expansion in planet 
eccentricities and inclinations, once one employs the algorithm of \citep{Laskar1995} and a computer algebra system like TRIP \citep{Gastineau2011,TRIP2020} 
(Sect. \ref{sect:ham_expansion}). We compare its orbital solution to a reference N-body integration over short (secular) timescales 
(Sect. \ref{sect:comparison}) and compute the corresponding maximal Lyapunov exponent in a statistical fashion (Sect. \ref{sect:lyapunov}). 
We then determine the percentages of the high Mercury eccentricities over the next 5 billion years and highlight 
the destabilizing role of the $g_1-g_5$ secular resonance (Sect. \ref{sect:evolution_5Gyr}). 
Finally, we perform a new numerical simulation involving a thousand orbital solutions over one hundred billion years, 
to characterize the effective stochastic process which drives the destabilization of the inner solar system, and discuss 
its implications on the conjecture of marginal stability of a secularly evolving planetary system formulated in \citep{Laskar1996} 
(Sect. \ref{sect:evolution_100Gyr}).  

\section{Dynamical model}
\label{sect:dyn_model}
We model the dynamics of the largest bodies in the solar system, by considering the Sun and the $N=8$ planets as point masses $m_0$, $(m_k)_{k=1,N}$, 
indexed in order of increasing semi-major axis. The barycentric coordinates of the bodies and the corresponding momenta are denoted by 
$(\vec{R}_0,\tilde{\vec{R}}_0 = m_0 \dot{\vec{R}}_0)$, $(\vec{R}_k,\tilde{\vec{R}}_k = m_k \dot{\vec{R}}_k)_{k=1,N}$. 
By employing the canonical heliocentric variables of \citep{Poincare1896}, the Hamiltonian of the Newtonian gravitational interactions 
among the bodies reads
\begin{equation}
\label{eq:NbodyHam}
H = \sum_{k=1}^{N} \left( 
\frac{\Vert \tilde{\vec{r}}_k \Vert^2}{2 \mu_k} - G \frac{m_0 m_k}{\Vert \vec{r}_k \Vert} 
\right)
+ \sum_{1 \leq k < \ell \leq N} \left( 
\frac{\tilde{\vec{r}}_k \cdot \tilde{\vec{r}}_\ell}{m_0} - G \frac{m_k m_\ell}{\Vert \vec{r}_k - \vec{r}_\ell \Vert}
\right),
\end{equation}
where $\vec{r}_k = \vec{R}_k - \vec{R}_0$ are the planet heliocentric coordinates and $\tilde{\vec{r}}_k = \tilde{\vec{R}}_k$ their
conjugated momenta, $\mu_k = m_0 m_k / (m_0 + m_k)$ are the reduced masses of the planets and $G$ is the gravitational constant 
\citep{Laskar1991,Laskar1995}. The Hamiltonian $H$ is a perturbation to the union of disjoint Kepler problems, so that it is useful to introduce 
a set of canonical variables which trivially integrates the unperturbed problems \citep[e.g.,][]{Morbidelli2002}. By adopting 
from now on the momentum-coordinate ordering of conjugate pairs, appropriate variables are $(\Lambda_k,\lambda_k;$ $x_k, -j \bar{x}_k;$ $y_k, -j \bar{y}_k)_{k=1,N}$, defined as 
\begin{equation}
\label{eq:poincare_vars}
\begin{aligned}
&\Lambda_k = \mu_k \sqrt{G(m_0 + m_k) a_k}, \\
&x_k = \sqrt{\Lambda_k} \sqrt{1 - \sqrt{1- e_k^2}} \, \E^{j \varpi_k}, \\
&y_k = \sqrt{2 \Lambda_k} \left(1- e_k^2\right)^{\frac{1}{4}} \sin(i_k/2) \, \E^{j \Omega_k}, 
\end{aligned}
\end{equation}
where $a_k$ are the planet semi-major axes, $\lambda_k$ the mean longitudes, $e_k$ the eccentricities, $i_k$ the inclinations, 
$\varpi_k$ the longitudes of the perihelia and $\Omega_k$ the longitudes of the nodes \citep{Laskar1991,Laskar1995}. Throughout the paper, 
$j = \sqrt{-1}$ stands for the imaginary unit, $\E$ represents the exponential operator and the overbar denotes the conjugate of 
a complex variable. The variables $x_k$ and $y_k$ are the \Poincare's rectangular coordinates in complex form; we shall refer to 
them throughout the paper as \Poincare's complex variables, or simply \Poincare's variables. With such a choice of canonical 
variables, the integrable part of the Hamiltonian \eqref{eq:NbodyHam} reads 
\begin{equation}
\label{eq:Kepler_energy}
H_0 = 
\sum_{k=1}^{N} \left( \frac{\Vert \tilde{\vec{r}}_k \Vert^2}{2 \mu_k} - G \frac{m_0 m_k}{\Vert \vec{r}_k \Vert} \right)
= - \sum_{k=1}^{N} \frac{G^2 (m_0 + m_k)^2 \mu_k^3}{2 \Lambda_k^2} ,
\end{equation}
so that the \Poincare's complex variables are constants of motion for the Kepler problem.

In the regime of small orbital eccentricities and inclinations, which characterizes the solar system, Eqs. \eqref{eq:poincare_vars} give 
$x_k = \sqrt{\Lambda_k/2} e_k \E^{j \varpi_k} + \mathcal{O}(e_k^3)$ and $y_k = \sqrt{\Lambda_k/2} i_k \E^{j \Omega_k} + \mathcal{O}(e_k^2 i_k, i_k^3)$, 
and the \Poincare's variables are also small. The principal part of the two-body perturbation in Eq. \eqref{eq:NbodyHam} 
can be thus expanded as a Fourier series in the planet mean longitudes, with polynomial coefficients depending 
on the \Poincare's variables \citep[e.g.,][]{Laskar1990c,Laskar1991,Laskar1995}, 
\begin{equation}
\label{eq:potential_expansion}
\begin{aligned}
&\mathcal{U}_1 = - G \frac{m m^\prime}{\Vert \vec{r} - \vec{r}^\prime \Vert} = 
- G \frac{m m^\prime}{a^\prime} \sum_{\ell, \ell^\prime \in \mathbb{Z}} 
\widetilde{\mathcal{U}}_{\ell, \ell^\prime} 
\E^{j(\ell \lambda + \ell^\prime \lambda^\prime)}, \\
&\widetilde{\mathcal{U}}_{\ell, \ell^\prime} = 
\sum \Gamma^{\ell, \ell^\prime}_{\mathcal{N}}\!(\alpha) 
\calX^n {\calX^\prime}^{n^\prime} \bar{\calX}^{\bar{n}} {\bar{\calX}}^{\prime \bar{n}^\prime} 
\calY^m {\calY^\prime}^{m^\prime} \bar{\calY}^{\bar{m}} {\bar{\calY}}^{\prime \bar{m}^\prime},
\end{aligned}
\end{equation}
where $\mathcal{N} = (n,n^\prime,\bar{n},\bar{n}^\prime,m,m^\prime,\bar{m},\bar{m}^\prime)$ is
a tuple of non-negative integers\footnote{The symmetries of the planetary Hamiltonian \eqref{eq:NbodyHam} imply 
constraints on the exponents $(\ell,\ell^\prime,\mathcal{N})$ known as D'Alembert rules \citep[e.g.,][]{Laskar1995,Morbidelli2002}. 
The rotational invariance requires $n + n^\prime + m + m^\prime - \bar{n} - \bar{n}^\prime - \bar{m} - \bar{m}^\prime + \ell + \ell^\prime = 0$, 
while from the planar symmetry it follows that the terms of the series $\widetilde{\mathcal{U}}_{\ell, \ell^\prime}$ are even with respect to the 
variables $(\mathcal{Y},\mathcal{Y}^\prime,\bar{\mathcal{Y}},\bar{\mathcal{Y}}^\prime)$.}. 
Following \citep{Laskar1995}, we have defined the dimensionless \Poincare's variables $\calX = x \sqrt{2/\Lambda}$, 
$\calY = y/\sqrt{2\Lambda}$ and the semi-major axis ratio $\alpha = a/a^\prime$, with $a < a^\prime$. 
The analytical expression of the coefficients $\Gamma^{\ell, \ell^\prime}_{\mathcal{N}}\!(\alpha)$, only depending on 
the semi-major axis ratio, is given in \citep{Laskar1995} in terms of Laplace coefficients. 
The indirect part of the two-body perturbation can also be expanded as a Fourier series in the mean longitudes, 
\begin{equation}
\label{eq:kinetic_expansion}
\mathcal{T}_1 = \frac{\tilde{\vec{r}} \cdot \tilde{\vec{r}}^\prime}{m_0} = 
\sum_{\ell, \ell^\prime \in \mathbb{Z}} 
\widetilde{\mathcal{T}}_{\ell, \ell^\prime}(\Lambda,\Lambda^\prime,\mathcal{X},\mathcal{X}^\prime,\mathcal{Y},\mathcal{Y}^\prime) 
\, \E^{j(\ell \lambda + \ell^\prime \lambda^\prime)},
\end{equation}
The computation of the Fourier coefficients $\widetilde{\mathcal{T}}_{\ell, \ell^\prime}$ is outlined in Appendix 
\ref{appendix:1}. Equations \eqref{eq:potential_expansion} and \eqref{eq:kinetic_expansion} allow to explicitly compute 
the Fourier expansion of the Hamiltonian perturbing function,
\begin{equation}
\label{eq:perturbation_expansion}
H_1 = 
\sum_{1 \leq k < \ell \leq N} \left( 
\frac{\tilde{\vec{r}}_k \cdot \tilde{\vec{r}}_\ell}{m_0} - G \frac{m_k m_\ell}{\Vert \vec{r}_k - \vec{r}_\ell \Vert}
\right) = 
\sum_{\vec{\ell} \in \mathbb{Z}^N} 
\widetilde{H}_{\vec{\ell}} \, \E^{j \vec{\ell} \cdot \vec{\lambda}},
\end{equation}
where $\vec{\lambda}$ stands for the vector of the planet mean longitudes, $\vec{\lambda} = (\lambda_1,\dots,\lambda_8)$, 
and the coefficients $\widetilde{H}_{\vec{\ell}}$ depend on all the remaining canonical variables. 

\subsection{Secular dynamics}
\label{subsect:sec_dyn}
The long-term dynamics of the solar system planets, in particular that of the inner ones, essentially consists of 
the slow precession of their perihelia and nodes, driven by secular, i.e. orbit-averaged, gravitational interactions 
\citep{Laskar1990,Laskar2004}. A secular Hamiltonian, describing such long-term motion, can be introduced in its simplest
form by searching for a change of variables that eliminate, at first order in the planet masses, all the harmonics with non-null 
wavevectors $\vec{\ell}$ in the Fourier expansion of the perturbation \eqref{eq:perturbation_expansion}. 
In canonical perturbation theory \citep{Hori1966,Deprit1969,Morbidelli2002}, this elimination is achieved through a 
canonical transformation defined as the time-1 flow of a generating Hamiltonian $S$ satisfying the homologic equation 
\begin{equation}
\label{eq:homologic_eq}
H_1 + \{S,H_0\} = \langle H_1 \rangle ,
\end{equation}
where the braces represent the Poisson bracket. The angle-bracket operator stands for averaging over 
the mean longitudes,
\begin{equation}
\langle \cdot \rangle = \frac{1}{(2\pi)^N} \int_{\mathbb{T}^N} d\vec{\lambda} \,\, \cdot \,\, ,
\end{equation}
with the integration defined over the hypertorus $\mathbb{T}^N$, at fixed values of all the
remaining canonical variables. This means that $\langle H_1 \rangle$ is the Fourier coefficient $\widetilde{H}_{\vec{0}}$ 
corresponding to the null harmonic in the expansion \eqref{eq:perturbation_expansion}. The homologic equation 
\eqref{eq:homologic_eq} gives the generating Hamiltonian $S$ as a formal Fourier series,
\begin{equation}
\label{eq:gen_func}
S = 
-j \sum_{\vec{\ell} \in \mathbb{Z}^N \backslash \{\vec{0}\}} 
\frac{\widetilde{H}_{\vec{\ell}}}{\vec{\ell} \cdot \vec{n}} 
\E^{j \vec{\ell} \cdot \vec{\lambda}} ,
\end{equation}
where $\vec{n} = \partial H_0/\partial\vec{\Lambda}$ is the vector of the planet mean motions. The secular Hamiltonian $\Hsec$ is 
formally given by the Lie transform generated by the function $S$ and applied to the Hamiltonian $H$, 
\begin{equation}
\label{eq:ham_sec_0}
\Hsec = \E^{L_S} H 
\Big|_{\hat{\vec{\Lambda}},\hat{\vec{\lambda}},\hat{\vec{x}},\hat{\vec{y}}} , \quad
\E^{\pm L_S} \, \cdot = \sum_{n=0}^{+\infty} \frac{(\pm 1)^n}{n!} L_S^n \, \cdot 
\end{equation}
where $L_S \cdot = \{S,\cdot\}$ is the Lie derivative associated to the generating Hamiltonian, $L_S^0$ is defined as the identity operator and 
$L^n_S \cdot = L_S L_S^{n-1} \cdot $ for $n \geq 1$. The Hamiltonian $\Hsec$ in Eq. \eqref{eq:ham_sec_0} is expressed in the 
new canonical variables $(\hat{\Lambda}_k,\hat{\lambda}_k;$ $\hat{x}_k, -j \hat{\bar{x}}_k;$ $\hat{y}_k, -j \hat{\bar{y}}_k)_{k=1,N}$, 
which we shall call the secular variables. They are related to the original ones via the Lie transforms 
\begin{equation}
\label{eq:sec_vars}
\Lambda_k = \E^{L_S} \hat{\Lambda}_k, 
\quad \lambda_k = \E^{L_S} \hat{\lambda}_k, 
\quad x_k = \E^{L_S} \hat{x}_k, 
\quad y_k = \E^{L_S} \hat{y}_k .
\end{equation}
The original variables are therefore the superposition of the secular ones and short-time oscillations generated by the $n\geq1$ terms 
of the Lie transforms. Since in the present study we are interested in the long-term dynamics of the planets, we shall only focus on 
the secular variables. Therefore, to keep simpler notation, we shall omit the hat on the secular variables from now on. 

Differently from the outer planets, the inner ones are not currently involved in strong mean-motion 
resonances. More precisely, a maximal contribution of only 0.07\arcsecyr to the fundamental 
precession frequencies of the inner orbits in the Laplace-Lagrange solution arises from the second order in the 
planet masses \citep[Table 8]{Laskar1985}. This same contribution is 0.9\arcsecyr for the outer planets, mainly due to the 
Great Inequality. 
Building on this fact, in the present work we choose to truncate the series \eqref{eq:ham_sec_0} at first order in the planet 
masses, i.e. we neglect quadratic and higher-order terms with respect to the Fourier coefficients $\widetilde{H}_{\vec{\ell}}$. 
Indeed, we expect the main contribution to the precession frequencies of the inner orbits to come from the linear terms, 
given the small masses of the inner planets. In absence of strong mean-motion resonances, in Eq. \eqref{eq:gen_func} 
the denominators $\vec{\ell} \cdot \vec{n}$ involving at least one inner planet are 
sufficiently far from zero. Under these assumptions, the 
higher-order terms only produce small corrections to the dynamics generated by the leading ones. 
Using the homologic equation \eqref{eq:homologic_eq}, one thus obtains 
\begin{equation}
\label{eq:ham_sec_1}
\Hsec = H_0 + \langle H_1 \rangle .
\end{equation}
The resulting secular Hamiltonian $\Hsec$ is simply the average of the N-body Hamiltonian $\eqref{eq:NbodyHam}$ over 
the planet mean longitudes. The averaging process is mathematically equivalent to replacing each planet by its 
instantaneous Keplerian orbit, with the corresponding mass distributed along it in a way that is inversely proportional 
to the local orbital speed of the planet. The secular dynamics is thus the slow gravitational interaction of such 
Keplerian rings. This equivalence was pointed out by \citet{Gauss1818} and it is thus referred to as Gauss' dynamics. 
As a result of the averaging over the mean longitudes, the $\Lambda_k$ variables are constant of motion
in the secular dynamics, and so are the semi-major axes of the planets. The resulting Hamiltonian system consists of two 
degrees of freedom for each planet\footnote{The entire planetary system is therefore described by 16 degrees of freedom, 
even reduced to 15 when taking into account the conservation of the total angular momentum.}, corresponding 
to the two \Poincare's complex variables $x_k$ and $y_k$. 


\paragraph{General relativity and minor effects.}
\label{sect:minor_effects}
The dynamical interactions described by the Hamiltonian \eqref{eq:NbodyHam} are not sufficient to finely reproduce the precession 
frequencies of the inner orbits; to this end, additional physical effects must be taken into account \citep{Laskar1999}. 
Indeed, it is well-known that general relativity contributes with 0.430\arcsecyr to the secular precession of Mercury perihelion. 
Such a correction is critical for the statistics of the long-term destabilization of the inner orbits \citep{Laskar2008,Laskar2009}, 
since it moves the system away from the $g_1-g_5$ secular resonance, which is responsible for the very high eccentricities of Mercury 
\citep{Laskar2008,Batygin2008,Boue2012}. We shall therefore include in the Hamiltonian \eqref{eq:ham_sec_1} the leading secular 
contribution of general relativity, which reads 
\begin{equation}
\label{eq:GR}
\Hsec_\textrm{GR} = 
\sum_{k=1}^N \frac{G^2 m_0^2 m_k}{c^2 a_k^2} \left( \frac{15}{8} -\frac{3}{\sqrt{1-e_k^2}} \right) ,
\end{equation}
where $c$ is the speed of light \citep[e.g.,][]{Saha1992,Mogavero2017}. 
We point out that the first term in the summation only depends on the semi-major axis $a_k$ and is thus a constant quantity in the 
secular dynamics. 
The next largest effect to be taken into account would be the Earth-Moon gravitational interaction, accounting for a 0.077\arcsecyr 
contribution to the secular perihelion precession of the Earth \citep{Laskar1999}. However, this is the order of magnitude 
of the contribution coming from the Hamiltonian terms at second order in the planet masses, which are neglected in the present model. 
Indeed, at degree 2 in planet eccentricities and inclinations, \citep[Table 8]{Laskar1985} reported contributions of 0.073\arcsecyr 
and 0.062\arcsecyr to the Earth-dominated and Mars-dominated eccentricity modes $g_3$ and $g_4$, respectively, the other 
frequency corrections being at least ten times smaller. Since the denominators $\vec{\ell} \cdot \vec{n}$ appearing 
in Eq. \eqref{eq:gen_func} and involving the inner planets are not close to zero, we expect the contributions from higher degrees to 
be generally smaller than those from degree 2. We thus choose to not include the Earth-Moon interaction in the present model, 
preferring to deal with a Hamiltonian whose dependence on eccentricities and inclinations is exact, as compared to a non-decisive 
increase of precision on the orbit precession frequencies. Moreover, we know that the N-body Hamiltonian \eqref{eq:NbodyHam}, corrected 
for general relativity, reproduces the maximum Lyapunov exponent of the inner solar system \citep{Rein2018}. 

Therefore, apart from irrelevant constant terms only depending on the semi-major axes 
in Eqs. \eqref{eq:Kepler_energy} and \eqref{eq:GR}, the secular Hamiltonian of 
the $8$ solar system planets considered in the present study reads 
\begin{equation}
\label{eq:ham_sec}
\Hsec = - \sum_{k=1}^8 G \frac{m_0 m_k}{a_k}
\left( 
\sum_{\ell=1}^{k-1} \frac{m_\ell}{m_0} \left< \frac{a_k}{\| \vec{r}_k - \vec{r}_\ell \|} \right> 
+ 3 \frac{G m_0}{c^2 a_k} \frac{1}{\sqrt{1-e_k^2}}
\right) 
\end{equation}

\subsection{Forced inner planets}
\label{subsect:forced_model}
As discussed in the Introduction, the secular Hamiltonian \eqref{eq:ham_sec} would not correctly reproduce the frequencies of 
the Jupiter and Saturn dominated eccentricity modes $g_5$ and $g_6$, respectively. Fortunately enough, the very 
small variations over billions of years of the precession frequencies of the outer orbits, compared to those of the inner one 
\citep{Laskar1990,Laskar2004,Hoang2021}, naturally suggest a way to make out of the Hamiltonian \eqref{eq:ham_sec} a very effective model for 
the inner system. This is achieved by choosing, once for all, an explicit quasi-periodic time dependence for the orbits of the giant
planets, i.e. by expressing the corresponding \Poincare's complex variables as finite Fourier series,
\begin{equation}
\label{eq:qp_decomposition}
\left.
\begin{aligned}
&x_k(t) = \sum_{\ell=1}^{M_k} \tilde{x}_{k\ell} \, 
\E^{j \vec{m}_{k\ell} \cdot \vec{\phi}(t)} \\
&y_k(t) = \sum_{\ell=1}^{N_k} \tilde{y}_{k\ell} \, 
\E^{j \vec{n}_{k\ell} \cdot \vec{\phi}(t)} 
\end{aligned}
\;
\right \}
\;
k \in \{5,6,7,8\} \,,
\end{equation}
where $t$ denotes the time, $\tilde{x}_{k\ell}$ and $\tilde{y}_{k\ell}$ are complex amplitudes, $\vec{m}_{k\ell}$ 
and $\vec{n}_{k\ell}$ integer vectors, and $\vec{\phi}(t) = \out{\vec{\omega}} t$, with $\out{\vec{\omega}}$ the vector 
of the (constant) fundamental precession frequencies of the outer orbits, denoted as $\out{\vec{\omega}} = (g_5,g_6,g_7,g_8,s_6,s_7,s_8)$\footnote{The 
frequency of the Jupiter-dominated inclination mode $s_5$ is null because the total angular momentum of the system 
is conserved.} \citep{Laskar1990}. The number of harmonics $M_k, N_k$ appearing 
in the decompositions depends on the planet. 
When such predetermined time dependence is injected in Eq. \eqref{eq:ham_sec}, one obtains the Hamiltonian $\Hiss$ 
of a \emph{forced secular inner solar system},
\begin{equation}
\label{eq:ham_sec_inn}
\begin{aligned}
\Hiss[(x_k,y_k)_{k=1,4},t] \! = \! 
\Hsec[(x_k,y_k)_{k=1,4},(x_k \!\! = \!\! x_k(t),y_k \!\! = \!\! y_k(t))_{k=5,8}] .
\end{aligned}
\end{equation}
The explicit time dependence in the Hamiltonian $\Hiss$ physically means that the inner planets interact with each other 
while moving in an external gravitational potential generated by the giant planets\footnote{The terms in $\Hsec$ only 
involving outer planet variables can be discarded in Eq. \eqref{eq:ham_sec_inn}, as they just depend on time and do not 
affect the inner planet dynamics anymore.}. The inner planets thus constitute an open system and the corresponding dynamics 
does not possess any fundamental integral of motion, such as the energy or angular momentum. As a result of the predetermination 
of the outer orbits, in addition to the explicit time dependence, the Hamiltonian $\Hiss$ possesses 8 degrees of freedom. 
As already stated, compared to the second-order secular system of \citep{Laskar1984,Laskar1985}, 
the present model neglects terms of order higher than one in the planet masses. Nevertheless, the precession frequencies of the 
outer orbits can be set to very precise values in Eq. \eqref{eq:qp_decomposition}. Moreover, the Hamiltonian \eqref{eq:ham_sec_inn} is 
not truncated neither in eccentricities nor in inclinations, so that the range of validity of its dynamics extends to very excited states, 
provided that the Keplerian orbits of the planets do not cross each other. This makes the present model perfectly suited to explore 
the very long-term evolution of the inner system, even when high eccentric and inclined orbits become statistically recurrent. 

\begin{table*}
\caption{Average wall-clock time on an Intel(R) Core(TM) i7-7700T CPU @ 2.90GHz for the numerical integration 
of the forced inner system according to the total degree of truncation of the Hamiltonian in eccentricities and inclinations 
($\Hiss_{2n}$ is given in Eq. \eqref{eq:ham_iss_expansion}, while $\Hiss$ in Eq. \eqref{eq:ham_sec_inn}).} 
\label{tab:integ_times}
\centering
\begin{tabular}{c c c c c | c} 
\hline\hline 
\rule{0pt}{1em} 
$\Hiss_{2n}$ & 4 & 6 & 8 & 10 & $\Hiss$ \\ 
\hline
100-Myr orbital solution & 1.3 s & 7.5 s & 35 s & 4 m 01 s & 2 m 48 s  \\ 
\hline
\end{tabular}
\end{table*}

\subsection{Construction of the outer planet solution. Initial conditions}
\label{subsect:init_cond}
The quasi-periodic form of the outer orbits in Eq. \eqref{eq:qp_decomposition} is established numerically and explicitly reported in Appendix \ref{appendix:QPSO}. 
The full equations of motion of the main bodies of the solar system are numerically integrated over 30 Myr in the future, following the 
comprehensive model of \citep{Laskar2011}. The initial conditions of the integration were adjusted through least squares to the high 
precision planetary ephemeris INPOP13b \citep{Fienga2014a,Fienga2014b} extending over 1 Myr. Throughout the paper, we shall refer to this 
direct numerical integration as \LaX{}. Through frequency analysis \citep{Laskar1988,Laskar1992b,Laskar1993,Laskar2005} 
of the orbital solution, the leading secular harmonics of the dimensionless \Poincare's variables of the
outer planets $(\mathcal{X}_k,\mathcal{Y}_k)_{k=5,8}$ are extracted to construct the Fourier series \eqref{eq:qp_decomposition}, as in 
\citep{Laskar1988,Laskar1990}. This is performed up to a numerical precision that does not allow anymore to recognize in an unambiguous 
way new harmonics as linear combination of the fundamental frequencies $\out{\vec{\omega}}$. The precision of the outer planet 
solution is shown in Table \ref{table:QPSO_diffs} of Appendix \ref{appendix:QPSO}. 
We report there the root mean square of the dimensionless \Poincare's variables of the outer planets 
in the \LaX{} solution, and that of the corresponding residuals after subtraction of their quasi-periodic decomposition 
from Eq. \eqref{eq:qp_decomposition} and filtering out of the non-secular Fourier components. The effectiveness of 
the quasi-periodic approximation is illustrated in Figs. \ref{fig:resX} and \ref{fig:resY}. 

The choice of the secular semi-major axes and that of the initial conditions for the secular \Poincare's variables of the inner planets 
deserve a discussion. In principle, the initial secular variables have to 
be computed by inverting the Lie transforms in Eq. \eqref{eq:sec_vars}. They should not be set to the initial value of the 
corresponding original variable, as this could cause an offset in the secular frequencies of the motion, due to the short-time 
(high-frequency) oscillations generated by the Lie transform. This is a general rule when constructing averaged dynamical systems 
\citep[e.g.,][]{LaskarSimon1988}. To avoid the explicit computation of the generating function $S$ in Eq. \eqref{eq:gen_func} 
and the related Lie transforms, appropriate initial conditions for the secular solar system can be effectively computed 
by filtering out the short-time components of the numerical solution of the non-averaged system, as it is done in the present study.
The constant term in the frequency analysis of the variables $\Lambda_k$, performed on the \LaX{} solution, provides the 
value of the secular semi-major axes of the planets (Table \ref{table:CI}). Moreover, a polynomial expansion in time 
over the first few thousand years of the same solution, once filtered, provides the nominal initial conditions for the secular 
\Poincare's complex variables of the inner planets (Table \ref{table:CIXY}). 

\section{Numerical integration. Gauss's method in Hamiltonian formalism}
\label{sect:gauss_method}
The dimensionless Hamilton's equations for the forced inner system \eqref{eq:ham_sec_inn} read
\begin{equation}
\label{eq:eom_iss}
\dot{\mathcal{X}}_k = - \frac{2j}{\Lambda_k} \frac{\partial \Hiss}{\partial \bar{\mathcal{X}}_k}, \quad
\dot{\mathcal{Y}}_k = - \frac{j}{2 \Lambda_k} \frac{\partial \Hiss}{\partial \bar{\mathcal{Y}}_k} ,
\end{equation}
for $k \in \{1,2,3,4\}$. Because of the averaging over the mean longitudes in Eq. \eqref{eq:ham_sec}, 
the above equations include double integrals of the form 
\begin{align}
\label{eq:gauss_double_integral_1}
&\frac{\partial}{\partial \bar{\mathcal{Z}}} 
\left< \frac{1}{\| \vec{r} - \vec{r}^\prime \|} \right> 
= \frac{1}{2\pi} \int_0^{2\pi} \vec{f} \cdot \frac{\partial \vec{r}}{\partial \bar{\mathcal{Z}}} d\lambda , \\
\label{eq:gauss_double_integral_2}
&\vec{f} = \frac{1}{2\pi} \int_0^{2\pi} - \frac{\vec{r} - \vec{r}^\prime}{\| \vec{r} - \vec{r}^\prime \|^3} d\lambda^\prime ,
\end{align}
where $\mathcal{Z}$ stands for $\mathcal{X}$ or $\mathcal{Y}$, alternatively.
The equations of motion \eqref{eq:eom_iss} are thus non-algebraic, and the numerical computation of the derivatives 
is in principle much more complex than in N-body dynamics or in polynomial secular systems as in \citep{Laskar1985,Laskar1990}. 
Nevertheless, it is well-known that the Eqs. \eqref{eq:eom_iss} can be integrated numerically in a very 
efficient way through the so-called Gauss's method \citep{Gauss1818,Bour1855,Hill1882}. Indeed, the vector $\vec{f}$ 
in Eq. \eqref{eq:gauss_double_integral_2} is proportional to the gravitational force exerted on a test particle by a Keplerian 
ring, and can be analytically expressed in terms of complete Legendre's elliptic integrals \citep[e.g.,][Chapter 19]{NIST:DLMF}. This 
was first shown by \citet{Gauss1818}, who introduced at same time the arithmetic-geometric mean to numerically evaluate 
such special functions in a few elementary iterations and with high precision. Building on its modern derivation in 
\citep{Musen1970}, we implemented Gauss's method into the present complex Hamiltonian formalism \eqref{eq:eom_iss}. 
This allows, in particular, to eliminate the degeneracy typically appearing at $e=0$ (circular orbits) and $i=0$ (equatorial 
orbits) \citep[e.g.,][]{Fouvry2020}, which is fundamental to guarantee high numerical precision.
We employ a state-of-the-art algorithm, based on piecewise minimax rational function approximation, to numerically compute 
the complete elliptic integrals in double floating-point precision at the cost of elementary functions \citep{Fukushima2015}. 
The remaining simple integral in Eq. \eqref{eq:gauss_double_integral_1} is then effectively evaluated via the 
trapezoidal rule, which converges exponentially fast with the number of function evaluations, because of the periodicity 
of the integrand \citep[e.g.,][]{Touma2009}. We apply the trapezoidal rule in an adaptive way, by doubling the number of function evaluations 
until the estimated relative error on the integral is smaller than $10^{-12}$. It should be noted that, as the convergence 
of the numerical integral is exponential, the resulting error is often orders of magnitude smaller than this tolerance.

We integrate the Eqs. \eqref{eq:eom_iss} using an Adams PECE method of order 12 and a conservative timestep of 250 years, 
as in \citep{Laskar1994}. In absence of any fundamental integral of motion (see Sect. \ref{subsect:forced_model}), we 
estimate the integration error after a time $T$ following \citep{Laskar1994}. Starting from the nominal initial conditions, 
we integrate the dynamics over the time interval [0, $T$/2] and then backwards to the initial time. By denoting the deviations 
from the initial coordinates of the system in the phase space as $(\delta\mathcal{X}_k(T), \delta\mathcal{Y}_k(T))_{k=1,4}$, 
we define the relative integration error as 
\begin{equation}
\label{eq:integ_error}
\begin{aligned}
\delta(T) = \max_{k \in \{1,\dots,4\}} \Bigg\{ 
&\left| \frac{\operatorname{Re}[\delta\mathcal{X}_k(T)]}{\operatorname{Re}[\mathcal{X}_k(0)]} \right|, 
\left| \frac{\operatorname{Im}[\delta\mathcal{X}_k(T)]}{\operatorname{Im}[\mathcal{X}_k(0)]} \right|, \\
&\left| \frac{\operatorname{Re}[\delta\mathcal{Y}_k(T)]}{\operatorname{Re}[\mathcal{Y}_k(0)]} \right|, 
\left| \frac{\operatorname{Im}[\delta\mathcal{Y}_k(T)]}{\operatorname{Im}[\mathcal{Y}_k(0)]} \right|  
\Bigg\} ,
\end{aligned}
\end{equation}
where $(\mathcal{X}_k(0), \mathcal{Y}_k(0))_{k=1,4}$ are the initial conditions of the system. Over the time interval $T \in$ 
[0, 20] Myr, the average integration error is well described by the power law $\delta(T) = \delta_0 [1+(T/T_0)^\alpha]$, 
with $\delta_0 = 9 \cdot 10^{-15}$, $T_0 = 0.6$ Myr and $\alpha = 1.2$. Therefore, one has $\delta(10 \, \mathrm{Myr}) \approx 3 
\cdot 10^{-13}$, and $\delta(10 \, \mathrm{Gyr}) \approx 10^{-9}$ if one extrapolates by ignoring the chaotic behaviour of 
the solution. The integration error thus turns out to be similar to that of the secular system in \citep{Laskar1994}. 
We emphasize that the present fixed-timestep integration scheme allows to reach very highly excited orbital states. Indeed, 
violations of the precision goal on the integral $\eqref{eq:gauss_double_integral_1}$ typically occurs only when the system is  
already close to the first intersection of the instantaneous Keplerian orbits of two planets (see Sect. \ref{sect:evolution_5Gyr}), and are 
related to an intrinsic discontinuity in the equations of motion \eqref{eq:eom_iss} at orbit crossing \citep[e.g.,][]{Touma2009}. 

From a computational perspective, the choice of a multistep method is well adapted to the present dynamical system, as 
the derivative evaluation constitutes by far the most expensive step of the integration scheme. 
Once implemented in the C programming language, Gauss's method allows to compute 
100-Myr orbital solutions of the forced inner planets in a few minutes on a PC, as shown in Table \ref{tab:integ_times}. 
Therefore, the computational cost turns out to be comparable to that of the secular system of \citet{Laskar1990}, 
in spite of the different complexity of the corresponding Hamiltonians. 

The rather involved analytical derivations adapting Gauss's method to the present complex Hamiltonian formalism shall 
be the subject of a forthcoming paper \citep{Mogavero2021}, along with the release of the C program implementing it.

\section{Analytical expansion. Computer algebra}
\label{sect:ham_expansion}
In addition to its fast numerical integration, the present dynamical model allows for a systematic 
analytical development of its Hamiltonian by means of computer algebra. In this study we employ TRIP, 
a computer algebra system dedicated to perturbation series, specially those of celestial 
mechanics \citep{Laskar1990b,Gastineau2011,TRIP2020}. The main objects of its symbolic kernel are the Poisson series, 
i.e. multivariate Fourier series whose coefficients are multivariate Laurent series, 
\begin{equation}
\label{eq:poisson_series}
S(z_1, \dots, z_n, \varphi_1, \dots, \varphi_m) = \sum C_{k,\ell} z_1^{k_1} \cdots z_n^{k_n} 
\E^{j \left(\ell_1 \varphi_1 + \dots + \ell_m \varphi_m \right)} ,
\end{equation}
where $(z_p)_{p=1,n}$ and $(\varphi_p)_{p=1,m}$ are complex and real variables, respectively, 
$k = (k_p)_{p=1,n} \in \mathbb{Z}^n$, $\ell = (\ell_p)_{p=1,m} \in \mathbb{Z}^m$ and $C_{k,\ell}$ are
complex coefficients. 

Following \citep{Laskar1995}, the smallness of eccentricities and inclinations in the solar 
system can be exploited to develop the averages appearing in Eq. \eqref{eq:ham_sec} as formal series in  
\Poincare's variables $(x_k,\bar{x}_k,y_k,\bar{y}_k)_{k=1,8}$. Indeed, Eq. \eqref{eq:potential_expansion} gives 
\begin{equation}
\label{eq:disturb_func_expansion}
\left< \frac{a^\prime}{\| \vec{r}^\prime - \vec{r} \|} \right> = 
\sum \Gamma^{0,0}_{\mathcal{N}}(\alpha) 
\calX^n {\calX^\prime}^{n^\prime} \bar{\calX}^{\bar{n}} {\bar{\calX}}^{\prime \bar{n}^\prime} 
\calY^m {\calY^\prime}^{m^\prime} \bar{\calY}^{\bar{m}} {\bar{\calY}}^{\prime \bar{m}^\prime} .
\end{equation}
The rotational invariance of the secular Hamiltonian \eqref{eq:ham_sec} requires 
$n + n^\prime + m + m^\prime - \bar{n} - \bar{n}^\prime - \bar{m} - \bar{m}^\prime = 0$, 
while from the planar symmetry it follows that $m + m^\prime + \bar{m} + \bar{m}^\prime$ is an even integer. 
These relations imply that the monomials in Eq. \eqref{eq:disturb_func_expansion} are even 
with respect to both $(x,x^\prime,\bar{x},\bar{x}^\prime)$ and $(y,y^\prime,\bar{y},\bar{y}^\prime)$ variables. 
The relativistic terms in Eq. \eqref{eq:ham_sec} are readily expanded, 
\begin{equation}
\label{eq:relat_expansion}
\frac{1}{\sqrt{1-e^2}} = \frac{1}{1 - x \bar{x}/\Lambda} = \sum_{p=0}^\infty \left( \frac{x \bar{x}}{\Lambda} \right)^p .
\end{equation}
By truncating the series \eqref{eq:disturb_func_expansion} and \eqref{eq:relat_expansion} at a given total degree 
$2n$ ($n \in \mathbb{N}_0$) in \Poincare's variables $(x_k,\bar{x}_k,y_k,\bar{y}_k)_{k=1,8}$, and after 
substitution in Eq. \eqref{eq:ham_sec}, one obtains a polynomial secular Hamiltonian for the ensemble of the solar 
system planets, 
\begin{equation}
\label{eq:ham_sec_expansion}
\Hsec_{2n} = 
\sum_{p = 0}^n \Hsec_{(2p)} \,,
\end{equation}
where $\Hsec_{(2p)}$ groups all the monomials of same total degree $2p$. 
Practically, the expansion \eqref{eq:ham_sec_expansion} is readily obtained in TRIP, which implements 
the algorithm of \citep{Laskar1995} to compute a truncation of the series \eqref{eq:disturb_func_expansion}. 
The truncated Hamiltonian \eqref{eq:ham_sec_expansion} is in the form of a Poisson series \eqref{eq:poisson_series} 
in \Poincare's complex variables, with no angular dependencies. The crucial point here is that its coefficients only depend on the 
planet masses and secular semi-major axes, which are constant parameters in the present dynamics. They can be thus 
numerically evaluated once for all, to obtain very compact series. This is shown in Table \ref{tab:number_monomials}, 
which gives the total number of monomials in \eqref{eq:ham_sec_expansion} with respect to the degree of truncation, 
along with the disk usage of the corresponding series when stored in a plain-text file, and their typical computation time in TRIP. 

The expansion \eqref{eq:ham_sec_expansion} straightforwardly provides a truncated Hamiltonian for the forced 
inner system \eqref{eq:ham_sec_inn}, 
\begin{equation}
\label{eq:ham_iss_expansion}
\begin{aligned}
&\Hiss_{2n} = 
\sum_{p = 1}^n \Hiss_{(2p)} \,, \\
&\Hiss_{(2p)}[(x_k,y_k)_{k=1,4},t] \! = \! 
\Hsec_{(2p)}[(x_k,y_k)_{k=1,4},(x_k \!\! = \!\! x_k(t),y_k \!\! = \!\! y_k(t))_{k=5,8}] .
\end{aligned}
\end{equation}
The computation of the equations of motion for the truncated Hamiltonian \eqref{eq:ham_iss_expansion} only requires 
to take the derivatives of a multivariate polynomial, and can be thus systematically performed in TRIP. We put the resulting 
polynomials in Horner form, to ensure speed and stability of their numerical evaluation, and wrap them into a C code to 
achieve the best computational performance\footnote{Numerical integration of ODEs is also available directly in 
TRIP.}. The equations of motion are then integrated through the Adams PECE method of order 12 employed for Gauss's dynamics, 
with the same timestep of 250 years. The typical computational cost of a 100-Myr orbital solution is shown in Table 
\ref{tab:integ_times}, according to the degree of truncation. In light of the pertinence of the forced inner system 
that will be shown in Sect. \ref{sect:comparison}, the truncated Hamiltonian \eqref{eq:ham_iss_expansion}, at degrees 4 
and 6 in particular, constitutes the state of the art of very fast, still realistic, dynamical models of the inner solar system. 

\begin{table*}
\caption{Number of monomials of variables $(x_k,\bar{x}_k,y_k,\bar{y}_k)_{k=1,8}$ in $\Hsec_{2n}$ 
(Eqs.~\ref{eq:poincare_vars}, \ref{eq:ham_sec_expansion}) 
and of variables $\{(u_k, \bar{u}_k, v_k, \bar{v}_k)_{k=1,4},(\E^{j \phi_k(t)})_{k=1,7}\}$ in $\Huv_{2n}$ 
(Eqs.~\ref{eq:proper_modes_u}, \ref{eq:proper_modes_v}, \ref{eq:ham_sec_inn_uv}) 
according to the truncation degree. It also shows the disk space in Bytes occupied by 
the series, their computation time on an Intel(R) Xeon(R) CPU E5-2698 v3 @ 2.30GHz (CPU time in parentheses) and 
the number of Fourier harmonics in $\Huv_{2n}$ according Eq. \eqref{eq:fourier_expansion}.} 
\label{tab:number_monomials}
\centering
\begin{tabular}{c c c c c c} 
\hline\hline 
\rule{0pt}{0.9em} & 2 & 4 & 6 & 8 & 10 \\ 
\hline
\multicolumn{1}{c}{\rule{0pt}{1.075em} $\Hsec_{2n}$} \\
\hline
Monomials & 129 & 1\,345 & 6\,561 & 23\,213 & 66\,253 \\ 
Disk usage & 21 kB & 215 kB & 1.0 MB & 3.7 MB & 10.6 MB \\ 
Wall-clock time & 2.4 s (137\%) & 2.5 s (129\%) & 3.0 s (154\%) & 5.7 s (187\%) & 17.1 s (272\%) \\ 
\hline
\multicolumn{1}{c}{\rule{0pt}{0.9em}$\Huv_{2n}$} \\
\hline
Monomials & 8 & 6\,304 & 188\,024 & 3\,394\,892 & 42\,817\,100 \\ 
Disk usage & -- & 0.9 MB & 29.5 MB & 547.6 MB & 7.1 GB \\ 
Wall-clock time & -- & 0.15 s (447\%) & 3.4 s (773\%) & 1 m 16 s (1038\%) & 20 m 37 s (1343\%) \\
Harmonics & 1 & 2\,748 & 69\,339 & 1\,029\,137 & 10\,279\,581 \\ 
\hline
\end{tabular}
\end{table*}

\subsection{Forced Laplace-Lagrange dynamics}
For small eccentricities and inclinations, the secular Hamiltonian  \eqref{eq:ham_sec} is a perturbation to the integrable 
Laplace-Lagrange (LL) problem. Indeed, its truncation at degree 2 in \Poincare's variables gives the 
quadratic form 
\begin{equation}
\label{eq:ham_LL}
\HLL = \Hsec_{(2)} = \vec{x}^\dagger \vec{M} \vec{x} + \vec{y}^\dagger \vec{N} \vec{y} ,
\end{equation}
where we have defined the column vectors $\vec{x} = (x_1, \dots, x_8)^T$ and $\vec{y} = (y_1, \dots, y_8)^T$,
$T$ is the transposition operator and the dagger stands for Hermitian transposition, i.e. $\vec{x}^\dagger = \bar{\vec{x}}^T$. The matrices
$\vec{M}$ and $\vec{N}$ are real and symmetric. Since the variables $\vec{x}$ and $\vec{y}$
are not coupled in the LL Hamiltonian, the following derivations focus on the degrees of freedom related to $\vec{x}$.
A similar treatment holds for the $\vec{y}$ variables.

After substitution of the time dependence of the outer planet variables given in Eq. \eqref{eq:qp_decomposition}, one obtains
\begin{equation}
\HissLL{x} =  
\vec{x}^\dagger \vec{M} \vec{x} \Big|_{(x_k=x_k(t))_{k=5,8}} = 
\left( \inn{\vec{x}}^\dagger, \out{\vec{x}}^\dagger(t) \right) 
\begin{pmatrix}
\ii{\vec{M}} & \subscript{\vec{M}}{io} \\
\subscript{\vec{M}}{io}^T & \oo{\vec{M}}
\end{pmatrix} 
\begin{pmatrix}
\inn{\vec{x}} \\
\out{\vec{x}}(t)
\end{pmatrix} ,
\end{equation}
where we have defined the inner and outer planet column vectors, $\inn{\vec{x}} = (x_1,x_2,x_3,x_4)^T$ and 
$\out{\vec{x}}(t) = (x_5(t),x_6(t),x_7(t),x_8(t))^T$, respectively, and the real matrix $\vec{M}$ has been 
written as a block matrix, with $\ii{\vec{M}}$ and $\oo{\vec{M}}$ symmetric $4 \times 4$ matrices. By discarding terms
only depending on time, the Hamiltonian reads 
\begin{equation}
\label{eq:hamiss_LL}
\HissLL{x} = 
\inn{\vec{x}}^\dagger \ii{\vec{M}} \inn{\vec{x}} 
+  2 \operatorname{Re}\left( \inn{\vec{x}}^\dagger \subscript{\vec{M}}{io} \out{\vec{x}}(t) \right) ,
\end{equation}
where $\operatorname{Re}$ stands for the real part of a complex quantity. The Hamiltonian \eqref{eq:hamiss_LL} corresponds to a forced 
Laplace-Lagrange system. Since it is quadratic in the inner vector $\inn{\vec{x}}$, one can analytically 
solve the corresponding equations of motion and introduce appropriate action-angle variables.

\subsubsection{Solution of the equations of motion}
The matrix $\ii{\vec{M}}$ is real and symmetric and can thus be diagonalized through an orthogonal 
matrix $\matrOD{O}{M}$,
\begin{equation}
\ii{\vec{M}} = \matrOD{O}{M} \matrOD{D}{M} \matrOD{O}{M}^T .
\end{equation}
The columns of $\matrOD{O}{M}$ are the eigenvectors of $\ii{\vec{M}}$, while the diagonal entries of 
$\matrOD{D}{M} = -\mathrm{diag}(\vec{\gLL})$ are the corresponding real eigenvalues\footnote{The minus sign in the 
definition of the matrix $\matrOD{D}{M}$ is such that the precession frequencies of the perihelia $\dot{\vec{\varpi}}$ have the 
same sign as $\vec{\gLL}$.}, $\vec{\gLL}$ being a column vector. The orthogonal matrix $\matrOD{O}{M}$ induces the 
canonical change of variables\footnote{In the present notation, the canonical momenta 
constitute column vectors, while the coordinates are wrapped in row vectors.} 
$(\inn{\vec{x}},-j {\inn{\vec{x}^\dagger}}) \rightarrow (\innxprime,-j {\innxprime}^\dagger)$ defined as 
\begin{equation}
\innxprime = \matrOD{O}{M}^T \inn{\vec{x}} .
\end{equation}
The transformed Hamiltonian reads
\begin{equation}
\label{eq:hamiss_LL_xprime}
\HissLL{x} =
{\innxprime}^\dagger \matrOD{D}{M} \innxprime 
+ 2 \operatorname{Re} \left( {\innxprime}^\dagger \subscript{\vec{M}}{io}^\prime \out{\vec{x}}(t) \right) ,
\end{equation}
with $\subscript{\vec{M}}{io}^\prime = \matrOD{O}{M}^T \subscript{\vec{M}}{io}$.
The corresponding Hamilton's equations are given by
\begin{equation}
\label{eq:forcLLeqs}
\inn{\vec{\dot{x}}}^\prime = -j \frac{\partial \HissLL{x}}{\partial {\innxprime}^\dagger} = 
-j \left( \matrOD{D}{M} \innxprime + \subscript{\vec{M}}{io}^\prime \out{\vec{x}}(t) \right) ,
\end{equation}
and constitute a first-order inhomogeneous matrix ordinary differential equation. The general solution can 
be written as  
\begin{equation}
\label{eq:LLsolution}
\innxprime(t) = \free{\vec{x}}^\prime(t) + \forc{\vec{x}}^\prime(t) .
\end{equation}
The free solution $\free{\vec{x}}^\prime(t)$ is the general integral to the associated homogeneous equation 
$\inn{\vec{\dot{x}}}^\prime = -j \matrOD{D}{M} \innxprime$, representing the autonomous perihelia precession 
of the inner orbits, while the forced solution $\forc{\vec{x}}^\prime(t)$ is a particular integral 
of the complete Eq. \eqref{eq:forcLLeqs}, which arises from the gravitational forcing of the outer planets.
We define, once for all, the forced solution to be:
\begin{equation}
\label{eq:forc_sol_def}
\forc{\vec{x}}^\prime(t) = 
-j \, \E^{-j t \matrOD{D}{M}} \int^t d\tau \, \E^{j \tau \matrOD{D}{M}} \subscript{\vec{M}}{io}^\prime \out{\vec{x}}(\tau) .
\end{equation}
By employing the decomposition given in Eq. \eqref{eq:qp_decomposition}, the components of the forced solution are 
\begin{equation}
\label{eq:forc_sol}
(\forc{\vec{x}}^\prime)_k = 
\sum_{\ell=5}^8 \sum_{p=1}^{M_\ell}
\frac{(\subscript{\vec{M}}{io}^\prime)_{k \ell} \tilde{x}_{\ell p}}{(\vec{\gLL})_k - \vec{m}_{\ell p} \cdot \out{\vec{\omega}}} 
\E^{j \vec{m}_{\ell p} \cdot \vec{\phi}}
\quad k \in \{1,2,3,4\} .
\end{equation}
The (constant) denominators appearing in Eq. \eqref{eq:forc_sol} are far from zero, as the inner planets are not 
involved in the corresponding secular resonances. The forced solution is thus well defined.

\subsubsection{Proper modes}
The following derivation shows that there exists a canonical transformation depending on time,
$(\innxprime,-j {\innxprime}^\dagger) \rightarrow (\vec{u},-j \vec{u}^\dagger)$, 
such that the transformed LL Hamiltonian reads $\HuvLL{u} = \vec{u}^\dagger \matrOD{D}{M} \vec{u}$. 
As in the case of an autonomous Laplace-Lagrange system, the new canonical variables $\vec{u}$ will 
physically correspond to the free part $\free{\vec{x}}^\prime(t)$ of the solution in 
Eq. \eqref{eq:LLsolution} \citep[e.g.,][]{Morbidelli2002}. We thus begin by defining a new set of variables 
$\vec{u}$ such that 
\begin{equation}
\label{eq:u_vars}
\innxprime = \vec{u} + \forc{\vec{x}}^\prime(t) .
\end{equation}
By using the fact that $\forc{\vec{x}}^\prime(t)$ is a solution to Eq. \eqref{eq:forcLLeqs}, and 
discarding terms only depending on time, the forced LL Hamiltonian in Eq. \eqref{eq:hamiss_LL_xprime} 
can be written as 
\begin{equation}
\label{eq:hamiss_LL_u}
\HissLL{x} = 
\vec{u}^\dagger \matrOD{D}{M} \vec{u}
+ j \vec{u}^\dagger \forc{\dot{\vec{x}}}^\prime(t)
- j {\forc{\dot{\vec{x}}}}^{\prime \dagger}(t) \innxprime .
\end{equation}
We now ask that the change of variables 
$(\innxprime,-j {\innxprime}^\dagger) \rightarrow (\vec{u},-j \vec{u}^\dagger)$ 
derive from a time-depending generating function $F(\innxprime, -j \vec{u}^\dagger, t)$, satisfying
\begin{equation}
\frac{\partial F}{\partial t} = 
- j \vec{u}^\dagger \forc{\dot{\vec{x}}}^\prime(t)
+ j {\forc{\dot{\vec{x}}}}^{\prime \dagger}(t) \innxprime ,
\end{equation}
so that the last two terms in Eq. \eqref{eq:hamiss_LL_u} shall be killed in the transformed Hamiltonian.
By integrating the above equation with respect to time, one obtains
\begin{equation}
F(\innxprime, -j \vec{u}^\dagger, t) = 
- j \vec{u}^\dagger \forc{\vec{x}}^\prime(t)
+ j \forc{\vec{x}^{\prime \dagger}}(t) \innxprime 
+ f(\innxprime, \vec{u}^\dagger) ,
\end{equation}
where $f(\innxprime, \vec{u}^\dagger)$ is an unknown function, which has to be determined. 
Since the generating function $F$ depends on the old momenta $\innxprime$ and the new coordinates 
$- j \vec{u}^\dagger$, it must verify the relations \citep[e.g.,][]{Landau} 
\begin{equation}
\frac{\partial F}{\partial \innxprime} = j {\innxprime}^\dagger, 
\quad \frac{\partial F}{\partial \vec{u}^\dagger} = j \vec{u} ,
\end{equation}
which imply 
\begin{equation}
j \forc{\vec{x}^{\prime \dagger}}(t) + \frac{\partial f}{\partial \innxprime} = j {\innxprime}^\dagger, 
\quad - j \forc{\vec{x}}^\prime(t) + \frac{\partial f}{\partial \vec{u}^\dagger} = j \vec{u} .
\end{equation}
With the choice $f(\innxprime, \vec{u}^\dagger) = j \vec{u}^\dagger \innxprime$, these relations are both 
equivalent to Eq. \eqref{eq:u_vars}. 

The previous derivations imply that the variable transformation 
$(\inn{\vec{x}},-j {\inn{\vec{x}}}^\dagger) \rightarrow 
(\innxprime,-j {\innxprime}^\dagger) \rightarrow 
(\vec{u},-j \vec{u}^\dagger)$ 
is canonical, with 
\begin{align}
\label{eq:proper_modes_u}
&\vec{u} = \matrOD{O}{M}^T \inn{\vec{x}} - \forc{\vec{x}}^\prime(t), \\ 
&\HuvLL{u} = 
\HissLL{x} + \frac{\partial F}{\partial t} = \vec{u}^\dagger \matrOD{D}{M} \vec{u} .
\end{align}
We call the complex variables $\vec{u}$ the \emph{proper modes} of the \Poincare's variables $\inn{\vec{x}}$. 
In the forced Laplace-Lagrange dynamics, they simply rotate in the complex plane, 
with constant angular frequencies given by $\vec{\gLL}$. According to Eq. \eqref{eq:proper_modes_u}, 
the corresponding \Poincare's variables thus result in a superposition of such independent harmonic oscillations 
and those arising from the forcing of the outer planets. When higher-degree terms of the Hamiltonian \eqref{eq:ham_iss_expansion} 
are taken into account, the dynamics of the proper modes becomes coupled, and their frequency spectrum is no more monochromatic. 
Physically, the proper modes $\vec{u}$ correspond to the variables $(z^*_k)_{k=1,4}$ defined in \citep{Laskar1990}. 
Nevertheless, it is important to emphasize that their mathematical definitions do not coincide. Indeed, the transformation 
matrix $\matrOD{O}{M}$ do not include contributions at the second order in planet masses, as that considered in 
\citep{Laskar1990} does. Moreover, the transformation \eqref{eq:proper_modes_u} is nonlinear in the forcing 
of the outer planets, in the sense that harmonics of order\footnote{We define the order of a harmonic 
as the 1-norm of its wavevector, i.e. $|\vec{m}| = \sum_{\ell=1}^7 |m_\ell|$.} 
$|\vec{m}|$ greater than one are present in the decomposition \eqref{eq:qp_decomposition} (see Table \ref{table:tabX}) 
and thus in the forced solution \eqref{eq:forc_sol}. 

Similar derivations allow to define the proper modes $\vecv$ of the \Poincare's variables $\inn{\vec{y}}$, by means of a 
time-dependent canonical change of variables 
$(\inn{\vec{y}},-j {\inn{\vec{y}}}^\dagger) \rightarrow (\vecv,-j \vecv^\dagger)$, 
\begin{align}
\label{eq:proper_modes_v}
&\vecv = \matrOD{O}{N}^T \inn{\vec{y}} - \forc{\vec{y}}^\prime(t), \\
&\HuvLL{v} = \vecv^\dagger \matrOD{D}{N} \vecv .
\end{align}
We have defined 
\begin{equation}
\vec{N} = 
\begin{pmatrix}
\ii{\vec{N}} & \subscript{\vec{N}}{io} \\
\subscript{\vec{N}}{io}^T & \oo{\vec{N}}
\end{pmatrix} , \quad 
\ii{\vec{N}} = \matrOD{O}{N} \matrOD{D}{N} \matrOD{O}{N}^T ,
\end{equation}
the columns of $\matrOD{O}{N}$ being the eigenvectors of $\ii{\vec{N}}$, while the diagonal entries of 
$\matrOD{D}{N} = -\mathrm{diag}(\vec{\sLL})$ are the corresponding real eigenvalues. Using the quasi-periodic decomposition in Eq. \eqref{eq:qp_decomposition}, the components of the forced solution $\forc{\vec{y}}^\prime(t)$ are 
\begin{equation}
\label{eq:forc_sol_y}
(\forc{\vec{y}}^\prime)_k = 
\sum_{\ell=5}^8 \sum_{p=1}^{N_\ell}
\frac{(\subscript{\vec{N}}{io}^\prime)_{k \ell} \tilde{y}_{\ell p}}{(\vec{\sLL})_k - \vec{n}_{\ell p} \cdot \out{\vec{\omega}}} 
\E^{j \vec{n}_{\ell p} \cdot \vec{\phi}}
\quad k \in \{1,2,3,4\} .
\end{equation}
where $\subscript{\vec{N}}{io}^\prime = \matrOD{O}{N}^T \subscript{\vec{N}}{io}$.

\begin{table*}
\caption{Frequency in arcsec yr$^{-1}$ of the largest-amplitude Fourier harmonic for each proper mode of the \Poincare's variables $(x_k,y_k)_{k=1,4}$ 
over the time interval [0, 20] Myr. For the solutions \LaX{} and La90 the proper modes $(\vec{z}_k^\star,\vec{\zeta}_k^\star)_{k=1,4}$ 
defined in \citep{Laskar1990} are employed, while for the present solutions we use the proper modes $(\vec{u}_k,\vec{v}_k)_{k=1,4}$.}
\label{tab:fund_freqs}
\centering
\begin{tabular}{r c c c c c c c c} 
\hline\hline
& $g_1$ & $g_2$ & $g_3$ & $g_4$ & $s_1$ & $s_2$ & $s_3$ & $s_4$ \\ 
\hline
\LaX{}  & 5.546 & 7.457 & 17.388 & 17.928 & -5.617 & -6.983 & -18.844 & -17.759 \\ 
La90    & 5.531 & 7.460 & 17.379 & 17.927 & -5.620 & -7.011 & -18.846 & -17.758 \\
$\Hiss$ & 5.577 & 7.446 & 17.390 & 17.932 & -5.588 & -6.993 & -18.901 & -17.817 \\
\hline
\rule{0pt}{0.9em} 
$\Hiss_4$    & 5.631 & 7.457 & 17.371 & 17.923 & -5.602 & -7.071 & -18.922 & -17.817 \\ 
$\Hiss_6$    & 5.560 & 7.453 & 17.396 & 17.935 & -5.628 & -7.002 & -18.894 & -17.816 \\ 
$\Hiss_8$    & 5.576 & 7.446 & 17.390 & 17.932 & -5.589 & -6.993 & -18.900 & -17.816 \\ 
$\Hiss_{10}$ & 5.577 & 7.446 & 17.390 & 17.932 & -5.587 & -6.993 & -18.901 & -17.817 \\ 
\hline
\end{tabular}
\end{table*}

Once the proper modes $\vecu,\vecv$ have been defined, it is very useful to establish a corresponding truncated Hamiltonian in the form 
\begin{equation}
\label{eq:ham_sec_inn_uv}
\Huv_{2n} = 
\sum_{p = 1}^{n} \Huv_{(2p)} ,
\end{equation}
where $\Huv_2 = \HuvLL{u} + \HuvLL{v} = \vecu^\dagger \matrOD{D}{M} \vecu + \vecv^\dagger \matrOD{D}{N} \vecv$ is the 
LL Hamiltonian. To define higher-degree truncations, we first consider the non-quadratic part $\delta\Hsec$ of the 
secular Hamiltonian \eqref{eq:ham_sec}, truncated at a given total degree in \Poincare's variables $(x_k,\bar{x}_k,y_k,\bar{y}_k)_{k=1,8}$,
\begin{equation}
\label{eq:non-quad_sec_ham}
\delta\Hsec_{2n} = \sum_{p = 2}^n \Hsec_{(2p)} .
\end{equation}
Then, the proper modes $\vecu,\vecv$ are injected in \eqref{eq:non-quad_sec_ham} by means of Eqs. \eqref{eq:proper_modes_u} and \eqref{eq:proper_modes_v}, 
and the \Poincare's variables of the outer planets replaced by their quasi-periodic decompositions given in Eq. \eqref{eq:qp_decomposition}. 
However, such substitutions do not conserve the degree of the terms in the expansion, since non-linear 
harmonics are present in Eqs. \eqref{eq:qp_decomposition}, \eqref{eq:forc_sol} and \eqref{eq:forc_sol_y}, i.e. 
harmonics of order $|\vec{m}|$ or $|\vec{n}|$ greater than one. Such harmonics are in principle much smaller than the 
linear ones, and generate higher-degree terms in the substitution process. To define the terms $\Huv_{(2p)},p\geq2$ in a 
consistent way, we thus introduce a fictitious real variable $\epsilon$ to redefine the quasi-periodic decompositions \eqref{eq:qp_decomposition} as 
\begin{equation}
\label{eq:qp_decomposition_epsilon}
\left.
\begin{aligned}
&x_k^{(\epsilon)}(t) = \sum_{\ell=1}^{M_k} \epsilon^{|\vec{m}_{k \ell}|} \tilde{x}_{k\ell} \, 
\E^{j \vec{m}_{k\ell} \cdot \vec{\phi}(t)} \\
&y_k^{(\epsilon)}(t) = \sum_{\ell=1}^{N_k} \epsilon^{|\vec{n}_{k \ell}|} \tilde{y}_{k\ell} \, 
\E^{j \vec{n}_{k\ell} \cdot \vec{\phi}(t)} 
\end{aligned}
\;
\right \}
\;
k \in \{5,6,7,8\} ,
\end{equation}
meaning that each harmonic is treated as of degree $|\vec{m}_{k \ell}|$ or $|\vec{n}_{k \ell}|$ for the purpose 
of the expansion\footnote{Indeed, this would be the case if such terms came from some analytical non-linear secular theory.}. 
The forced solutions \eqref{eq:forc_sol} and \eqref{eq:forc_sol_y} are modified accordingly. 
The series resulting from the substitution of variables in \eqref{eq:non-quad_sec_ham} can be thus truncated at total degree $2n$ with 
respect to $\{(u_k, \bar{u}_k, v_k, \bar{v}_k)_{k=1,4},\epsilon \}$. Finally, addition of the quadratic terms $\Huv_2$ gives, by definition, the truncated 
Hamiltonian \eqref{eq:ham_sec_inn_uv}. Truncation is effectively performed in TRIP in parallel with variable 
substitutions, thanks to a dedicated monomial truncated product \citep{Gastineau2011}, which allows to minimize the computational cost of the expansion. 
It is worthwhile to note that, at the end of the expansion process, the Hamiltonian terms only depending on time can be removed, 
since they are dynamically irrelevant. When numerical evaluations involving the terms of the Hamiltonian have to be performed, 
the fictitious variable $\epsilon$ is simply set to $1$. 
In Table \ref{tab:number_monomials} we show the total number of monomials of variables $\{(u_k, \bar{u}_k, v_k, \bar{v}_k)_{k=1,4},
(\E^{j \phi_k(t)})_{k=1,7}\}$ in the Hamiltonian $\Huv_{2n}$ according the degree of truncation. We also report the typical computation 
time of the series in TRIP, related to the transformation $(x_k,\bar{x}_k,y_k,\bar{y}_k)_{k=1,8} \rightarrow 
\{(u_k, \bar{u}_k, v_k, \bar{v}_k)_{k=1,4}, (\E^{j \phi_k(t)})_{k=1,7}\}$. One may note the dramatic increase of the 
number of terms due to such change of variables. We point out that the precise number of monomials in $\Huv_{2n}$ depends on the 
particular quasi-periodic decomposition \eqref{eq:qp_decomposition} we used in this work (see Appendix \ref{appendix:QPSO}). 

\begin{figure*}
\resizebox{\hsize}{!} {
\subfloat[Mercury]{
\includegraphics[]{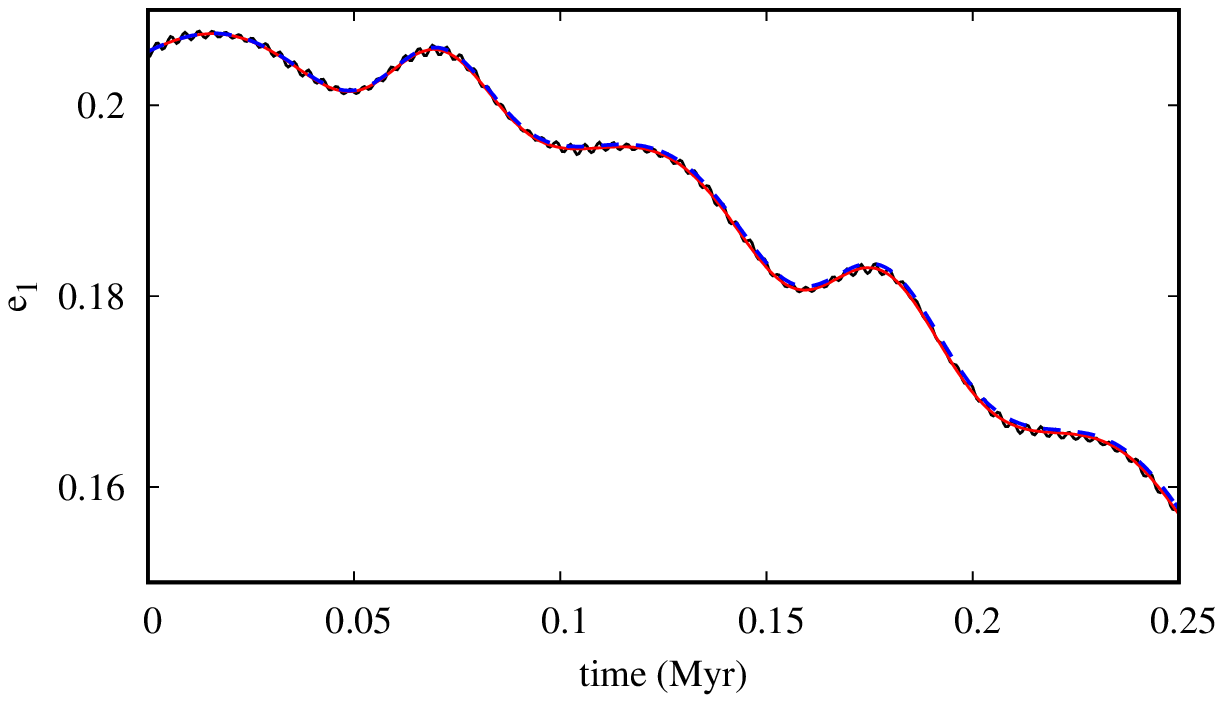}
\includegraphics[]{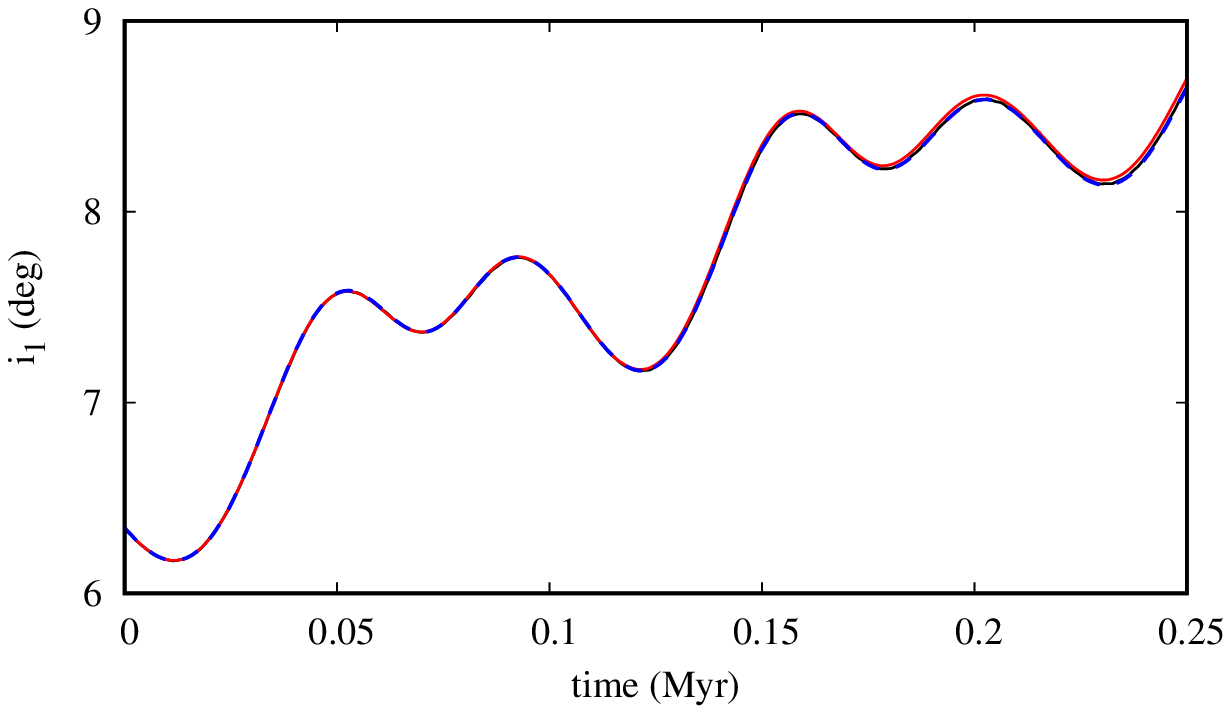}} }
\resizebox{\hsize}{!} {
\subfloat[Venus]{
\includegraphics[]{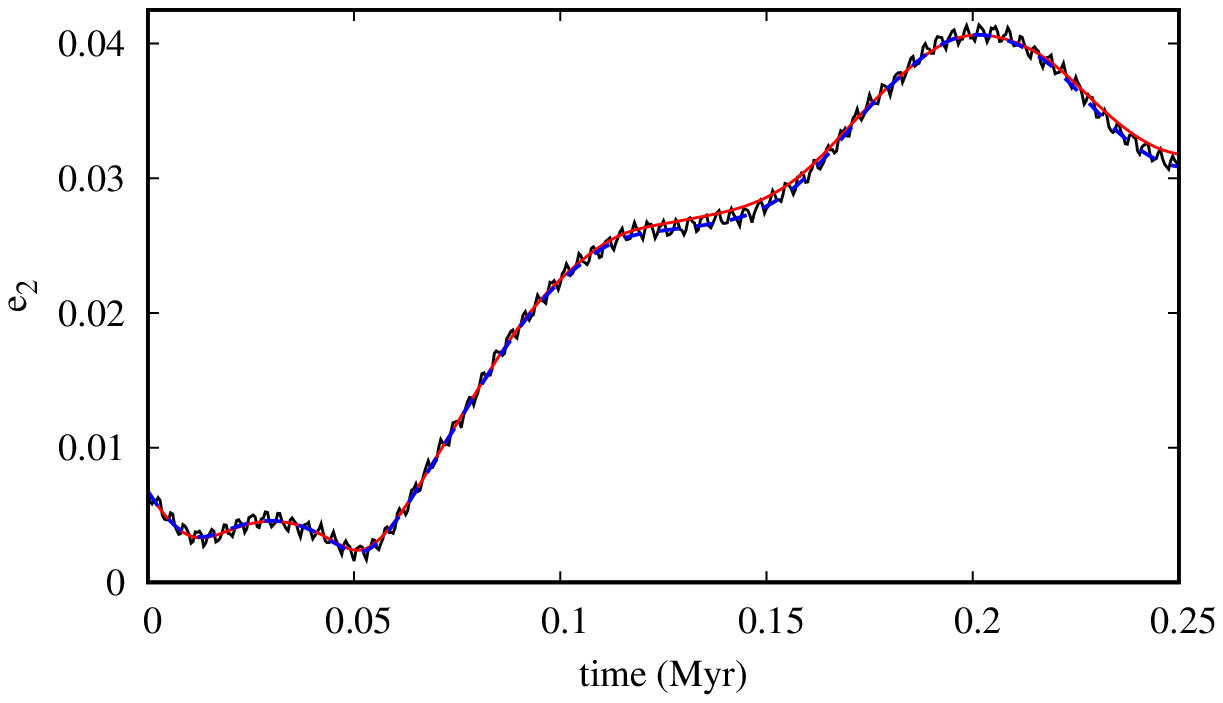}
\includegraphics[]{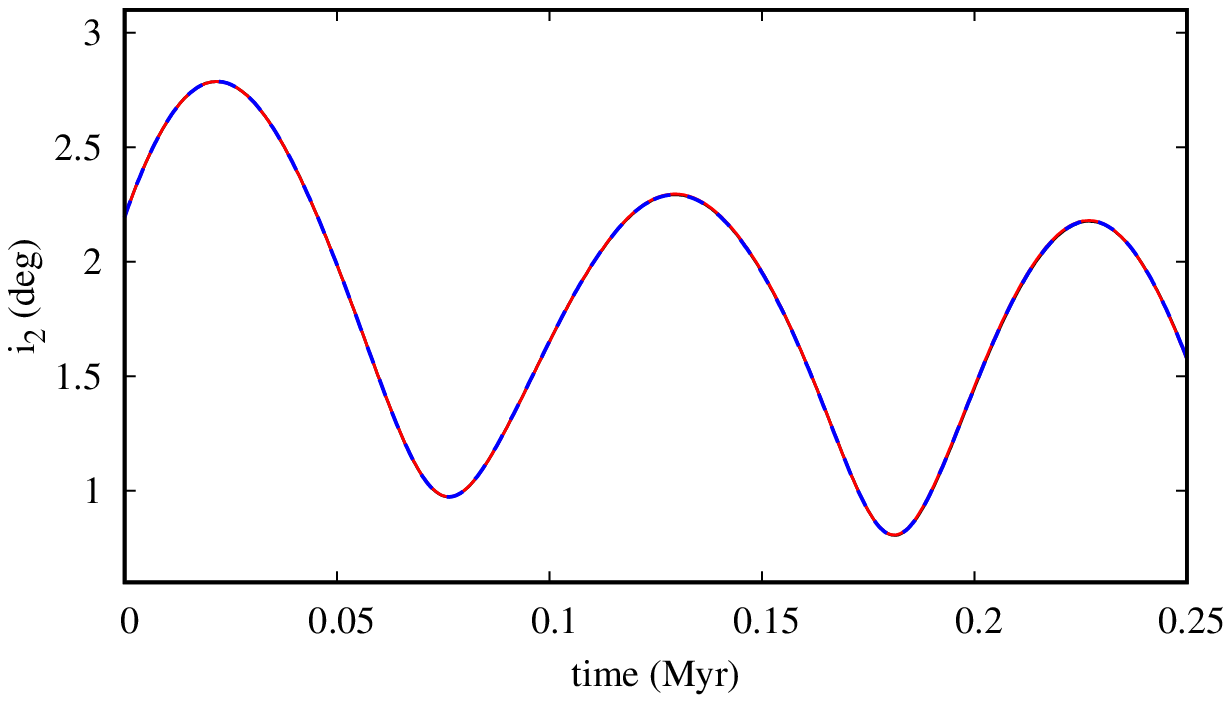}} }
\resizebox{\hsize}{!} {
\subfloat[Earth]{
\includegraphics[]{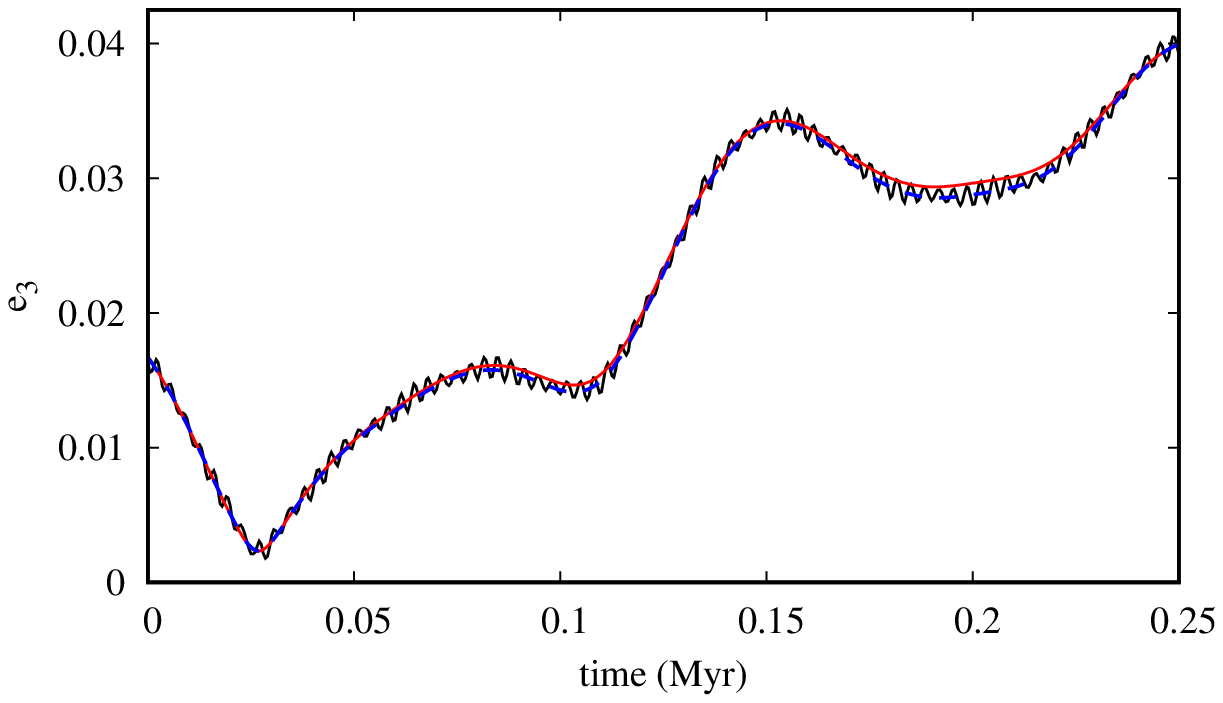}
\includegraphics[]{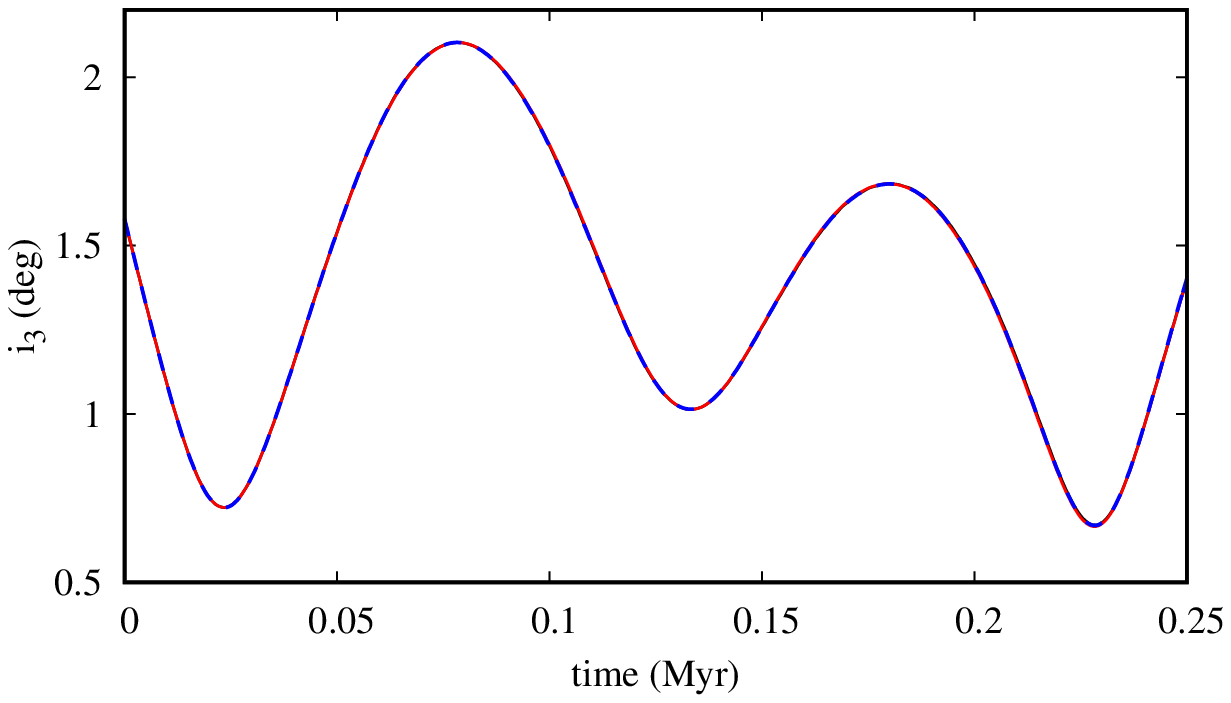}} }
\resizebox{\hsize}{!} {
\subfloat[Mars]{
\includegraphics[]{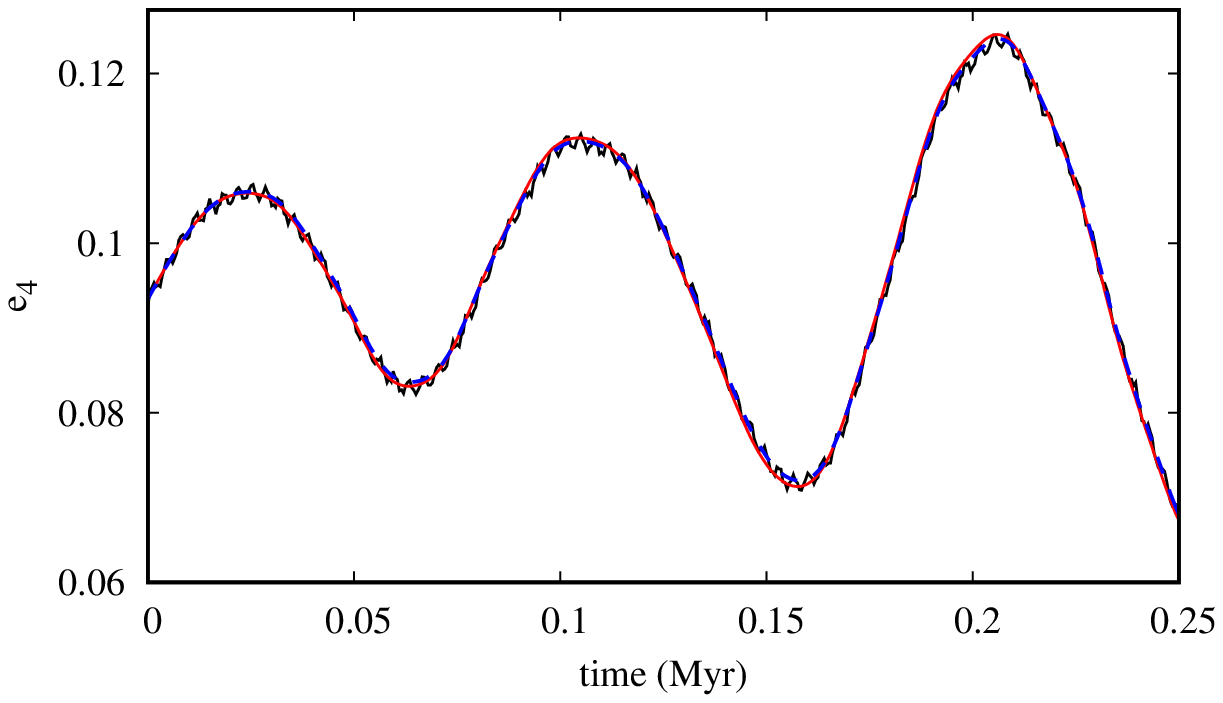}
\includegraphics[]{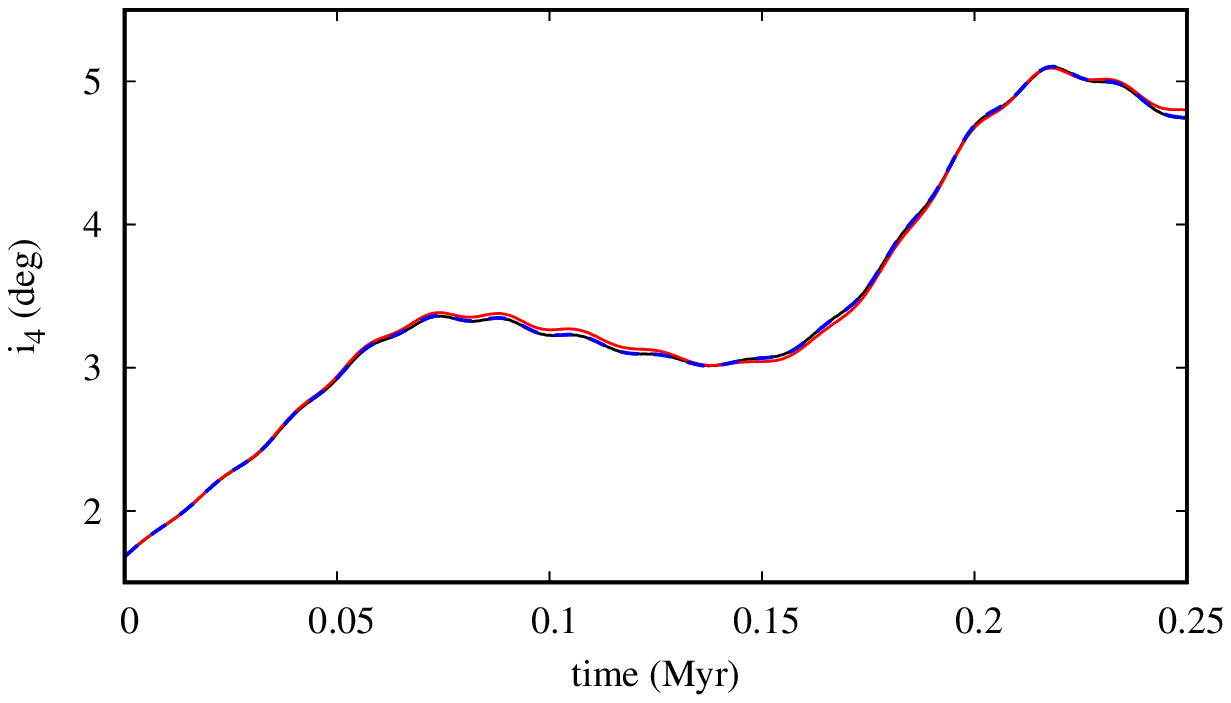}} }
\caption{Eccentricities (left side) and inclinations (right side) over the invariant plane J2000 
(see Appendix \ref{sec:inv_plane}) of the inner planets over 250\,000 yr in the future. 
The black solid line stands for the full (i.e. non-filtered) direct integration \LaX{}, while the 
red one is the solution of the present forced model in Eq. \eqref{eq:ham_sec_inn}. The secular 
integration La90 is represented by the dashed blue line. The maximal relative root mean square of 
the present model residuals is 2\% for the eccentricities and 0.7\% for the inclinations.}
\label{fig:short_ecc_inc}
\end{figure*}

\begin{figure*}
\resizebox{\hsize}{!} {
\subfloat[Mercury]{
\includegraphics[]{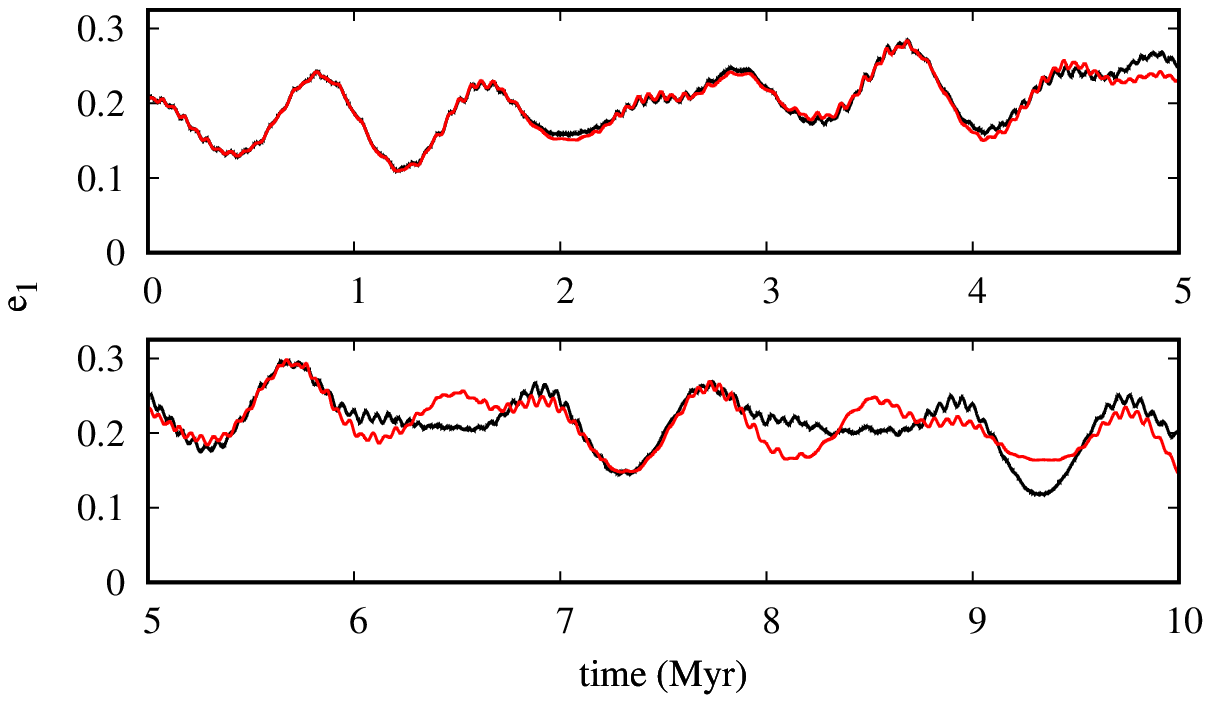}
\includegraphics[]{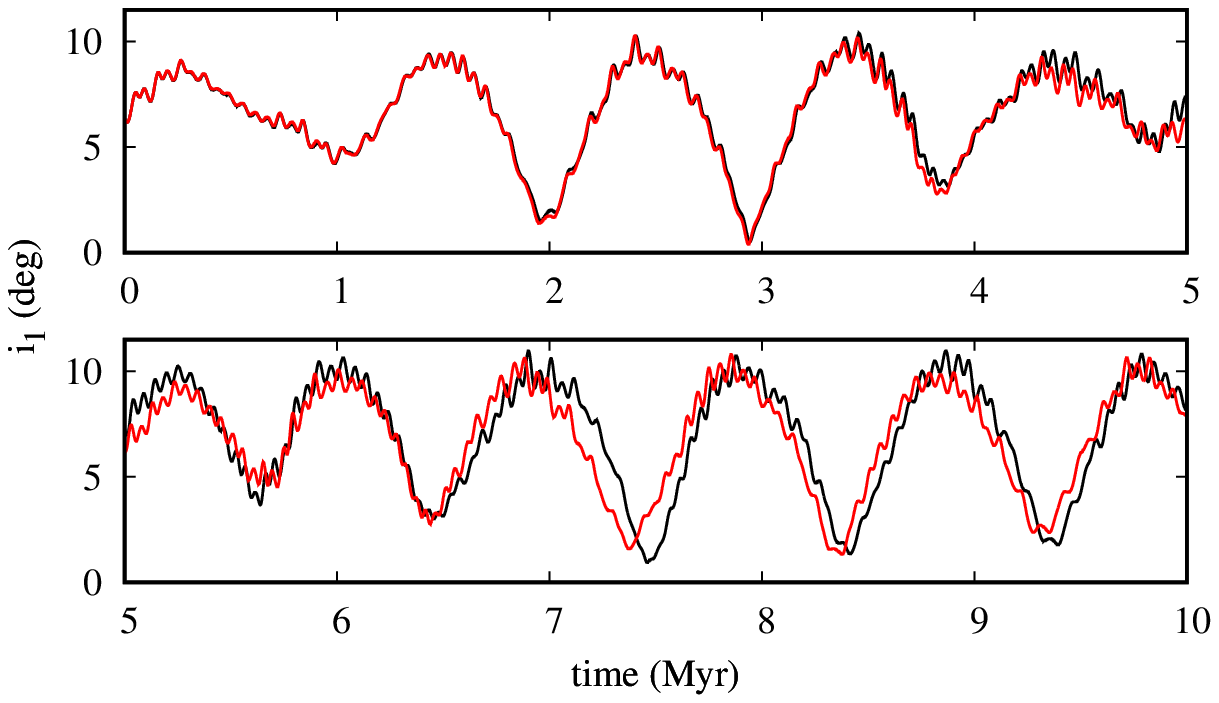}} }
\resizebox{\hsize}{!} {
\subfloat[Venus]{
\includegraphics[]{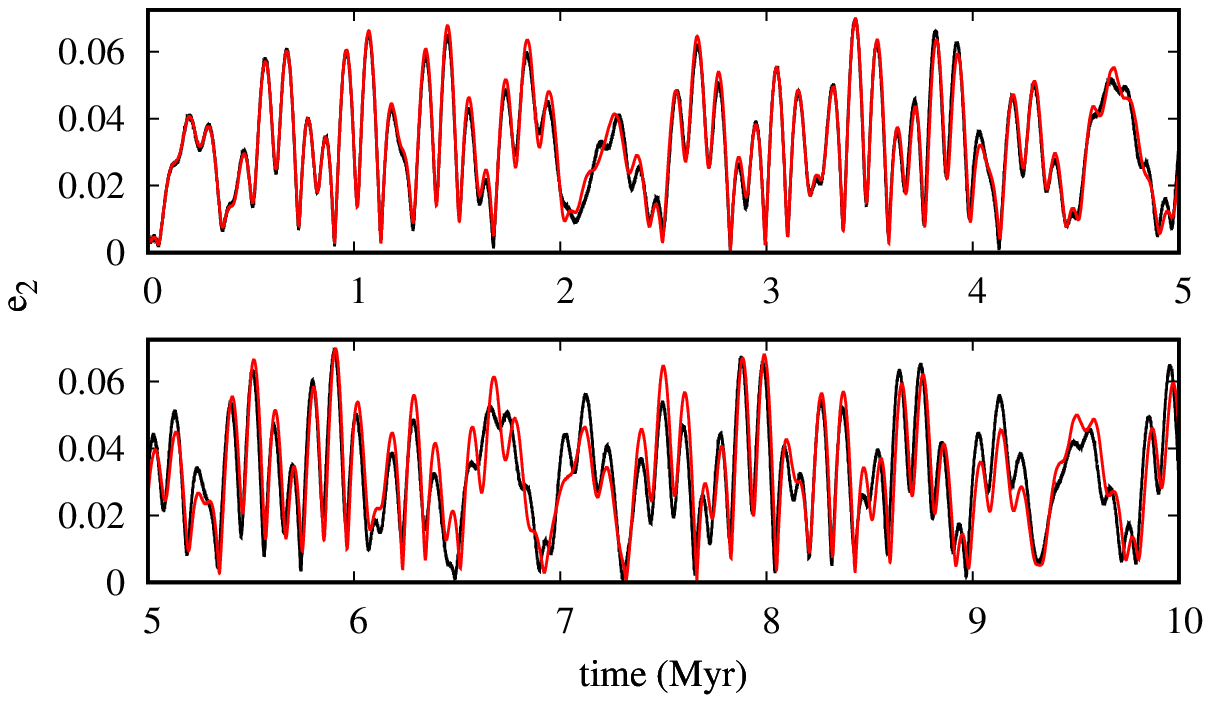}
\includegraphics[]{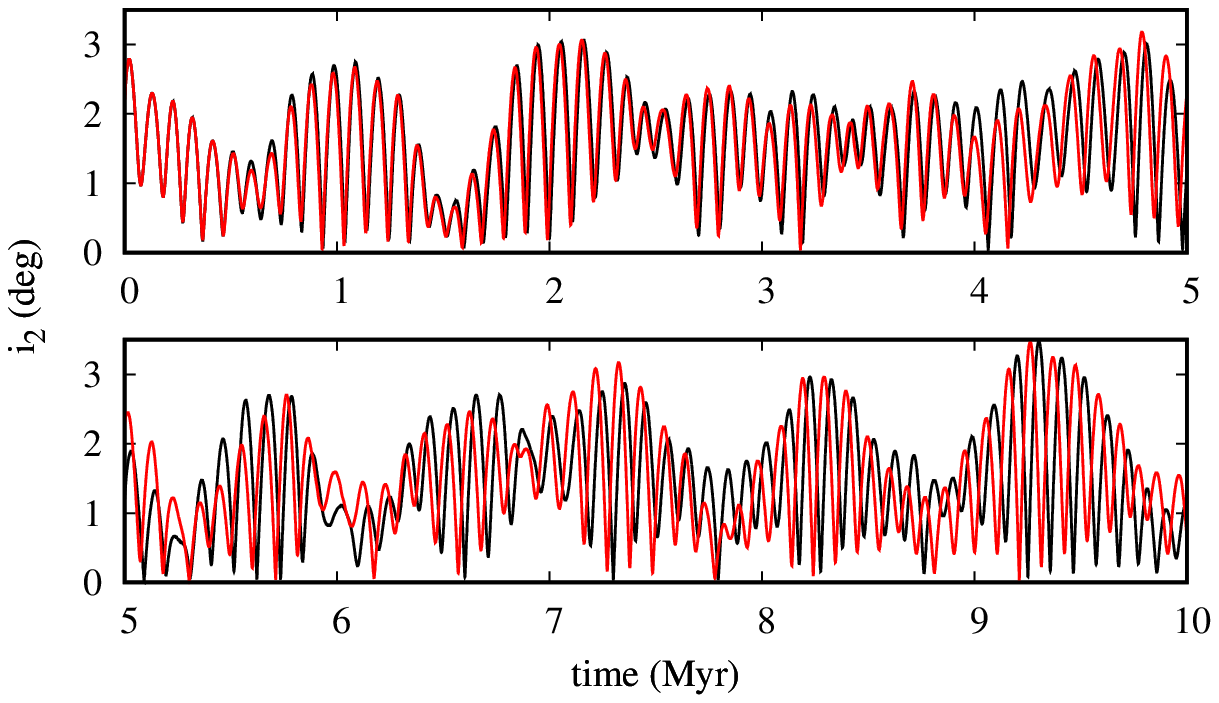}} }
\resizebox{\hsize}{!} {
\subfloat[Earth]{
\includegraphics[]{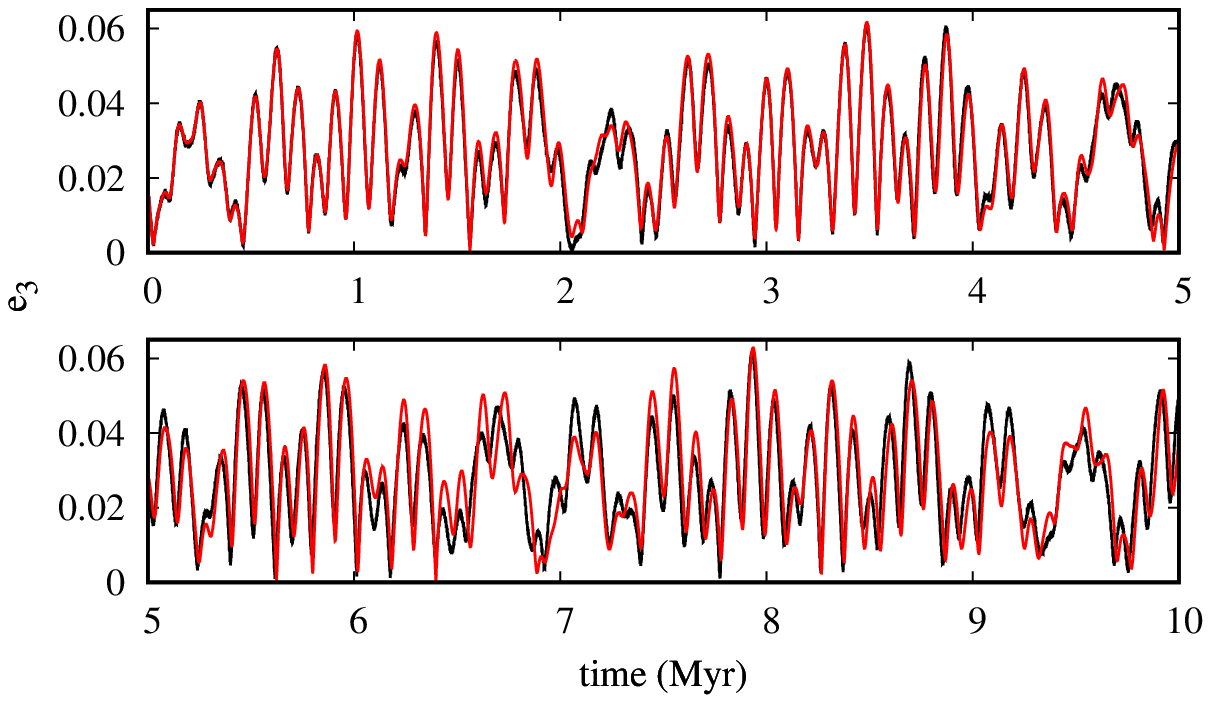}
\includegraphics[]{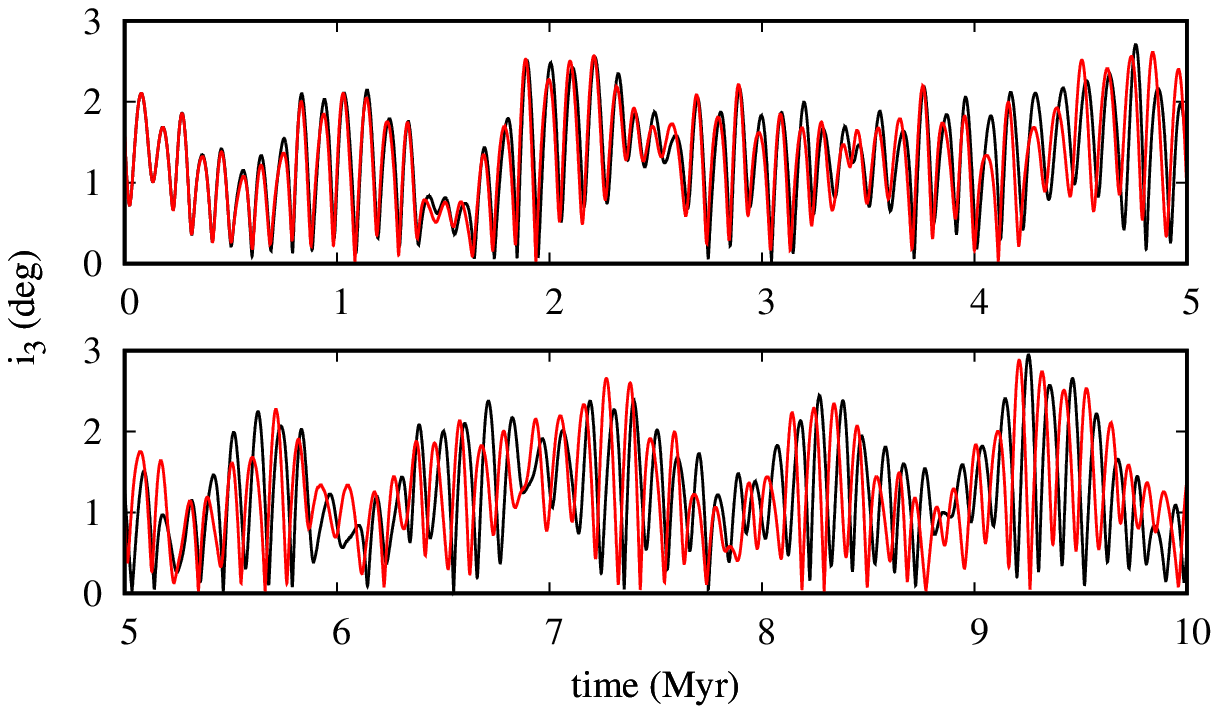}} }
\resizebox{\hsize}{!} {
\subfloat[Mars]{
\includegraphics[]{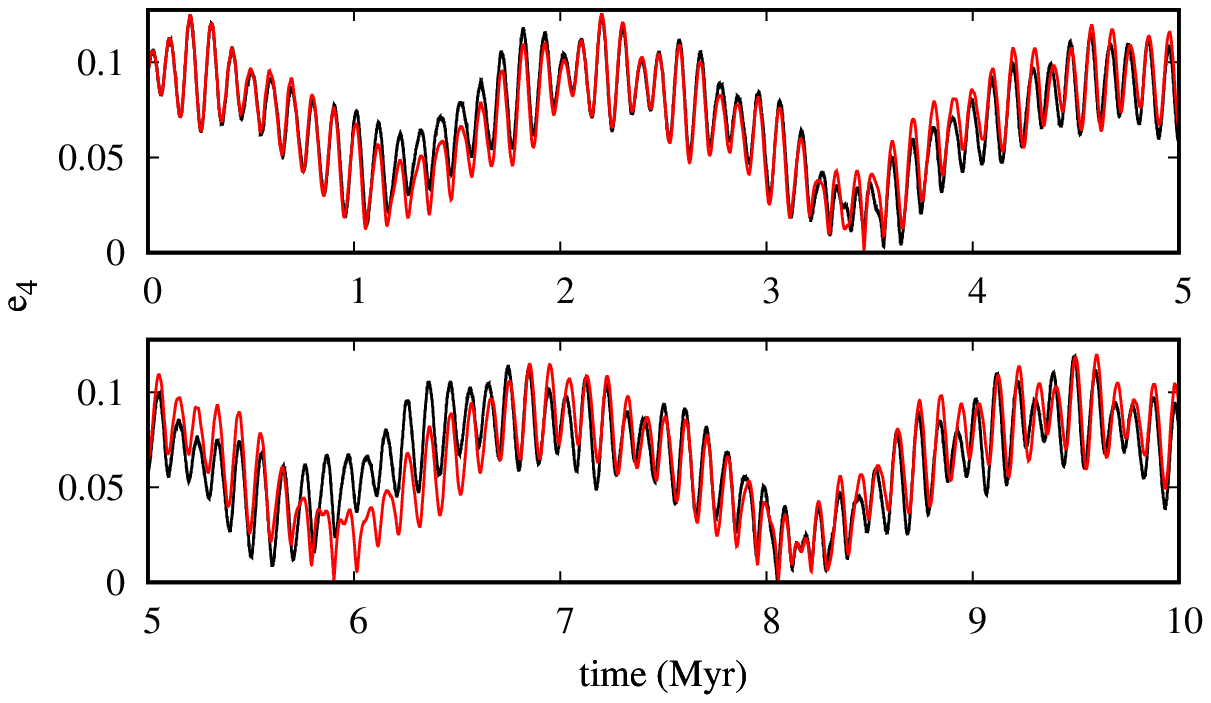}
\includegraphics[]{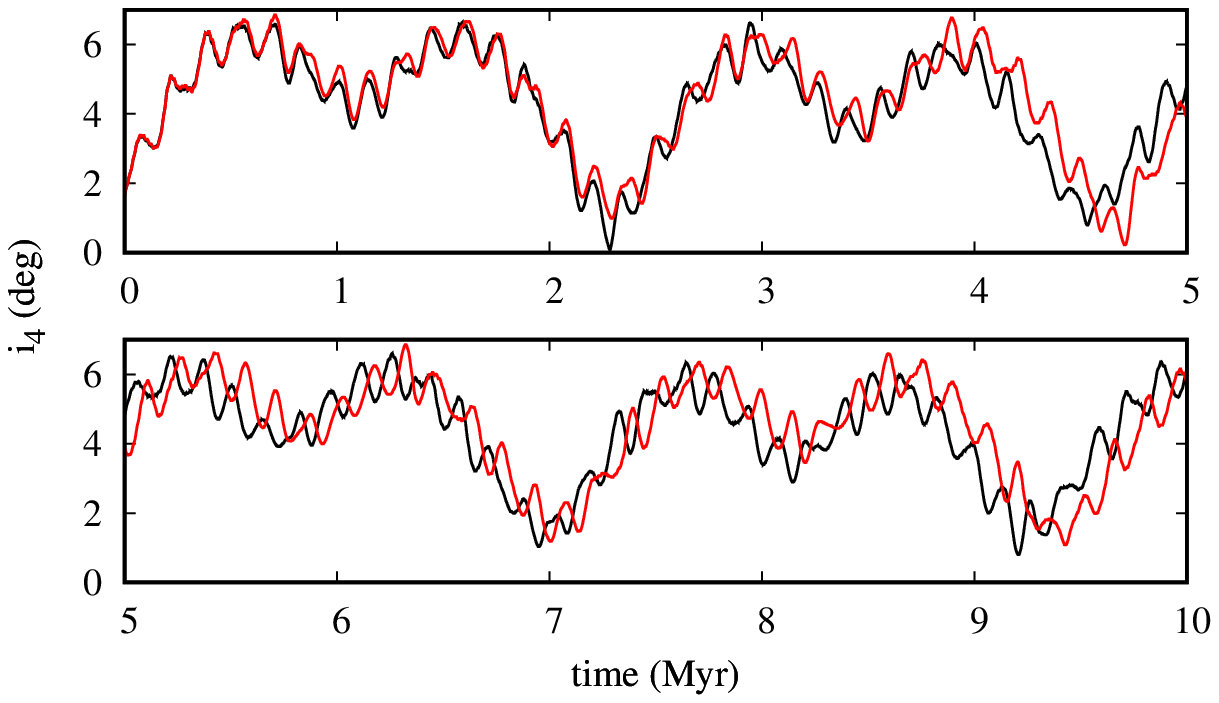}} }
\caption{Eccentricities (left side) and inclinations (right side) over the invariant plane J2000 
(see Appendix \ref{sec:inv_plane}) of the inner planets over 10 Myr in the future. The black solid 
line stands for the full (i.e. non-filtered) direct integration \LaX{}, while the red one 
is the solution of the present forced model in Eq. \eqref{eq:ham_sec_inn}. The secular solution La90 
practically coincides with the direct integration on these plots and it is thus not shown.}
\label{fig:ecc_inc}
\end{figure*}

\subsubsection{Action-angle variables}
Action-angle variables corresponding to the proper modes $\vecu,\vecv$ are introduced by standard canonical 
transformations $(\vecu,-j \vecu^\dagger) \rightarrow (\vec{\Chi},\vec{\chi})$ and 
$(\vecv,-j \vecv^\dagger) \rightarrow (\vec{\Psi},\vec{\psi})$ defined as 
\begin{equation}
\label{eq:proper_modes_AA}
\begin{aligned}
&u_k = \sqrt{\Chi_k} \, \E^{-j \chi_k}, \\
&v_k = \sqrt{\Psi_k} \, \E^{-j \psi_k},  
\end{aligned}
\end{equation}
for $k \in \{1,2,3,4\}$. The Laplace-Lagrange Hamiltonian thus reads $\Huv_2 = -\vec{g}_{\text{LL}} \cdot \vec{\Chi} 
- \vec{s}_{\text{LL}} \cdot \vec{\Psi}$ and it is trivially integrable. Such transformations allow to expand the Hamiltonian 
$\Huv_{2n}$ as a Fourier series of the angle variables $\vecchi$, $\vecpsi$ and $\vec{\phi}(t) = \out{\vec{\omega}} t$, 
\begin{equation}
\label{eq:fourier_expansion}
\Huv_{2n} = 
\sum_{\vec{k} \in \mathbb{Z}^8}
\sum_{\vec{\ell} \in \mathbb{Z}^7} 
\widetilde{\Huv}_{2n}^{\vec{k},\vec{\ell}}(\vecI) 
\, \E^{j \left( \vec{k} \cdot \vec{\theta} + \vec{\ell} \cdot \out{\vec{\omega}} t \right)} ,
\end{equation}
where $\widetilde{\Huv}_{2n}^{\vec{k},\vec{\ell}}$ are complex amplitudes and we have employed a compact notation for the 
action-angle variables, 
\begin{equation}
\label{eq:compact_AA_notation}
\vecI = (\vecChi,\vecPsi), \quad 
\vectheta = (\vecchi,\vecpsi) .
\end{equation}
Only a finite number of harmonics have non-zero amplitude in Eq. \eqref{eq:fourier_expansion}, since we deal with a truncated Hamiltonian. 
We also recall that, the Hamiltonian being a real function, one has $\widetilde{\Huv}_{2n}^{-\vec{k},-\vec{\ell}} = 
\overline{\widetilde{\Huv}_{2n}^{\vec{k},\vec{\ell}}}$ for all $\vec{k},\vec{\ell}$. In Table \ref{tab:number_monomials} we show the 
number of harmonics in Eq. \eqref{eq:fourier_expansion} according the truncation degree of the Hamiltonian, i.e. 
the number of different wavevectors $(\vec{k},\vec{\ell})$ up to a global minus sign. 

It is worthwhile to mention that the explicit time dependency of the Hamiltonian $\Huv_{2n}$ can be easily absorbed in a phase-space 
extension, through the definition of action-angle variables for the trivial degrees of freedom of the outer orbits. Indeed, one introduces $(\vec{\Phi},\vec{\phi})$ such that the new Hamiltonian reads 
\begin{equation}
\label{eq:extended_phasespace}
\Huv^\star_{2n} = 
\out{\vec{\omega}} \cdot \vec{\Phi} 
+ \sum_{\vec{k} \in \mathbb{Z}^8}
\sum_{\vec{\ell} \in \mathbb{Z}^7} 
\widetilde{\Huv}_{2n}^{\vec{k},\vec{\ell}}(\vecI) 
\, \E^{j \left( \vec{k} \cdot \vec{\theta} + \vec{\ell} \cdot \vec{\phi} \right)} ,
\end{equation}
with $\vec{\Phi} = (\Phi_k)_{k=1,7}$. The dynamics of the additional angles is thus consistently given by $\dot{\vec{\phi}} = \out{\vec{\omega}}$, 
while that of the actions $\vec{\Phi}$ is irrelevant. Such an autonomous formulation is useful in the context of canonical 
perturbation theory. 

\section{Comparison with reference dynamical models}
\label{sect:comparison}
It is essential, when constructing averaged models, to compare the resulting trajectories to those 
of the nominal Hamiltonian, or of a more comprehensive dynamical model. 
This allows to validate both the underlying averaging approximations and the choice of the initial conditions 
for the secular variables (see Sect. \ref{subsect:init_cond} and Appendix \ref{appendix:QPSO}). 
In this Section, we compare the orbital solution of the 
forced inner system (Eq.~\ref{eq:ham_sec_inn}) with the direct numerical integration of the 
full equations of motion \LaX{}, which has been used to predetermine the secular motion of the outer planets 
(Sect. \ref{subsect:init_cond} and Appendix \ref{appendix:QPSO}). 
We shall also employ in the comparison the orbital solution of \citep[La90 from now on]{Laskar1990}, as it includes 
all the dynamical interactions taken into account in the present work and constitutes the most precise secular system to this day. 

The comparison between the different models begins at short secular timescales in Fig. \ref{fig:short_ecc_inc}, where we show the inner 
planet eccentricities and inclinations over 250\,000 years in the future. The plots indicate that the present model correctly 
reproduces the secular behaviour of the \LaX{} solution. This is particularly manifest in the eccentricity plots, where 
the forced model constitutes the average of the short-time (orbital) oscillations of the direct integration, as the La90 solution also does. 
In particular, the behaviour of the solutions at the origin of time shows that the initial conditions for the 
secular variables of this work have been correctly determined. Some small periodic differences between the two secular 
models appear on some of the plots, with the La90 solution generally being more precise in reproducing the long-term
average of the direct integration. However, such deviations are practically irrelevant at these scales. 

In Fig. \ref{fig:ecc_inc} the comparison is extended to a longer time interval of 10 Myr, to evaluate how long the present solution 
remains close to the direct integration. The solution La90 practically coincides with \LaX{} on these plots \citep[see][]{Laskar2004} and it is thus 
not shown. The general agreement shown over the first few million years by the curves in Fig. \ref{fig:ecc_inc} is still 
very satisfactory. For longer times, the divergence of the two models becomes important. First of all, the orbital oscillations show sometime 
different amplitudes. Nevertheless, as we are not interested in the construction of a secular ephemeris (the solution La90 
is clearly more adapted to this end), this is not the most relevant point. When considering that the inner solar 
system has a Lyapunov time of about 5 Myr (see Sect. \ref{sect:lyapunov}), the crucial aspect is rather the slow dephasing of the solutions 
appearing from this comparison, as it translates a difference in the fundamental frequencies of the motion. To ensure that the forced model 
faithfully reproduces the resonant structure of the inner system, it is essential to quantify such deviations. To this end, 
we perform a frequency analysis of the proper modes of the \Poincare's variables $(x_k,y_k)_{k=1,4}$ 
for the different orbital solutions over the [0, 20] Myr time interval. For the solutions \LaX{} and La90 we employ the proper modes 
$(\vec{z}_k^\star,\vec{\zeta}_k^\star)_{k=1,4}$ defined in \citep{Laskar1990}, while for the present solution the proper modes 
$(\vec{u}_k,\vec{v}_k)_{k=1,4}$ are used (the employ of $(\vec{z}_k^\star,\vec{\zeta}_k^\star)_{k=1,4}$ gives the same results 
at the numerical precision used in the comparison). 
Table \ref{tab:fund_freqs} shows the dominant frequency in the Fourier spectrum for each proper mode, according the notation of 
\citep{Laskar1990}. The average absolute difference between the \LaX{} solution and the present one amounts to a few hundredths of 
arc second per year, with a maximum of 0.06\arcsecyr in the case of $s_3$ and $s_4$. This is in agreement with the behaviour shown in 
Fig. \ref{fig:ecc_inc} and the general expectation from the discussion on the model precision in Sect. \ref{sect:minor_effects}. 
It is important to note that, when considering combinations of the fundamental frequencies (which determine the resonant structure), 
the differences between the two models can be even smaller. Indeed, in the case of the two leading resonances $(g_1-g_5)-(s_1-s_2)$ 
and $2(g_3-g_4)-(s_3-s_4)$ described in \citep{Laskar1990}, the deviations amount to just 0.01\arcsecyr and 0.004\arcsecyrNOSPACE, 
respectively. 

Table \ref{tab:fund_freqs} also reports the fundamental frequencies of the truncated model $\Hiss_{2n}$ 
(Eq.~\ref{eq:ham_iss_expansion}) for different total degrees of truncation. When considering the combinations of frequencies 
$(g_1-g_5)-(s_1-s_2)$ and $2(g_3-g_4)-(s_3-s_4)$, the models $\Hiss_4$ and $\Hiss_6$ practically show the same deviations as 
the full forced model $\Hiss$ with respect to the solution \LaX{}. This suggests that the truncated model should already realistically 
reproduce the resonant structure of the inner system at the lowest degrees. We also note that the orbital solutions 
of the models $\Hiss_8$ and $\Hiss_{10}$ practically coincide with that of the full forced model $\Hiss$ over the first 20 Myr. 

\begin{figure}
\centering
\includegraphics[width=\hsize]{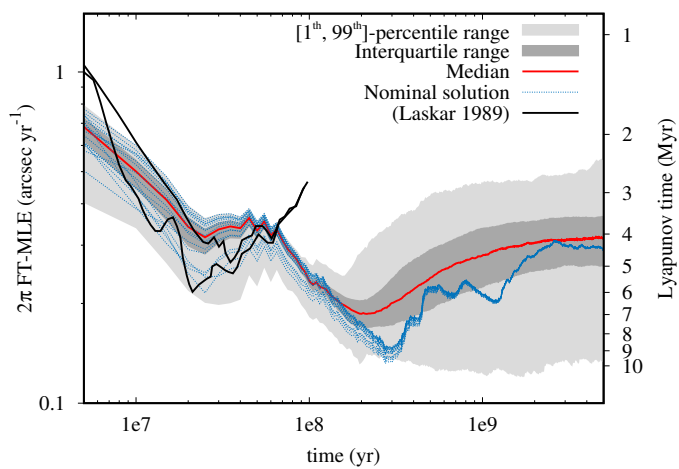}
\caption{Finite-time maximum Lyapunov exponent (FT-MLE) and corresponding Lyapunov time FT-MLE$^{-1}$ 
of the forced inner solar system over 5 Gyr, from an ensemble of 1\,148 stable orbital solutions with 
very close initial conditions.} 
\label{fig:MLE}
\end{figure}

\section{Finite-time maximum Lyapunov exponent}
\label{sect:lyapunov}
Once the adequacy of the present dynamical model has been established over a few million years, one has to assess if it is 
able to correctly reproduce the resonant structure of the inner solar system over longer times. Even if the fundamental 
frequencies in Table \ref{tab:fund_freqs} suggest this is actually the case, the computation of the maximum Lyapunov exponent (MLE) 
remains an essential test, as its value is related to the width of the leading resonant harmonics of the Hamiltonian, which are the 
dynamical sources of stochasticity \citep{Chirikov1979}. At this point, it is important to realize that the non-null probability 
of an unstable evolution of the inner planets \citep{Laskar1994,Laskar2008,Batygin2008,Laskar2009} prevents the existence of the MLE 
as an infinite-time limit, as in its usual mathematical definition \citep{Oseledec1968}; pertinent to the present case is the 
consideration of a finite-time MLE (FT-MLE). As we do not have at our disposal an efficient numerical algorithm to evaluate the 
variational equations of Gauss's dynamics, we consider the FT-MLE given by the two-particle algorithm proposed in \citep{Benettin1976}, 
which has been used in \citep{Laskar1989}. This method computes the 
divergence of close trajectories of a dynamical system, by considering the motion of reference and shadow particles, initially separated 
by a tiny vector $\vec{d}_0$ in the phase space. At time intervals $\tau$, a renormalization procedure applied to the trajectory
separation $\vec{d}(t)$ brings again the shadow particle at a distance $||\vec{d}_0||$ from the reference one. 
The resulting FT-MLE depends in a intricate way on the initial position of the reference particle in the phase space, so that its 
asymptotic evolution is chaotic (and independent of the choice of the vector $\vec{d}_0$ \citep{Benettin1976}). Therefore, its 
computation acquires full physical significance only for an ensemble of trajectories. 

\begin{figure}
\centering
\includegraphics[width=\hsize]{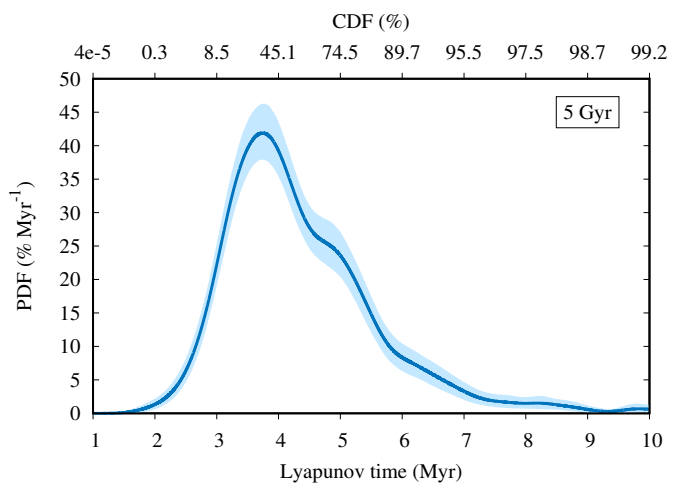}
\caption{Probability density function (PDF) of the Lyapunov time FT-MLE$^{-1}$ at 5 Gyr. 
Kernel density estimation in dark-blue line and corresponding cumulative distribution function (CDF) on the 
upper horizontal axis. Pointwise confidence interval at 98\% level from bootstrap as light-blue region.} 
\label{fig:PDF_MLE_5Gyr}
\end{figure}

In Fig. \ref{fig:MLE} we show the computation of the FT-MLE of the forced inner solar system over 5 Gyr in the future, 
for an ensemble of 1\,148 stable orbital solutions with initial conditions very close to the nominal ones 
(see Sect. \ref{sect:evolution_5Gyr} for the definition of an unstable solution in the framework of the present model). 
On the left vertical axis, we report the FT-MLE expressed as an angular frequency in arcsec yr$^{-1}$, while on the right one the
corresponding Lyapunov time, defined as FT-MLE$^{-1}$, is given in million years. The dark-grey region represents, at each 
renormalization time, the interquantile range of the observed probability density function (PDF) of the FT-MLE, which encloses by 
definition 50\% of the solutions around the median (shown by the red line). In the same manner, the light-grey region 
corresponds to the $[1^{\textrm{th}},99^{\textrm{th}}]$-percentile range of the PDF, enclosing the 98\% of the probability. 
The two FT-MLEs reported in \citep{Laskar1989} are also shown with black lines.

As stated above, the initial conditions of each solution in the ensemble are very close to the nominal ones. They are chosen by taking 
the relative variation of each coordinate of the nominal phase-space vector as a normal random variable with zero mean and a standard deviation 
of $10^{-9}$. The initial conditions thus follow a multivariate Gaussian distribution centred at the nominal phase-space position 
of the system, with a diagonal covariance matrix. The width of such distribution is of the same order of magnitude of that in \citep[Table 1]{Laskar2008}.
The initial position of each shadow particle, around the corresponding reference one, is then chosen according the same kind of 
Gaussian distribution, with a relative standard deviation of $10^{-8}$. Indeed, such a value should be close to minimize the 
accumulation over long timescales of numerical errors, due to the floating-point implementation of the algorithm of \citep{Benettin1976}, 
when working in double precision \citep{Tancredi2001,Mei2018}. In Fig. \ref{fig:MLE} the computation of the FT-MLE for the nominal 
initial conditions is shown by dotted blue lines for a set of 16 different initial tangent vectors $\vec{d}_0$. This manifestly 
shows that the present FT-MLEs are asymptotically independent of $\vec{d}_0$, the asymptotic regime being practically reached in a 
few hundred million years. In the same manner, our computation has been tested to be asymptotically independent of the renormalization 
time $\tau$, set to 5 Myr in our computation, and of the norm chosen for the phase-space vectors, taken here to be Euclidean as usual.

For times shorter than 100 Myr, Fig. \ref{fig:MLE} shows that the distribution of the FT-MLE only reflects the choice of different 
initial tangent vectors $\vec{d}_0$, around essentially the same reference trajectory, i.e. the nominal one. Therefore, this part 
of the plot can be discarded because of its non-asymptotic character. We also note that, for $t <$ 100 Myr, the PDF of the FT-MLE 
tends to shrinks with increasing time, as expected by the asymptotic behaviour of the method employed (see the blue lines and two curves 
from \citep{Laskar1989}). For longer times, the PDF begins to broaden, because each orbital solution acquires a macroscopically different
FT-MLE, which chaotically depends on its initial conditions. The time-asymptotic regions shown in Fig. \ref{fig:MLE} thus represent the 
probabilistic knowledge of the FT-MLE of the inner solar system, which arises from its chaotic behaviour and our determination of
the current planet positions and velocities. 

The dark-blue line in Fig. \ref{fig:PDF_MLE_5Gyr} shows the kernel density estimation \citep{Rosenblatt1956,Parzen1962} of the PDF of the
Lyapunov time FT-MLE$^{-1}$ at 5 Gyr, along with the corresponding cumulative distribution function (CDF) on the upper horizontal axis. 
We employed the standard Gaussian kernel and Silverman's rule of thumb for the selection of the
optimal bandwidth \citep{Silverman1986}. We also show, through the light-blue region, the pointwise confidence interval at
98\% level on the estimated PDF, obtained via nonparametric bootstrap, i.e. by resampling the original data with replacement 
\citep{Efron1979}. One can appreciate that the value of about 5 Myr typically reported in literature \citep{Laskar1989,Sussman1992} is right in
the bulk of the present PDF, even though the distribution rather peaks at 3.7 Myr and the average value is 4.3 Myr. The value of 6.5 Myr 
found in \citep{Rein2018} can also be accounted for by our computation. The curves of \citep{Laskar1989} suggest that the asymptotic 
PDFs of the two secular models should largely overlap. 
Finally, we note that numerical value of 1.1 Myr reported in \citep{Batygin2015}, as the Lyapunov time of a simplified Mercury 
dynamics, does not agree with the present findings. Figures \ref{fig:MLE} and \ref{fig:PDF_MLE_5Gyr} show that, even considering the 
diffusion of the planet orbits over 5 billion years, such a high FT-MLE is practically never reached by the present 
dynamical model, even though the latter includes all the Hamiltonian harmonics taken into account in \citep{Batygin2015}. 

\section{Orbit excitation over 5 Gyr}
\label{sect:evolution_5Gyr}
In this Section, we perform a statistical study of the orbit excitation in the forced inner system over 5 Gyr in the future, 
the aim being to assess if the present model is able to reproduce the rate of the high Mercury eccentricities observed in 
\citep{Laskar2008,Laskar2009}. Following \citep{Laskar1994}, we characterize the unstable evolutions of the system by 
defining a \emph{secular collision} as the intersection of the instantaneous Keplerian orbits of a pair of planets. 
Numerically, such an event is easily detected by tracking the relative positions of their mutual nodes, 
i.e. the intersections of each orbit with the orbital plane of the other planet. When two orbits crosses, 
a pair of mutual nodes exchanges their positions along the line of nodes (i.e. the intersection of the two orbital 
planes). 
Such a computation is inexpensive in the context of Gauss's dynamics and can be performed at each timestep of
the integration scheme. When a secular collision occurs, the present averaged model is no longer a good long-term approximation 
of the original N-body dynamics, in which the involved planets can experience a close encounter, possibly leading to a physical collision 
or to the escape of one of two planets from the solar system \citep{Laskar1994}. We thus denote by $\tcoll$ the time of the \emph{first} 
secular collision for a given orbital solution. The statistics of the first-collision time $\tcoll$ shall provide an estimate 
of the rate of catastrophic events in the real inner system. Moreover, by stopping the numerical integration at the first secular collision, 
we will not be involved in extrapolating the present secular dynamics out of its validity range, and the use of a fixed-timestep 
integration scheme (see Sect. \ref{sect:gauss_method}) turns out to be sufficient end even suitable. We note that a secular collision 
between a planet and the Sun can be simply defined to occur whenever its pericenter distance $a(1-e)$ is smaller than the Sun radius. 

\begin{table}
\centering
\begin{tabular}{c c c c} 
\hline\hline 
\rule{0pt}{1em}
$e_\textrm{max}$ & LG09 & La08 & This work \\ 
\hline
\rule{0pt}{1em}
0.35 & (46.9, 51.5) & (37.5, 48.0) & (48.5, 50.8) \\ 
0.4 & (23.6, 27.6) & (20.0, 29.1) & (20.3, 22.1) \\ 
0.5 & (3.15, 4.97) & (1.90, 5.84) & (1.74, 2.38) \\ 
0.6 & (0.60, 1.53) & (0.39, 2.80) & (0.38, 0.71) \\ 
0.7 & (0.57, 1.48) & (0.39, 2.80) & (0.32, 0.62) \\ 
0.8 & (0.54, 1.43) & (0.28, 2.50) & (0.28, 0.57) \\ 
0.9 & (0.51, 1.38) & (0.03, 1.50) & (0.08, 0.26) \\ 
\hline
\end{tabular}
\caption{Confidence interval at 98\% level of the probability $P(\sup_{t \leq  5 \, \textrm{Gyr}}e_1(t) > e_\textrm{max})$ in percentage, 
where $e_1$ is the eccentricity of Mercury. LG09 represents the 2\,501 orbital solutions in \citep{Laskar2009}, while La08 stands for 
the 478 integrations in \citep{Laskar2008}. The statistics of this work derives from an ensemble of 10\,560 orbital solutions.} 
\label{tab:5Gyr}
\end{table}

We perform an ensemble of 10\,560 numerical integrations of the forced inner system over 5 Gyr.  The initial conditions 
are chosen according the same multivariate Gaussian distribution centred at the nominal values as in Sect. \ref{sect:lyapunov}. 
The relative standard deviation of each phase-space coordinate is $10^{-9}$, of the same order of magnitude of that in \citep[Table 1]{Laskar2008}. 
Table \ref{tab:5Gyr} shows the observed probability of having, over the entire time span, a maximum Mercury eccentricity greater 
than a given value. The statistics of the present dynamical model is compared to those resulting from \citep[Table 3]{Laskar2008} and 
\citep[see Table \ref{Tab:08} in Appendix \ref{appendix:LG2009} of the present paper]{Laskar2009}. 
The statistical bounds represent the \citet{Wilson1927} score interval 
corresponding to a 98\% confidence level. At the present statistical precision\footnote{As there is no reason for the true probabilities of the three different dynamical models to be exactly the same, for a very large number of orbital solutions their statistical intervals are not likely to overlap.}, the probabilities 
of the high Mercury eccentricities in the forced inner system are fully compatible with those arising in the secular model of \citep{Laskar1990}. 
They also represent a very good estimate of the probabilities resulting from N-body dynamics, even if somewhat lower. This is 
in agreement with the general expectations from an averaged model, which has less degrees of freedom \citep{Laskar1994,Laskar2008}. 
When comparing the midpoints of the statistical intervals, Table \ref{tab:5Gyr} could also suggest that the true probabilities 
of the present model are slightly lower than those in \citep{Laskar2008}. If true, such a fact would agree with the expectation 
of higher regularity of the forced model, as the degrees of freedom dominated by the outer planets (their fundamental frequencies 
in particular) are frozen into a quasi-periodic time dependency. Moreover, the Hamiltonian terms at second order in planet masses 
could also contribute to somewhat larger chaotic excursions of the inner orbits. 

\subsection{Secular collisions}
\label{subsect:sec_collisions}
Among the 10\,560 orbital solutions of our ensemble, we found 42 secular collisions, corresponding to a 98\% confidence interval 
$(0.28\%, 0.57\%)$ for the probability $P(\tcoll \leq 5\,\textrm{Gyr})$. Out of the 2501 integrated solutions of \citep{Laskar2009}, 
there result 6 collisions between Mercury and the Sun, 9 collisions between Mercury and Venus, and 1 collision between Mars and the 
Sun (see Appendix \ref{appendix:LG2009}). These 16 events corresponds to a 98\% confidence interval $(0.36\%, 1.13\%)$ of probability. 
Our definition of secular collision thus provides a very good estimate of the probability of a physical collision in the non-averaged 
inner solar system over the next 5 Gyr, probably somewhat lower in agreement with the previous discussion. 

All the secular collisions in our ensemble of orbital solutions involve the Mercury-Venus couple, 
with a maximum Mercury eccentricity ranging from 0.797 to 0.966. In particular, in 41 out of 42 collisional solutions it is larger 
than 0.8, while in 15 solutions it exceeds 0.9. This implies that the statistics corresponding to the cases $e_\textrm{max} = 0.8$ 
and $0.9$ in Table \ref{tab:5Gyr} is affected by the choice of stopping the numerical integration when the first 
secular collisions is detected, independently of the fact that this could not physically correspond to any unstable further evolution. 
We emphasize that, as a consequence, the statistics of the secular collisions other than the Mercury-Venus one cannot be correctly 
reproduced by the present numerical computation, since such events, as observed in \citep{Laskar2009}, always require a very large 
Mercury eccentricity at some previous time. 

\begin{figure}
\centering
\includegraphics[width=\hsize]{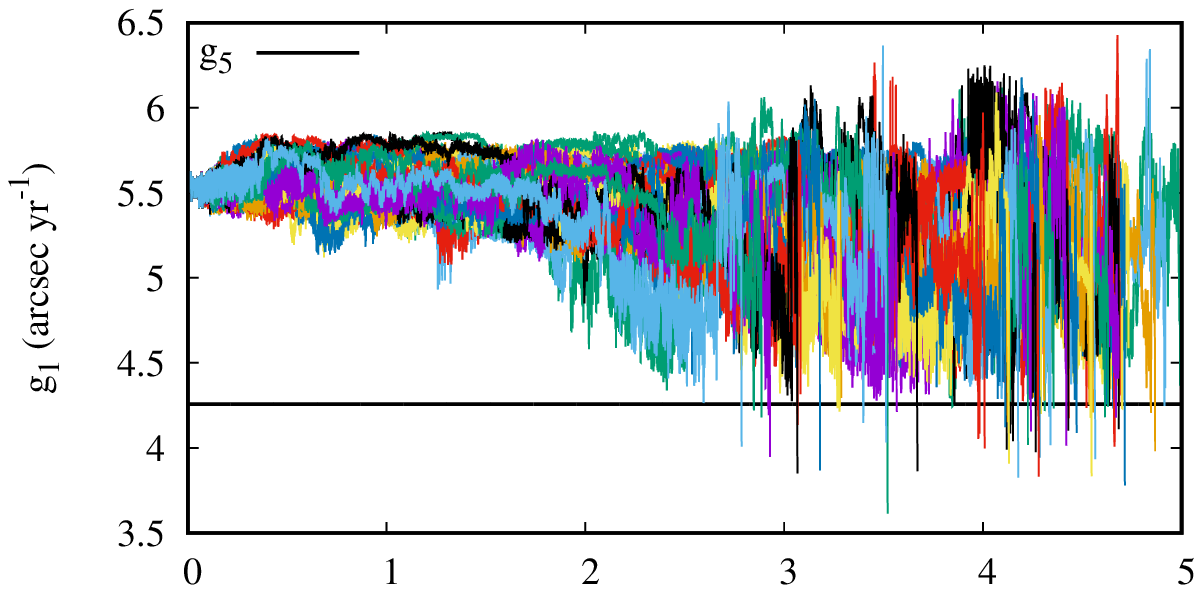}
\includegraphics[width=\hsize]{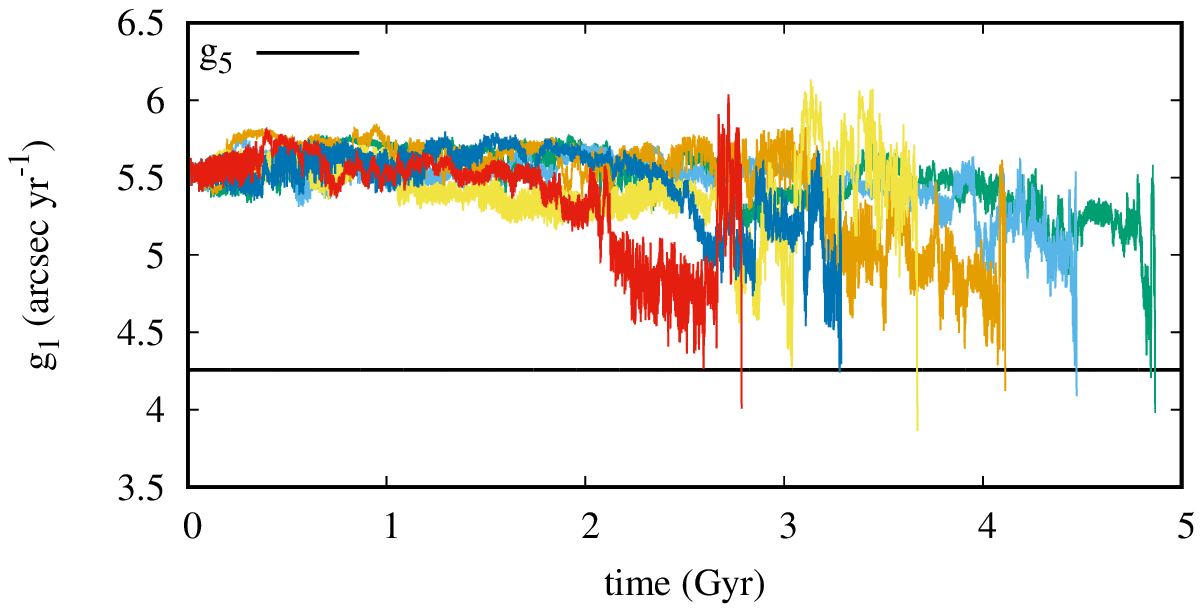}
\caption{Evolution of the fundamental frequency $g_1$, as defined by Eq. \eqref{eq:freq_g1}, 
along the 42 collisional orbital solutions (top panel). Six solutions are isolated in the bottom panel. 
The horizontal line stands for the constant fundamental frequency $g_5$.} 
\label{fig:g1_collisions}
\end{figure}

\subsection{Secular resonance $g_1 - g_5$}
\label{subsect:sec_res_g1_g5}
The destabilizing role of the secular resonance $g_1 - g_5$, able to drive Mercury orbit to very high eccentricities, has been first 
established numerically \citep{Laskar2008,Batygin2008} and then confirmed by an analytical study \citep{Boue2012}. We conclude this 
Section by showing that the $g_1 - g_5$ resonance is indeed reached by the collisional orbital solutions of our ensemble. While 
$g_5 \approx$ 4.257\arcsecyr is constant in the present dynamical model, the frequency $g_1$, which dominates the spectrum of the 
\Poincare's variable $x_1$ and thus the precession of Mercury perihelion, is numerically computed in the following way. 
Given an orbital solution, we sample the corresponding \Poincare's variables $\vec{x}$ with a timestep $\Delta t$ = 1\,000 yr, 
i.e. we consider the time series $\vec{x}(t_n)$ with $t_n = n \Delta t$ and $n \in \mathbb{N}_0$. The proper mode $u_1(t_n)$ is 
thus computed through Eq. \eqref{eq:proper_modes_u} and the corresponding angle variable $\chi_1(t_n)$, following the definition in 
Eq. \eqref{eq:proper_modes_AA}, retrieved as an unwrapped phase, i.e. as a continuous function of time. The instantaneous (angular) 
frequency $\omega_1$ of the proper mode $u_1$ is given by \citep{Cohen1995} 
\begin{equation}
\label{eq:insta_freq_g1}
\omega_1 = - \frac{d\chi_1}{dt} .
\end{equation}
We could analytically compute the derivative of Eq. \eqref{eq:proper_modes_u} to obtain the time series $\omega_1(t_n)$. In practice, 
we employ numerical differentiation via the Lagrange five-point formula \citep[Chapter 3]{NIST:DLMF}, as very high numerical 
precision is unnecessary here. As it is, the frequency $\omega_1(t_n)$ presents short-time oscillations which hide its long-term 
evolution. To suppress such high-frequency fluctuations, we thus define the time series $g_1(t_n)$ as the output of the low-pass 
Kolmogorov-Zurbenko (KZ) filter \citep[][and references therein]{Zurbenko2010} applied to $\omega_1(t_n)$, 
\begin{equation}
\label{eq:freq_g1}
g_1(t_n) = \textrm{KZ}[\omega_1(t_n)] .
\end{equation}
The KZ filter is defined as an iteration of the common moving average and has been comprehensively characterized in 
Appendix \ref{appendix:KZ}. For the present application, we use three iterations of the moving average and a cutoff 
frequency of 1 $\textrm{Myr}^{-1}$ (see Appendix \ref{appendix:KZ}). This particular choice is motivated by the typical 
duration of the libration episodes of the $g_1 - g_5$ resonance, which only last a few million years \citep{Laskar2008,Batygin2008}. 

In the top panel of Fig. \ref{fig:g1_collisions} we present the time evolution of the fundamental frequency $g_1$ along the 42 collisional 
integrations. We isolate in the bottom panel six solutions to show the typical evolution of a single curve more clearly. 
As previously discussed, the curves stop at the first secular collision. The horizontal line represents the constant 
fundamental frequency $g_5$ and thus the location of the $g_1-g_5$ resonance. Within the first 2 Gyr, the chaotic diffusion of 
$g_1$ along the solutions of Fig. \ref{fig:g1_collisions} is limited to no more than 0.25\arcsecyr with respect to its initial value. 
Nevertheless, when lower values of about 5\arcsecyr are eventually reached, the dynamics of $g_1$ starts to be characterized 
by large random variations. This seems to reproduce the large chaotic zone related to the resonance $g_1-g_5$, with a half-width of 
more than 1\arcsecyrNOSPACE, which has been reported in \citep[][Fig. 6(a2)]{Laskar2008}\footnote{Although no analytical computations 
of the $g_1-g_5$ half-width have been reported in literature, such a large value would not be completely unexpected for a linear secular resonance.}. 
Along all the curves, the fluctuations eventually lead to a crossing of the resonance (corresponding locally to a libration of the related 
argument), shortly before the first secular collision. 

\section{A numerical experiment over 100 Gyr}
\label{sect:evolution_100Gyr}
In this last Section, we present the results of a numerical experiment consisting in integrating an ensemble of orbital solutions of 
the forced inner system over 100 Gyr in the future. Through such simulations, we aim to characterize the kind of stochastic 
process which arises from the Hamiltonian \eqref{eq:ham_sec_inn} and drives the destabilization of the inner system, in a regime where the 
highly excited orbits do not represent rare events anymore, as they do in Sect. \ref{sect:evolution_5Gyr}. Obviously, this kind of
simulation does not correspond in anyway to the real future evolution of the orbits of the inner planets over much more than 5 Gyr. Indeed, 
it is well known that over the next 7 Gyr, the Sun will experience a significant mass loss, down to 0.54 $M_\odot$, as it leaves the main 
sequence \citep{Sackmann1993}. 
As a consequence, the semi-major axes of the planet will adiabatically inflate by a factor of 1.85. Even more critically for the fate 
of the inner system, it is likely that the innermost planets Mercury, Venus and the Earth, will be engulfed by the Sun as its radius 
expands along the red giant branch, marking the end of their existence \citep{Rybicki2001,Schroeder2008}. Therefore, this Section 
intends to investigate the very long-term stochastic behaviour of the dynamical system \eqref{eq:ham_sec_inn}, without inferring any 
conclusion on the real evolution of the inner system over such timescales. 

\begin{figure}
\centering
\includegraphics[width=\hsize]{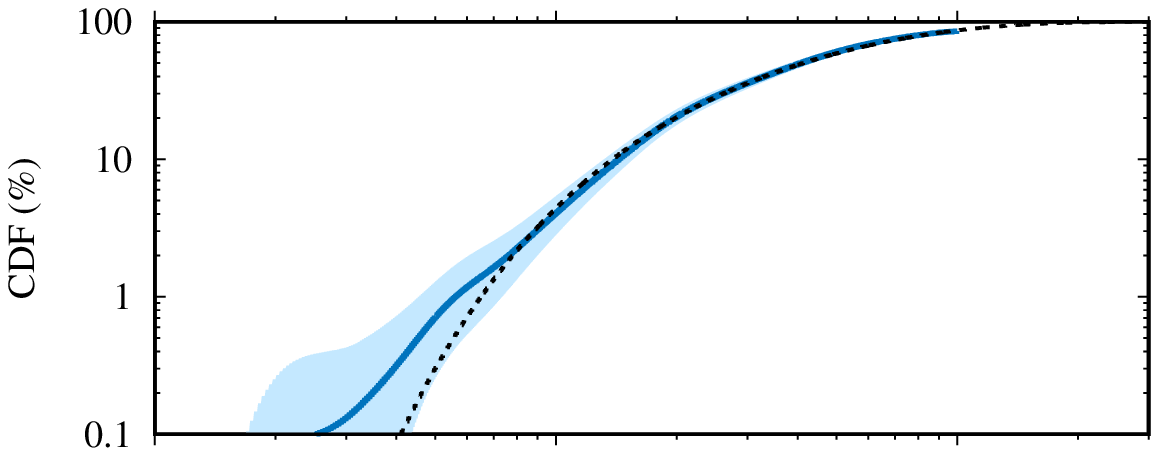}
\includegraphics[width=\hsize]{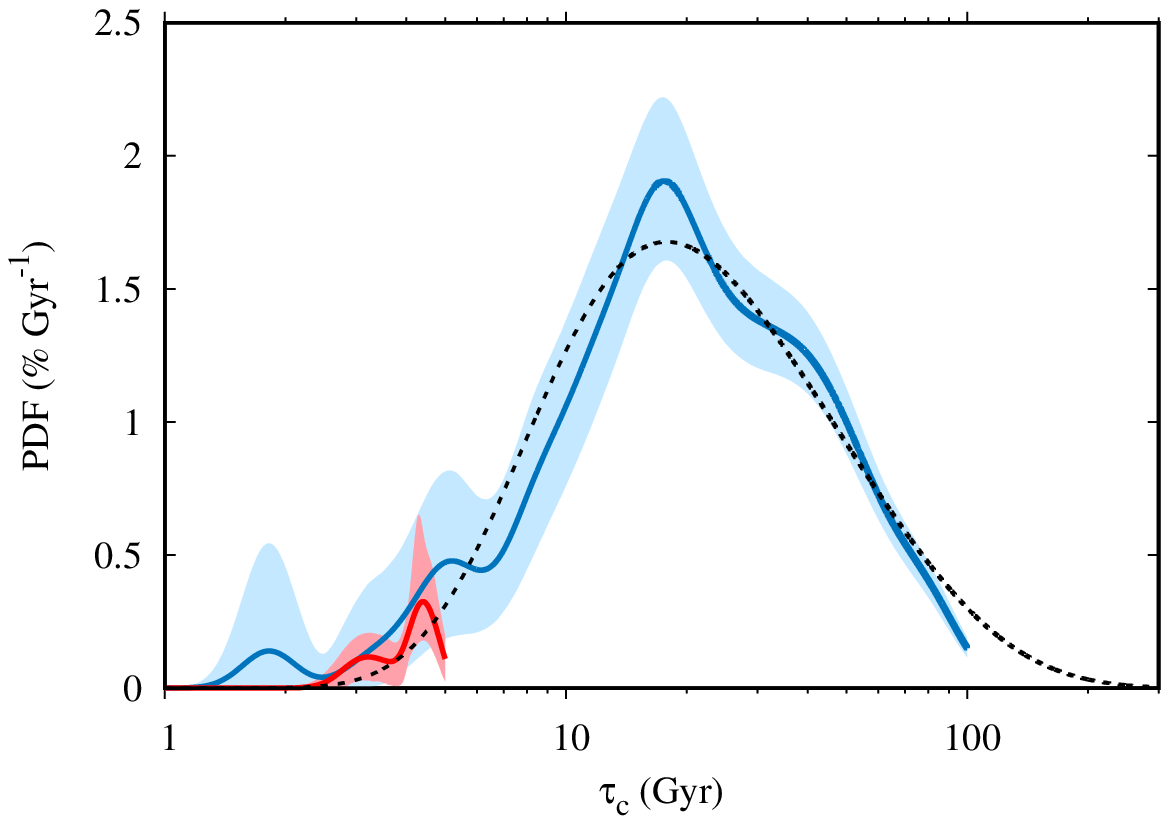}
\caption{Kernel density estimation of the PDF of the time $\tcoll$ of the first secular collision, 
from 1\,042 integrations of the forced inner system over 100 Gyr (blue line in the lower panel). 
The estimation of the cumulative distribution function (CDF) is shown as a blue line in the upper panel. 
The blue regions represent the pointwise confidence intervals at 98\% level from bootstrap.
In red colour the PDF estimation over 5 Gyr from the 10\,560 orbital solutions of Sect. \ref{sect:evolution_5Gyr}. 
The dashed black lines stand for the analytical PDF \eqref{eq:PDF_hitting_time} 
with $T_1$ = 27.6 Gyr and $\alpha$ = 0.9 and the corresponding CDF.} 
\label{fig:collisions_100Gyr}
\end{figure}

We numerically integrate an ensemble of 1\,042 orbital solutions of the forced inner system over 100 Gyr. The initial conditions are 
chosen according the same Gaussian distribution as in Sect. \ref{sect:evolution_5Gyr}, but with different realizations, to obtain different 
evolutions over the first 5 Gyr. As in the previous Section, the numerical integration is stopped at the first secular collision. 
In Fig. \ref{fig:collisions_100Gyr} we show the probability density function of the time $\tcoll$ of the first secular collision. 
In the context of stochastic processes, this represents the hitting time (or first hit time) corresponding to the subspace of the 
phase space realizing a secular collisions between planets. The blue curve in the lower panel of 
Fig. \ref{fig:collisions_100Gyr} represents the PDF estimated via the kernel density estimation (KDE) method 
\citep{Rosenblatt1956,Parzen1962}. We employed a standard Gaussian kernel and Silverman's rule of thumb for the 
selection of the optimal bandwidth \citep{Silverman1986}. The PDF is normalized in such a way that its integral 
over 100 Gyr is equal to the overall percentage of secular collisions obtained in the experiment, i.e. $890/1042 \approx 85.4\%$. 
The blue region in the lower panel of Fig. \ref{fig:collisions_100Gyr} represents the pointwise confidence interval for the PDF at 98\% level, 
obtained via nonparametric bootstrap, i.e. by resampling the original data with replacement \citep{Efron1979}. 
This region thus corresponds pointwisely to the [$1^\mathrm{th}$,$99^\mathrm{th}$]-percentile range of the 
bootstrapped estimated PDF. We emphasize that, for simplicity, we do not correct for the bias arising in the kernel density estimation, 
so that it constitutes, strictly speaking, a confidence region for the expectation of the KDE and not for the true subjacent PDF \citep{Chen2017}. 
When referred to the true PDF, such a confidence region typically represents an undercoverage at the points where the function bends, i.e. 
where its second derivative is large. Even if we could consider simple corrections for the KDE bias \citep{Chen2017}, this would not be 
completely satisfactory given the very small number of events which we deal with over the first few Gyr. Anyway, such corrections on the 
confidence intervals turn out to be sufficiently small to not affect at all the kind of conclusions we infer in this Section. 
To give a further idea of the statistical variance, in Fig. \ref{fig:collisions_100Gyr} we also show in red colour the PDF of $\tcoll$ over the 
first 5 Gyr estimated from the ensemble of 10\,560 orbital solutions presented in Sect. \ref{sect:evolution_5Gyr}. 
We finally report in the upper panel of Fig. \ref{fig:collisions_100Gyr} the estimated cumulative distribution function (CDF) 
as a function of time, which gives the probability of a collisional evolution within a given time, along with its pointwise 
confidence interval at 98\% obtained through bootstrap. 

\subsection{Marginal stability of the inner solar system}
The PDF of the time $\tcoll$ of the first secular collision peaks at 17.6 Gyr, with a 98\% confidence interval (15.4, 20.1) Gyr, while its 
median is 40.8 Gyr, with a 98\% confidence interval (38.2, 43.5) Gyr. The percentage of collisional evolutions over 100 Gyr has a 98\% 
confidence interval (82.8\%, 87.9\%). These findings clarify the state of matastability which characterizes the inner solar system, according to 
the definition of \citep{Laskar1996}. Over a timescale comparable to its age, i.e. over the next 5 Gyr, the inner system is statistically 
very stable, with a probability of only about 0.5\% of an unstable evolution (see Sect. \ref{subsect:sec_collisions}). Nevertheless, here 
we find that such probability increases very fast with time, so that the system rapidly becomes statistically unstable over longer timescales, 
meaning that unstable evolutions do not represent rare events anymore. In particular, the probability of an instability is about 20\% at 20 Gyr already, 
and 50\% at 40 Gyr. In the conjecture of \citep{Laskar1996}, this picture should characterize any secularly evolving planetary system: at 
each moment of its formation history, a planetary system should be in a state of marginal stability, that is practically stable over 
a timescale comparable to its age, while strong instabilities arise very fast over longer times, resulting in more stable configurations 
of the surviving planets. 

\subsection{Destabilization of the inner solar system as a stochastic process}
Recently, in the framework of the simplified model introduced in \citep{Batygin2015}, \citet{Woillez2020} applied the theory 
of the white noise limit for slow-fast dynamical systems \citep[e.g.,][]{Gardiner1985}, to describe the long-term dynamics of 
Mercury through an effective stochastic process. They show that the instability timescale of Mercury orbit is reproduced by the 
hitting time of a one-dimensional Wiener process with a reflecting barrier. The corresponding PDF peaks at 1.3 Gyr 
\citep[Figure 4]{Woillez2020}, in agreement with the analysis in \citep{Batygin2015}. Such a timescale is an order of magnitude 
smaller than that in Fig. \ref{fig:collisions_100Gyr}, and the simplified dynamics of Mercury predicts a large probability of 
instability over the solar system age, in contrast with the findings of realistic models. 

In this Section, we reconsider the stochastic process discussed in \citep{Woillez2020} to characterize the overall structure 
of the PDF in Fig. \ref{fig:collisions_100Gyr}. To fix ideas, it could describe in a crude way the long-term wanderings of 
the fundamental frequency $g_1$ towards the destabilizing secular resonance $g_1-g_5$ (see Sect. \ref{subsect:sec_res_g1_g5}). 
Indeed, numerically, the PDF of $g_1$ turns out to be asymmetric, with a longer tail diffusing towards low values, while 
its upper end is practically fixed at about 6\arcsecyrNOSPACE \citep[][see also Fig. \ref{fig:g1_collisions}]{Hoang2021}. 
Therefore, let suppose that $g_1$ performs a random walk described 
by a Wiener process starting at $g_{1,0} = g_1(t=0)$, and consider an upper reflecting barrier at 
$g_{1,\textrm{max}} > g_{1,0}$. We assume that a secular collisions rapidly occurs whenever the secular resonance 
$g_1 - g_5$ is reached, as shown in Fig. \ref{fig:g1_collisions}; this is taken into account by considering an absorbing 
barrier at $g_1 = g_5$. The PDF $\rho(\tau)$ of the hitting time $\tau = \inf_t\{g_1(t) \leq g_5\}$ is thus given by 
\citep[e.g.,][]{Schwarz1992,Woillez2020} 
\begin{equation}
\label{eq:PDF_hitting_time}
\rho(\tau) = \frac{\pi}{2 T_1} 
\sum_{n=0}^{\infty} \left(n+\frac{1}{2}\right) \sin \left[ \pi \left(n+\frac{1}{2}\right) \alpha \right] 
\E^{-\pi^2 \left(n+\frac{1}{2}\right)^2 \tau / 4T_1} ,
\end{equation}
where $\alpha = (g_{1,0} - g_5)/(g_{1,\textrm{max}} - g_5)$ and $T_1 = (g_{1,\textrm{max}} - g_5)^2/4D$, with $D$ being the diffusion coefficient 
of the Wiener process. After a least-squares search, we plot in Fig. \ref{fig:collisions_100Gyr} as a dashed black line the curve \eqref{eq:PDF_hitting_time} with 
parameters $T_1$ = 27.6 Gyr and $\alpha$ = 0.9. Even if the real random walk performed by the frequency $g_1$, and the destabilization 
mechanism in general, are likely to be much more complex, a Wiener process with a reflecting barrier is able to reproduce the overall 
behaviour of the observed PDF, at the level of the present statistical precision and apart from a certain excess of events in the 
tail of the distribution\footnote{We stress that such a result is independent of the hypothesis of which precise dynamical quantity 
(if any) actually undergoes the stochastic process considered here.}. On the one hand, this shows that at short times the PDF behaves as 
\begin{equation}
\label{eq:Levy}
\rho(\tau) = \left(\frac{T_0}{\pi \tau^3} \right)^\frac{1}{2} \E^{-\frac{T_0}{\tau}}, \quad \textrm{for } \tau \ll T_1 , 
\end{equation}
where $T_0 = \alpha^2 T_1$. 
In this regime, the PDF coincides with that of the hitting time of a standard Wiener process (i.e., without reflecting barrier). 
The very fast decaying of the density probability for $\tau \rightarrow 0$ in Eq. \eqref{eq:Levy} characterizes 
the rare destabilizations of Mercury orbit over the first few billions of years. In particular, the instanton phenomenology 
described in \citep{Woillez2020} could indeed apply to the real solar system (such a hint still needs to be tested through 
integration of realistic models, as the present one.). 
On the other hand, Eq. \eqref{eq:PDF_hitting_time} shows that at large times the decay of the observed PDF is at least exponentially fast, 
with a characteristic time equal to $4 T_1$. This behaviour results from the reflecting barrier taken into account in the stochastic 
process; it would not be reproduced by a standard Wiener process, as the corresponding PDF has a polynomial heavy tail proportional to 
$\tau^{-3/2}$, as shown by Eq. \eqref{eq:Levy}. In particular, the average of the hitting time as given by Eq. \eqref{eq:PDF_hitting_time} 
is finite, differently from Eq. \eqref{eq:Levy}. 

We note that \citep{Woillez2020} reported $T_0$ = 1.56 Gyr, a value which is more than ten times smaller 
than that in Fig. \ref{fig:collisions_100Gyr}, i.e. $T_0$ = 22.4 Gyr, in agreement with the previous discussion about 
the validity of the simplified dynamics of Mercury in \citep{Batygin2015}. Here we emphasize that the incompatibility with 
realistic models seems to be related to the limitations of some of the simplifying assumptions\footnote{Freezing the 
fundamental frequencies of the inner planet orbits other than $g_1$ and $s_1$ is probably one of the sources of these discrepancies.}. 
Indeed, as discussed in Sect. \ref{sect:evolution_5Gyr}, averaged models generally tends to produce slower instabilities with respect to the full 
dynamics, because the number of degrees of freedom is smaller. Nevertheless, the model of \citep{Batygin2015} produces 
instabilities which are ten times faster than in the present dynamics, even though the latter includes all 
the dynamical interactions considered in the simplified one. Such a lack of continuity in the destabilization time with 
respect to the complexity of the dynamical model seems rather severe. 

We conclude this Section by noting that a lower bound for the typical instability time of Mercury orbit can be estimated from the maximum 
Lyapunov exponent (Fig. \ref{fig:MLE}), if one assumes that no dynamical constraints prevent $g_1$ to diffuse at  
a rate determined by the leading secular resonances. In particular, the resonance $(g_1-g_5)-(s_1-s_2)$, with a 
libration frequency of 0.12$\arcsec$ yr$^{-1}$, is among the leading ones affecting the frequency $g_1$ \citep{Laskar1990}. 
We may estimate an upper bound on the diffusion coefficient $D$ as 
\begin{equation}
\label{eq:diffusion_coeff}
D_\textrm{max} = \frac{\left(2\pi \, \textrm{MLE}\right)^2}{\textrm{MLE}^{-1}} 
\approx 0.77 \left( \textrm{arcsec yr}^{-1} \right)^2 \textrm{Gyr}^{-1} ,
\end{equation}
with $2\pi \, \textrm{MLE}$ = 0.1\arcsecyr (this is also the lower limit for the MLE in Fig. \ref{fig:MLE}). 
A lower bound on the typical destabilization time of Mercury orbit is thus given by 
\begin{equation}
\label{eq:diffusion_time}
T_{0,\textrm{min}} = \frac{(g_{1,0} - g_5)^2}{4 D_\textrm{max}} 
\approx 0.56 \, \textrm{Gyr} ,
\end{equation}
with $g_{1,0}$ = 5.577\arcsecyr (Table \ref{tab:fund_freqs}) and $g_5$ = 4.257\arcsecyr. Eq. \eqref{eq:diffusion_time} 
gives essentially the same diffusion timescale as in \citep{Batygin2015,Woillez2020}. This is already two order of magnitude 
greater than the Lyapunov time of the system, but still insufficient to explain the findings of Fig. \ref{fig:collisions_100Gyr}. 
This means that some dynamical constraints effectively makes the diffusion towards the $g_1-g_5$ resonance slower 
than in the estimates \eqref{eq:diffusion_coeff}, \eqref{eq:diffusion_time}. Certain secondary resonant 
harmonics, in particular, must play a determinant role in the destabilization of the inner solar system. 

\section{Conclusions}
This work introduces a new secular model for the dynamics of the inner planets of the solar system. It exploits the practical 
constancy of the fundamental precession frequencies of the outer planet orbits over timescales of billion years, the smallness 
of the inner planet masses and the absence of strong mean-motion resonances in the inner system. 
The range of validity of the resulting dynamics extends to very high eccentricities and inclinations, up to a secular planetary 
collision, i.e. the geometric intersection of the instantaneous Keplerian orbits of two planets. The new model can be studied 
analytically by truncating the Hamiltonian at a given total degree in planet eccentricities and inclinations, with the aid of 
a computer algebra system like TRIP. It can also be integrated numerically in a very efficient way through Gauss's method. 
The orbital solution matches the predictions of a comprehensive model of the solar system at a very satisfactory level over timescales 
shorter than or comparable to the Lyapunov time. The new model also correctly reproduces the maximal Lyapunov exponent of 
the inner system and the statistics of the highly eccentric Mercury orbits over the next 5 billion years. Moreover, the 
destabilizing role of the secular resonance $g_1-g_5$ clearly stands out. 
We performed a numerical experiment consisting of a thousand orbital solutions of the inner solar system over one hundred billion years, 
to explore a regime in which unstable orbits are statistically common. We first pointed out the fast growth of the rate of orbit instabilities 
in the framework of planetary system formation through successive metastable states. We then showed that the 
PDF of the time of the first secular collision is reasonably well reproduced by a Wiener process with a 
reflecting barrier, which could be performed, for example, by the fundamental precession frequency $g_1$. 
Given the robustness of the statistical predictions of the present dynamical model over 5 Gyr, the main 
properties of this PDF, that is the characteristic time of its peak and the behaviours at short and 
long times, are likely to represent what would arise from the full dynamics of the planets. 
We finally argued that a dynamical mechanics is needed to explain the rarity of the large excursions of $g_1$ 
up to the $g_1-g_5$ secular resonance within the next 5 Gyr. 

We emphasize that the new dynamical model can be straightforwardly implemented once truncated at degree 4 in planet 
eccentricities and inclinations, by using the corresponding expression of the two-body secular Hamiltonian reported in 
\citep[Appendix]{Laskar1995} and the quasi-periodic secular solution for the outer planets given in Appendix \ref{appendix:QPSO}. 

\begin{acknowledgements}
FM acknowledges the invaluable support of M. Gastineau in the implementation of the algebraic computations in TRIP and 
the fruitful discussions with N. H. Ho\`{a}ng. FM has been supported by a PSL post-doctoral fellowship. 
This project has received funding from the European Research Council (ERC) under the European Union’s Horizon 2020 research 
and innovation program (Advanced Grant AstroGeo-885250) and from the French Agence Nationale de la Recherche (ANR) 
(Grant AstroMeso ANR-19-CE31-0002-01). This work was granted access to the HPC resources of MesoPSL financed by the Region 
Île-de-France and the project Equip@Meso (reference ANR-10-EQPX-29-01) of the programme Investissements d’Avenir supervised 
by the Agence Nationale pour la Recherche. 
\end{acknowledgements}

%
\bibliographystyle{bibtex/aa} 
\bibliography{iss} 
%

\begin{appendix}

\section{Indirect part of the two-body Hamiltonian perturbing function}
\label{appendix:1}
We outline the derivation of the coefficients $\widetilde{\mathcal{T}}_{\ell, \ell^\prime}$ of the indirect part of 
the two-body perturbation in the Fourier expansion \eqref{eq:kinetic_expansion}, by adapting the 
presentations of \citep{Laskar1991,Laskar1995,Laskar2010}. From Sect. \ref{sect:dyn_model}, one has 
\begin{equation}
\label{eq:kinetic_term}
\mathcal{T}_1 = 
\frac{\tilde{\vec{r}} \cdot \tilde{\vec{r}}^\prime}{m_0} = 
\frac{\mu \mu^\prime}{m_0} \vec{\varv} \cdot \vec{\varv}^\prime .
\end{equation}
where $\vec{\varv} = \tilde{\vec{r}} / \mu$ denotes the velocity vector tangent to the instantaneous Keplerian 
ellipse of a given planet. In the reference frame of the Keplerian orbit, with origin at the Sun, $x$-axis directed 
towards the pericenter and $z$-axis normal to the orbital plane, the components of the vector $\vec{\varv}$ read 
\begin{equation}
\vec{\varv} = 
\frac{n a^2}{r} 
\begin{pmatrix}
- \sin \eccanomaly \\ \sqrt{1-e^2} \cos \eccanomaly \\ 0
\end{pmatrix} ,
\end{equation}
where $\eccanomaly$ is the eccentric anomaly, $n = \sqrt{G(m_0+m)/a^3}$ the mean motion and $r = ||\vec{r}||$ 
the heliocentric orbital distance. After transformation to the fixed reference frame, the velocity components are 
given by 
\begin{equation}
\label{eq:velocity_2}
\vec{\varv} = 
\frac{n a^2}{r} 
\mathcal{R}(i,\Omega) \mathcal{R}_3(\varpi) 
\begin{pmatrix}
- \sin \eccanomaly \\ \sqrt{1-e^2} \cos \eccanomaly \\ 0
\end{pmatrix} ,
\end{equation}
where the rotation matrices $\mathcal{R}(i,\Omega) = \mathcal{R}_3(\Omega) \mathcal{R}_1(i) \mathcal{R}_3(-\Omega)$ 
and $\mathcal{R}_3(\varpi)$ are given in \citep{Laskar1991,Laskar1995}. 

We introduce the eccentric longitude $F = \eccanomaly + \varpi$, which obeys the modified Kepler's equation 
\begin{equation}
\label{eq:mod_Kepler}
\lambda = 
F + \operatorname{Im} \left( z \, \E^{-j F} \right) = 
F + \gamma \operatorname{Im} \left(\mathcal{X} \, \E^{-j F}\right) ,
\end{equation}
where $z = e \E^{j \varpi}$ and $\gamma = \sqrt{1 - \mathcal{X} \bar{\mathcal{X}}/4}$. By using the fact that 
$1 - r/a = \gamma \operatorname{Re} \left( \mathcal{X} \, \E^{-j F} \right)$, and after some algebra, one finds  
\begin{equation}
\mathcal{R}_3(\varpi) 
\begin{pmatrix}
- \sin \eccanomaly \\ \sqrt{1-e^2} \cos \eccanomaly \\ 0
\end{pmatrix} = 
\begin{pmatrix}
-\sin F + (2\gamma)^{-1} \operatorname{Im} \left( \mathcal{X} \right) \left(1 - r/a\right) \\ 
\phantom{-} \cos F - (2\gamma)^{-1} \operatorname{Re} \left( \mathcal{X} \right) \left(1 - r/a\right) \\ 
0
\end{pmatrix} .
\end{equation}
By employing the complex formalism of \citep{Laskar1995}, one can thus write 
\begin{equation}
\label{eq:variable_Z}
\begin{aligned}
&\frac{n a^2}{r}
\mathcal{R}_3(\varpi) 
\begin{pmatrix}
- \sin \eccanomaly \\ \sqrt{1-e^2} \cos \eccanomaly \\ 0
\end{pmatrix} = 
\begin{pmatrix}
\mathcal{Z} \\ 0
\end{pmatrix} , \\
&\mathcal{Z} = jna \left[ \frac{a}{r} \E^{jF} 
- \frac{\mathcal{X}}{2\gamma} \left( \frac{a}{r} - 1 \right)  \right] .
\end{aligned}
\end{equation}
The application of matrix $\mathcal{R}(i,\Omega)$ in Eq. \eqref{eq:velocity_2} then gives 
\begin{equation}
\mathcal{R}(i,\Omega) 
\begin{pmatrix}
\mathcal{Z} \\ 0
\end{pmatrix} = 
\operatorname{Re} \left( \mathcal{Z} \vec{\mu} \right) , \quad
\vec{\mu} = \delta^{-1}
\begin{pmatrix}
\eta^2 + \bar{\mathcal{Y}}^2 \\ 
-j \left( \eta^2 - \bar{\mathcal{Y}}^2 \right) \\
-2 j \eta \bar{\mathcal{Y}}
\end{pmatrix} ,
\end{equation}
with $\delta = 1 - \mathcal{X} \bar{\mathcal{X}}/2$ and 
$\eta = \sqrt{1 - \mathcal{X} \bar{\mathcal{X}}/2 -\mathcal{Y} \bar{\mathcal{Y}} }$.
The scalar product in Eq. \eqref{eq:kinetic_term} thus reads 
\begin{equation}
\vec{\varv} \cdot \vec{\varv}^\prime = 
\frac{1}{2} \operatorname{Re} 
\left( 
\mathcal{Z} \mathcal{Z}^\prime \vec{\mu} \cdot \vec{\mu}^\prime + 
\mathcal{Z} \bar{\mathcal{Z}}^\prime \vec{\mu} \cdot \bar{\vec{\mu}}^\prime
\right).
\end{equation}
The coefficients $\widetilde{\mathcal{T}}_{\ell, \ell^\prime}$ are readily derived once 
the quantities $a/r$ and $(a/r) \E^{jF}$, appearing in Eq. \eqref{eq:variable_Z}, are expanded 
in Fourier series of the mean longitude $\lambda$. One has 
\begin{equation}
\frac{a}{r} = \sum_{\ell=-\infty}^{+\infty} c_\ell \E^{j \ell \lambda}, \quad \textrm{with} \quad
c_\ell = \frac{1}{2\pi} \int_0^{2\pi} \frac{a}{r} \E^{-j \ell \lambda} d\lambda .
\end{equation}
By using Eq. \eqref{eq:mod_Kepler} and the fact that $d\lambda/dF = r/a$, a classical calculation gives 
$c_\ell = J_{\ell}(\ell e) \, \E^{-j \ell \varpi}$, where the $J_{\ell}$ are the Bessel functions of the first kind. The following alternative expression of the Fourier coefficients, 
\begin{equation}
c_\ell = \sum_{m=-\infty}^{+\infty} (-j)^m J_m(\ell \gamma \operatorname{Im}\mathcal{X}) 
\, J_{\ell-m}(\ell \gamma \operatorname{Re}\mathcal{X}) ,
\end{equation}
can be straightforwardly expanded in series of $\mathcal{X}$ and $\bar{\mathcal{X}}$, 
and is well-defined at zero eccentricity. In a similar manner, one obtains 
\begin{equation}
\frac{a}{r} \E^{jF}  = \sum_{\ell=-\infty}^{+\infty} 
\left[ \sum_{m=-\infty}^{+\infty} (-j)^m J_m(\ell \gamma \operatorname{Im}\mathcal{X}) 
\, J_{\ell-1-m}(\ell \gamma \operatorname{Re}\mathcal{X}) \right]
\E^{j \ell \lambda} .
\end{equation}

\section{Kolmogorov-Zurbenko filter}
\label{appendix:KZ}
Low-pass filters are employed in time-series analysis to extract the long-term (low-frequency) components of a given signal. 
A dedicated low-pass filter has been constructed in \citep{Carpino1987}, to recover the secular changes of the giant 
planet orbits from the numerical output of a N-body integration. Here we propose the use of the Kolmogorov-Zurbenko filter 
\citep{Zurbenko2010} as an out-of-the-box, computationally advantageous and still effective choice. 

Let consider the real-valued time series 
\begin{equation}
\xi_n = \xi(n \Delta t), \quad n \in \mathbb{Z} ,
\end{equation}
resulting from the discrete sampling of a continuous signal $\xi(t)$, with constant sampling rate $\omega_s = 2\pi/\Delta t$. In the frequency domain, the time series is characterized by its discrete-time Fourier transform 
\begin{equation}
\tilde{\xi}(\omega) = \sum_{n=-\infty}^{+\infty} \xi_n \E^{-j \omega n \Delta t} ,
\end{equation}
which is a periodic function with period $\omega_s$ and can be thus restricted to the interval $|\omega| \leq \omega_s/2$.
The action of a finite-impulse-response filter F on the time series is defined as the discrete convolution of the signal $\xi_n$ 
with a given finite sequence $d_n$, 
\begin{equation}
\textrm{F}[\xi]_n = (d \ast \xi)_n = \sum_{m=-M}^{M} d_m \xi_{n-m} ,
\end{equation}
$M$ being a non-negative integer. The sequence $d_n$ is the impulse response function of the filter, and $L_M=2M+1$ is the filter length. A liner 
filter is characterized by the discrete-time Fourier transform $\tilde{d}(\omega)$ of its impulse response, which is called the frequency 
response of the filter and gives the spectrum of the output signal as 
\begin{equation}
\widetilde{\textrm{F}}[\xi](\omega) = \tilde{d}(\omega) \tilde{\xi}(\omega) .
\end{equation}
Ideally, a low-pass filter should be characterized by a null response above a given cutoff frequency, 
i.e. $\tilde{d}(\omega)=0$ for $\omega > \omega_c$, while the response should be unitary at lower frequencies, 
i.e. $\tilde{d}(\omega)=1$ for $\omega \leq \omega_c$. Moreover, the filter must not change the phase of the Fourier 
components of the signal, i.e. $\operatorname{Im}(\tilde{d}(\omega)) = 0$. This is achieved by requiring 
$d_n = d_{-n} \in \mathbb{R}$ \citep{Carpino1987}. Nevertheless, a real filter is rather characterized by: a passband, 
$|\hat{d}(\omega)-1| \leq \rho$ for $|\omega| \leq \omega_p$, $\rho$ being a given maximum 
loss; a stopband, $|\hat{d}(\omega)| \leq \alpha$ for $|\omega| \geq \omega_c$, $\alpha$ being a given maximum gain; 
a transition band between $\omega_p$ and $\omega_c$, where the response function smoothly decreases from $\approx$1 to $\approx$0.

\begin{figure}
\centering
\includegraphics[width=\hsize]{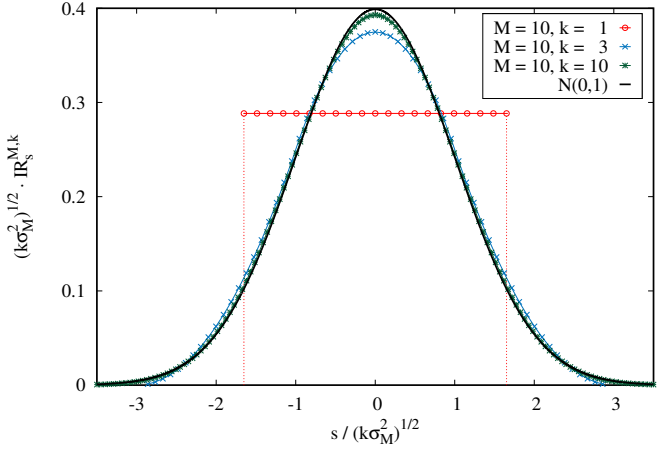}
\caption{Impulse response of the Kolmogorov-Zurbenko filter (the moving average corresponds 
to the case $k=1$). The axes have been normalized to show the asymptotic Gaussian behaviour of the 
function, the standard normal distribution $N(0,1)$ being represented by the black solid curve.} 
\label{fig:KZ_IRF}
\end{figure}

\begin{figure}
\centering
\includegraphics[width=\hsize]{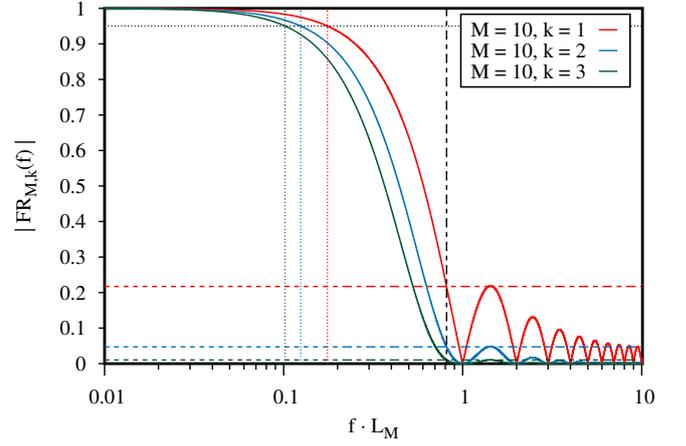}
\caption{Absolute value of the frequency response for the Kolmogorov-Zurbenko filter 
(the moving average corresponds to the case $k=1$). The vertical black dashed line stands for the cutoff 
frequency $f_c$ given by Eq. \eqref{eq:cutoff_freq_2}, while the horizontal coloured ones represent the maximum 
gains in the stopband $\alpha_k = \eta^k$. The vertical coloured dotted lines show the passband frequencies 
$f_p$ given by Eq. \eqref{eq:passband_2} for a loss level of 5\%, which is represented by the horizontal black 
dotted line.} 
\label{fig:KZ_FR}
\end{figure}

One of the simplest low-pass filters is the moving average (MA), 
\begin{equation}
\label{eq:moving_average}
\textrm{MA}_M[\xi]_n = \frac{1}{2M+1} \sum_{m=-M}^{M} \xi_{n-m} ,
\end{equation}
which simply constitutes the local unweighted average of the signal over a time window $L_M \Delta t$. 
Computationally, the moving average is very advantageous; however, it is well-known that it provides a poor attenuation 
of the Fourier components of the signal in the stopband (see Fig. \ref{fig:KZ_FR}). Kolmogorov proposed to 
bypass such a problem by applying the moving average iteratively \citep[and reference therein]{Zurbenko2010}. 
The Kolmogorov-Zurbenko filter is thus defined as 
\begin{equation}
\label{eq:KZ_filter}
\begin{aligned}
&\textrm{KZ}_{M,k=1}[\xi] = \textrm{MA}_M[\xi], \\
&\textrm{KZ}_{M,k}[\xi] = \textrm{MA}_M[\textrm{KZ}_{M,k-1}[\xi]], \quad k \geq 2 .
\end{aligned}
\end{equation}
The output signal of such a filter is given by 
\begin{equation}
\label{eq:KZ_impulse_response}
\begin{aligned}
&\textrm{KZ}_{M,k}[\xi]_n = \frac{1}{(2M+1)^k} \sum_{m_1=-M}^M \cdots \sum_{m_k=-M}^M \xi_{n-\sum_{\ell=1}^k m_\ell} = \\
&= \sum_{s=-kM}^{kM} \frac{C_s^{M,k}}{(2M+1)^k} \xi_{n-s} ,
\end{aligned}
\end{equation}
where $C_s^{M,k}$ is the number of ways of choosing $k$ integers in the interval $[-M,M]$ such that their sum is equal to $s$. 
The numbers $C_s^{M,k}$ can be expressed as the coefficients of the finite Laurent series $\left( \sum_{\ell=-M}^M z^\ell \right)^k$ \citep{Zurbenko2010}, i.e. 
\begin{equation}
\sum_{s=-kM}^{kM} C_s^{M,k} z^s = 
\left(z^{-M} + \ldots + 1 + \ldots + z^M \right)^k .
\end{equation}
The impulse response (IR) of the KZ filter given in Eq. \eqref{eq:KZ_impulse_response}, $\text{IR}^{M,k}_s = C_s^{M,k} / L_M^k$, 
can be interpret as a discrete probability distribution function (PDF) over the interval $[-kM,kM]$, resulting from the convolution of $k$ 
uniform distribution over $[-M,M]$. The central limit theorem implies that, for $k \gg 1$, such PDF is asymptotically Gaussian, 
with zero mean and variance equal to $k\sigma^2_M$, $\sigma^2_M = \left(L_M^2 - 1\right)/12$ being the variance of the discrete uniform 
distribution over $[-M,M]$. This is shown in Fig. \ref{fig:KZ_IRF}. Therefore, the width of the impulse response scales as $\sqrt{k} M$ 
for $M,k \gg 1$, even if the filter length $L_{kM}$ is linear in $k$.  

\subsection{Frequency response}

\begin{table*}
\caption{Frequency response of the KZ filter for $M\gg1$. Maximum gain in the stopband $\alpha_k = \eta^k$. 
Ratio of the cutoff frequency $f_c$ to the passband frequency $f_p$ for two different maximum loss $\rho$.} 
\label{tab:KZ_filter}
\centering
\begin{tabular}{c c c c c c} 
\hline\hline 
\rule{0pt}{1em}
$k$ & 1 & 2 & 3 & 4 & 5 \\ 
\hline
\rule{0pt}{1em}
$\alpha_k$ & 2.17 $\cdot$ 10$^{-1}$ & 4.72 $\cdot$ 10$^{-2}$ & 1.03 $\cdot$ 10$^{-2}$ & 2.23 $\cdot$ 10$^{-3}$ & 4.84 $\cdot$ 10$^{-4}$ \\ 
$f_c/f_p$, $\rho = 5\%$ & 4.63 & 6.53 & 7.99 & 9.22 & 10.3 \\ 
$f_c/f_p$, $\rho = 1\%$ & 10.4 & 14.7 & 18.0 & 20.8 & 23.3 \\ 
\hline
\end{tabular}
\end{table*}

The following derivations describe the behaviour of the KZ filter in the $M\gg1$ regime. Its frequency response (FR) is simply 
the $k^{\textrm{th}}$ power of that of the moving average, 
\begin{equation}
\textrm{FR}_{M,k}(f) =  
\left[\frac{\sin(\pi f L_M)}{L_M \sin(\pi f)}\right]^k, 
\quad |f| \leq \frac{1}{2} ,
\end{equation}
where $f = \omega \Delta t / 2\pi$ is the angular frequency in units of the sampling rate $\omega_s$. The zeros of the 
frequency response are given by 
\begin{equation}
\textrm{FR}_{M,k}(f) = 0 \iff f L_M \in \mathbb{Z} \setminus \{0\} ,
\end{equation}
as shown in Fig. \ref{fig:KZ_FR}. 
Following \citep{Carpino1987}, we define the dimensionless cutoff frequency $f_c$ as 
\begin{equation}
\label{eq:cutoff_freq}
|\textrm{FR}_{M,k}(f)| \leq \alpha_k 
\quad \textrm{for } |f| \geq f_c ,
\end{equation}
where we have defined the maximum gain in the stopband $\alpha_k = |\textrm{FR}_{M,k}(f^\star_M)|$, 
$f^\star_M$ being the frequency of the first local maximum of $|\textrm{FR}_{M,k}(f)|$ for $f>0$. In the regime 
$M\gg1$, such frequency is given by the first positive solution $x^\star$ of the equation 
\begin{equation}
x = \tan(x) ,
\end{equation}
where $x = \pi f L_M$. Numerically, one finds $x^\star \approx 4.49$. Therefore, we obtain 
$f^\star_M L_M = x^\star/\pi \approx 1.43$, independently of $k$, and the gain $\alpha_k = \eta^k$, with 
$\eta = \lvert\sin(x^\star)\rvert/x^\star \approx 0.217$, which is thus asymptotically independent of $M$. These results are 
shown in Fig. \ref{fig:KZ_FR}. The cutoff frequency $f_c$ can be computed from Eq. \eqref{eq:cutoff_freq} through 
the first positive solution $x_c$ of the equation 
\begin{equation}
\eta x = \lvert \sin(x) \rvert , 
\end{equation}
which is found numerically to be $x_c \approx 2.55$. The cutoff frequency of the KZ filter is thus given by 
\begin{equation}
\label{eq:cutoff_freq_2}
f_c L_M = x_c/\pi \approx 0.813 , 
\end{equation}
independently of $M$ 
and $k$ in the asymptotic regime $M\gg1$. 
Figure \ref{fig:KZ_FR} and Table \ref{tab:KZ_filter} show that, with a small number $k$ of iterations, the KZ filter is already able to provide a strong attenuation
in the stopband. Given its straightforward numerical implementation through Eq. \eqref{eq:KZ_filter}, the KZ filter turns out 
to be an effective general choice for a low-pass filter. However, choosing a very large number $k$ of iterations would reduce 
the passband in the frequency response of the filter. Indeed, we define the dimensionless passband $f_p$ as 
\begin{equation}
\label{eq:passband}
\textrm{FR}_{M,k}(f) \geq 1 - \rho,
\quad \textrm{for } |f| \leq f_p ,
\end{equation}
where $0 < \rho \ll 1$ is a given maximum loss. In the regime $M\gg1$, by developing the frequency response in a 
Taylor series about $x = 0$, one has 
\begin{equation}
1 - \frac{x_p^2}{6} + \frac{x_p^4}{120} + \mathcal{O}(x_p^6) = (1-\rho)^{1/k}, 
\quad x_p \ll 1 .
\end{equation}
By neglecting terms of order $\mathcal{O}(x_p^6)$, the pertinent solution is given by 
\begin{equation}
\label{eq:passband_2}
f_p L_M = \frac{\sqrt{10}}{\pi} \sqrt{1 - \sqrt{1 - \frac{6}{5} \bigg[1 - (1-\rho)^{1/k} \bigg]}} .
\end{equation}
At first order in $\sqrt{\rho / k}$, one obtains 
\begin{equation}
f_p L_M = \frac{\sqrt{6}}{\pi} \sqrt{\frac{\rho}{k}} + \mathcal{O}\left(\frac{\rho}{k}\right) ,
\end{equation}
which shows that the passband shrinks as $k^{-1/2}$, as shown in Fig. \ref{fig:KZ_FR}. Such a 
dependence on $k$ is fortunately sublinear. Generally speaking, a reasonable choice of the number of iterations 
should be restricted in our opinion to $3 \leq k \lesssim 5$, as suggested by Table \ref{tab:KZ_filter}. 
We note that the resulting transition band can easily span more than one decade of frequency, and it is thus not narrow as in more complex filters \citep[e.g.,][]{Carpino1987}. This is the major drawback of the simplicity of the KZ filter.

\section{Additional statistics for \citep{Laskar2009}}
\label{appendix:LG2009}

In \citep{Laskar2009}, 2501 numerical integrations of the full solar system were performed over 5 Gyr 
using the SABA4 high-order symplectic integrator \citep{Laskar2001}. In most cases, the stepsize was constant, 
but for the integrations that suffered close planetary encounters or effective collisions, 
the stepsize was very much reduced in order to conserve the accuracy of the integration. This led to very long 
integration times for some of the runs and all the solutions were not completed at the time of the publication of the paper. 
Indeed, all 2501 runs were started on August 7, 2008. Among them, 2472 were finished in 2008, but 22 ended in 2009 
and still 7 in 2010, the last one ending on  June, 15, 2010. We present here the full statistics resulting from these 
numerical integrations, which have since been presented in many conferences, but never published. 
In Table \ref{Tab:08} we give the total number of solutions for which the eccentricity of Mercury reaches a given 
value over 500 Myr to 5000 Myr. This table is similar to the Table S1 of \citep{Laskar2009} but provides the total 
number of solutions instead of the percentage. In the last column, we also provide 
the values reached after full completion of the 
numerical integrations, in June 2010. All the other numbers were unchanged. 
 
\begin{table}[h]
\begin{tabular}{rrrrrrrr}
\hline\hline
 $e_{m0}$      &500   &        1000    &     1500&     2000& 3000 & 4000& 5000\phantom{/22}\\
 \hline
  0.35  &    75 &   227 &  358 &  504  &   795  &   1046 &    1231\phantom{/22} \\
  0.40  &     7 &   49 &   97 &   167  &   315  &   472 &      639\phantom{/22} \\
  0.50  &     0 &    0 &    3 &    7  &    24  &    51 &       99\phantom{/22}  \\
  0.60  &     0 &    0 &    1 &    2  &     4  &     12 &      24\phantom{/22}  \\
  0.70  &     0 &    0 &    1 &    2  &     3  &     11 &      22/23  \\
  0.80  &     0 &    0 &    1 &    2  &     2  &     10 &      21/22  \\
  0.90  &     0 &    0 &    1 &    2  &     2  &     8 &       19/21  \\
\hline
  \end{tabular}
  \caption{Number of solutions that reach a given   
  eccentricity of Mercury ($e_{m0}$)  over a given time (500, 1000, 1500, 
  2000, 3000, 4000, 5000 Myr). The statistics  are made over 2501 orbital solutions as published in \citep{Laskar2009}. In the 3 bottom rows of the last column (5 Gyr), 
  the second number is the number reached after full completion of all the integrations, in June 2010.
  } 
  \label{Tab:08}
\end{table}

In the last column of Table \ref{Tab:08}, the number of solutions for which the eccentricity of Mercury extends beyond 0.9 increased from 19 to 21. 
Among these 21 solutions, 6 present a collision of Mercury with Venus (i.e. their center of mass distance is smaller than the sum of their radii),
9 a collision of Mercury with the Sun, 5 reach 5 Gyr without collision, but with very close encounters 
(for example, one solution presents an encounter of Mercury with Venus with less than 
1800 km between the two surfaces), and one solution has a very close encounter of Mars with the Earth (surface distance of less than 794 km) followed by a collision of Mars with the Sun.

\section{Quasi-periodic secular solution for the outer planets}
\label{appendix:QPSO}

\begin{table*}[ht!]
\centering
{\small
\begin{tabular}{C|C C C C}
\hline\hline 
\rule{0pt}{1.0em} 	
 k   &     \langle a_k \rangle \ (\mathrm{au}) &  \langle \!\!\sqrt{a_k} \rangle \  (\mathrm{au}^{1/2})  & N_k \  (\mathrm{rad} \ \mathrm{yr}^{-1}) &  \lambda_{0k} \ (\mathrm{rad})\\
\hline 
1 & +3.8709826346795750\mathrm{e-}1& 	+6.2217220269433726\mathrm{e-}1& 	+2.6087903147673952\mathrm{e+}1& 	-1.9409256207851238\mathrm{e+}0	\\
2 & +7.2332656745540558\mathrm{e-}1& 	+8.5048604874990819\mathrm{e-}1& 	+1.0213285580208819\mathrm{e+}1& 	+3.1147692710442865\mathrm{e+}0	\\	
3 & +9.9999548642649372\mathrm{e-}1& 	+9.9999769746809641\mathrm{e-}1& 	+6.2830757971490190\mathrm{e+}0& 	+1.6916552864570649\mathrm{e+}0	\\
4 & +1.5236795439419937\mathrm{e+}0& 	+1.2343740652952826\mathrm{e+}0& 	+3.3406125732385012\mathrm{e+}0& 	-1.4103263138988523\mathrm{e-}1	\\
5 & +5.1927031282775076\mathrm{e+}0& 	+2.2787503028934468\mathrm{e+}0& 	+5.2968797602323869\mathrm{e-}1& 	+5.4918813383105469\mathrm{e-}1	\\
6 & +9.5493731166789555\mathrm{e+}0& 	+3.0902059273326929\mathrm{e+}0& 	+2.1329780126024636\mathrm{e-}1& 	+8.2272979387628475\mathrm{e-}1	\\
7 & +1.9216488649834947\mathrm{e+}1& 	+4.3836614414294299\mathrm{e+}0& 	+7.4781213910514538\mathrm{e-}2& 	-8.6189037349800601\mathrm{e-}1	\\
8 & +3.0106792749722608\mathrm{e+}1& 	+5.4869659219790110\mathrm{e+}0& 	+3.8132752280987267\mathrm{e-}2& 	-1.0322632668252640\mathrm{e+}0	\\
\hline 
\end{tabular}
}
\caption{Secular average of the semi-major axes ($\langle a_k \rangle$), mean mean motions $(N_k)$ and mean longitudes at the origin $(\lambda_{0k})$ along the full solution LaX13b in the La2004 invariant reference frame (see Sec. \ref{sec:inv_plane}) derived by frequency analysis. 
The secular average of the action-like variable $\!\!\sqrt{a_k}$ is also provided, as it cannot be directly obtained from $\langle a_k \rangle$: due to the contribution of the short-period terms, the average $\langle f(x) \rangle$ of $f(x)$ is not $f(\langle x \rangle)$.} 
\label{table:CI} 
\end{table*}

\begin{table*}[ht!]
\centering
{\small
\begin{tabular}{C|C C C C}
\hline\hline 
 k   &     \operatorname{Re}(\cX_k) &   \operatorname{Im}(\cX_k)  &  \operatorname{Re}(\cY_k) &   \operatorname{Im}(\cY_k) \\
\hline 
1 & +5.6990505842292892\mathrm{e-}2	 & +1.9872894591036927\mathrm{e-}1 & 	+4.6165713227763444\mathrm{e-}2	 & +2.9410664010632515\mathrm{e-}2	\\
2 & -4.1735654068702976\mathrm{e-}3	 & +5.3327342759766343\mathrm{e-}3 & 	+1.1928053604473932\mathrm{e-}2	 & +1.4987173725293589\mathrm{e-}2	\\	
3 & -2.7278550885769647\mathrm{e-}3	 & +1.6484645103357756\mathrm{e-}2 & 	+3.3422496351974010\mathrm{e-}3	 & -1.3363750560862579\mathrm{e-}2	\\
4 & +8.2969437680297631\mathrm{e-}2	 & -4.3116239799066092\mathrm{e-}2 & 	+1.4518563462960753\mathrm{e-}2	 & -1.7430317715928134\mathrm{e-}3	\\
5 & +4.7651130651270288\mathrm{e-}2	 & +9.0836135513906891\mathrm{e-}3 & 	+1.9689371814761709\mathrm{e-}3	 & -2.0725403616100885\mathrm{e-}3	\\
6 & +4.7464039310316251\mathrm{e-}4	 & +5.5529067938940795\mathrm{e-}2 & 	-4.1232378462891421\mathrm{e-}3	 & +7.0205383101661738\mathrm{e-}3	\\
7 & -4.5530988000471201\mathrm{e-}2	 & +8.4408377405346107\mathrm{e-}3 & 	+5.5957559447492129\mathrm{e-}3	 & -7.0021269706505322\mathrm{e-}3	\\
8 & +6.4028803643053234\mathrm{e-}3	 & +6.3248118090214632\mathrm{e-}3 & 	-6.2175356759488895\mathrm{e-}3	 & -1.2331754395710286\mathrm{e-}3	\\
\hline 
\end{tabular}
}
\caption{Initial conditions for the secular eccentricity $(\cX_k)$ and inclination $(\cY_k)$ variables at time J2000 in the La2004 invariant  reference frame (see Sec. \ref{sec:inv_plane}), derived from the full LaX13b solution. In the first column, $k$ is the index of the planet, in columns 2 and 3, $\operatorname{Re}(\cX_k)$ and  $\operatorname{Im}(\cX_k)$ are the real part and imaginary part of the eccentricity variables $\cX_k$. In columns 4 and 5, $\operatorname{Re}(\cY_k)$ and  $\operatorname{Im}(\cY_k)$ are the same quantities for the inclination variables.}      
\label{table:CIXY} 
\end{table*}

\begin{table}[ht!]
\centering
\begin{tabular}{R C}
\hline\hline 
    &        \mathrm{arcsec} \ \mathrm{yr}^{-1} \\
\hline 
g_5 &  +4.2574706495769208\mathrm{e+}0  \\
g_6 &  +2.8245402509991674\mathrm{e+}1  \\ 
g_7 &  +3.0879599203482901\mathrm{e+}0  \\
g_8 &  +6.7303498995492750\mathrm{e-}1  \\
s_6 &  -2.6347830405033751\mathrm{e+}1  \\
s_7 &  -2.9925307659382381\mathrm{e+}0  \\
s_8 &  -6.9173578620513787\mathrm{e-}1  \\
\hline 
\end{tabular}
\caption{Values of the fundamental secular frequencies of the outer solar system derived by frequency analysis of the \LaX{} solution over 30 Myr. These are the values actually used in the QPSO model. They may slightly differ from the corresponding values given in \citep{Laskar2004,Laskar2011}.}      
\label{table:fsec} 
\end{table}

\begin{table}[ht!]
\centering
\begin{tabular}{R R R R R}
\hline\hline
         & \multicolumn{1}{c}{Jupiter} & \multicolumn{1}{c}{Saturn} & \multicolumn{1}{c}{Uranus} & \multicolumn{1}{c}{Neptune} \\
\hline
\max(\abs{\cX})            \times 10^6 &      63668  &      87244  &      74248  &      16993 \\  
\max(\abs{\Delta\cX})      \times 10^6 &       3433  &       3993  &       4617  &       2308 \\  
\max(\abs{\Delta_2\cX})    \times 10^6 &        932  &       2817  &       1463  &        364 \\  
\hline 
\rms(\cX)                  \times 10^6 &      46990  &      58506  &      47647  &      10128 \\ 
\rms(\Delta\cX)            \times 10^6 &       2130  &       1876  &       2497  &       1404 \\ 
\rms(\Delta_2\cX)          \times 10^6 &        341  &       1066  &        323  &         89 \\ 
\hline 
\max(\abs{\cY})            \times 10^6 &       4262  &       8908  &      10120  &       7054 \\ 
\max(\abs{\Delta\cY})      \times 10^6 &         47  &         92  &        108  &         51 \\ 
\hline
\rms(\cY)                  \times 10^6 &       3237  &       7870  &       8905  &       5905 \\ 
\rms(\Delta\cY)            \times 10^6 &         19  &         35  &         33  &         20 \\ 
\hline
\end{tabular}
\caption{Maximum and root mean square (rms) of the variables $\cX,\cY$ along the full \LaX{} solution and 
of the differences with their QPSO model $\cX_s, \cY_s$ ($\Delta\cX=\cX-\cX_s$, $\Delta\cY=\cY-\cY_s$). 
For the eccentricity variable $\cX$, the same values are given after removing the main short-period terms 
$\cX_c$ ($\Delta_2\cX=\cX-\cX_s-\cX_c$) (see Figs. \ref{fig:resX} and \ref{fig:resY}).} 
\label{table:QPSO_diffs} 
\end{table}

In this work, we use a quasi-periodic solution for the secular evolution of the orbits of the outer planets (Jupiter to Uranus), denoted hereafter as QPSO. 
In order to achieve a realistic modelling, this solution is deduced from a numerically integrated full solution of the solar system. 
The solution of reference (LaX13b) uses an identical model as in \citep{Laskar2011}, but which has been initially fitted to 
the improved INPOP13b high precision planetary ephemeris \citep{Fienga2014a,Fienga2014b},  extended over 1 Myr. 
The model is then integrated over 30 Myr to derive a quasi-periodic approximation through frequency analysis 
\citep{Laskar1988,Laskar2005}. The variables considered here are the Poincaré complex canonical variables in Eq. (\ref{eq:poincare_vars}), scaled to 
suppress the semi-major axis (or $\Lambda$) dependence, that is  
\begin{equation}
\label{eq:vXY}
\begin{aligned}
\cX &=x \sqrt{{\frac{2}{\Lambda}}} =\sqrt{2} \sqrt{1 - \sqrt{1- e^2}} \, \E^{j \varpi}, \\
\cY &=y \sqrt{{\frac{1}{2\Lambda}}} =  \left(1- e^2\right)^{\frac{1}{4}} \sin(i/2) \, \E^{j \Omega} \ . 
\end{aligned}
\end{equation}
It should be noted that $\cX = z + O(e^3)$ and $\cY = \zeta + O(e^2\sin(i/2))$ where $z=e\E^{j \varpi}$ and 
$\zeta=\sin(i/2) \E^{j \Omega}$ are the classical, non canonical, complex elliptic elements. 

The derivation  of the QPSO model needs some care. Indeed, we want to obtain a model that contains only the 
secular frequencies pertaining to the outer planet system. The secular terms related to the inner planets are thus discarded. 
Because of the presence of the 5:2 close mean-motion resonance among Jupiter and Saturn, and the proximity of the 2:1 
mean motion resonance in the Uranus-Neptune system, several short period terms (with arguments involving the planetary mean motions) 
have also to be discarded. The solution comprises a small number of terms and is fully given in Tables \ref{table:tabX} and \ref{table:tabY}. 
It should be noted that only the terms for which the angular argument is recognized in an unambiguous way as a combination 
of the fundamental frequencies (Table \ref{table:fsec}) are selected, as we want to derive an analytic model for the outer planets variables. These are expressed in the form 
\begin{equation}
\cZ = \sum_{n=1}^N \widetilde{\cZ}_n \, \E^{j \vec{k}_n \cdot \out{\vec{\omega}} t}
\label{eq:qpXY}
\end{equation}
where $\cZ$ stands for $\cX$ or $\cY$, alternatively, $\out{\vec{\omega}} = (g_5,g_6,g_7,g_8,s_6,s_7,s_8)$ is the vector of the fundamental secular frequencies of the outer planet system, given in Table \ref{table:fsec}, and $\vec{k}_n \in \mathbb{Z}^7$ is a 7-uple of integers.

\begin{figure*}[ht!]
    \centering
    \includegraphics[width=17cm]{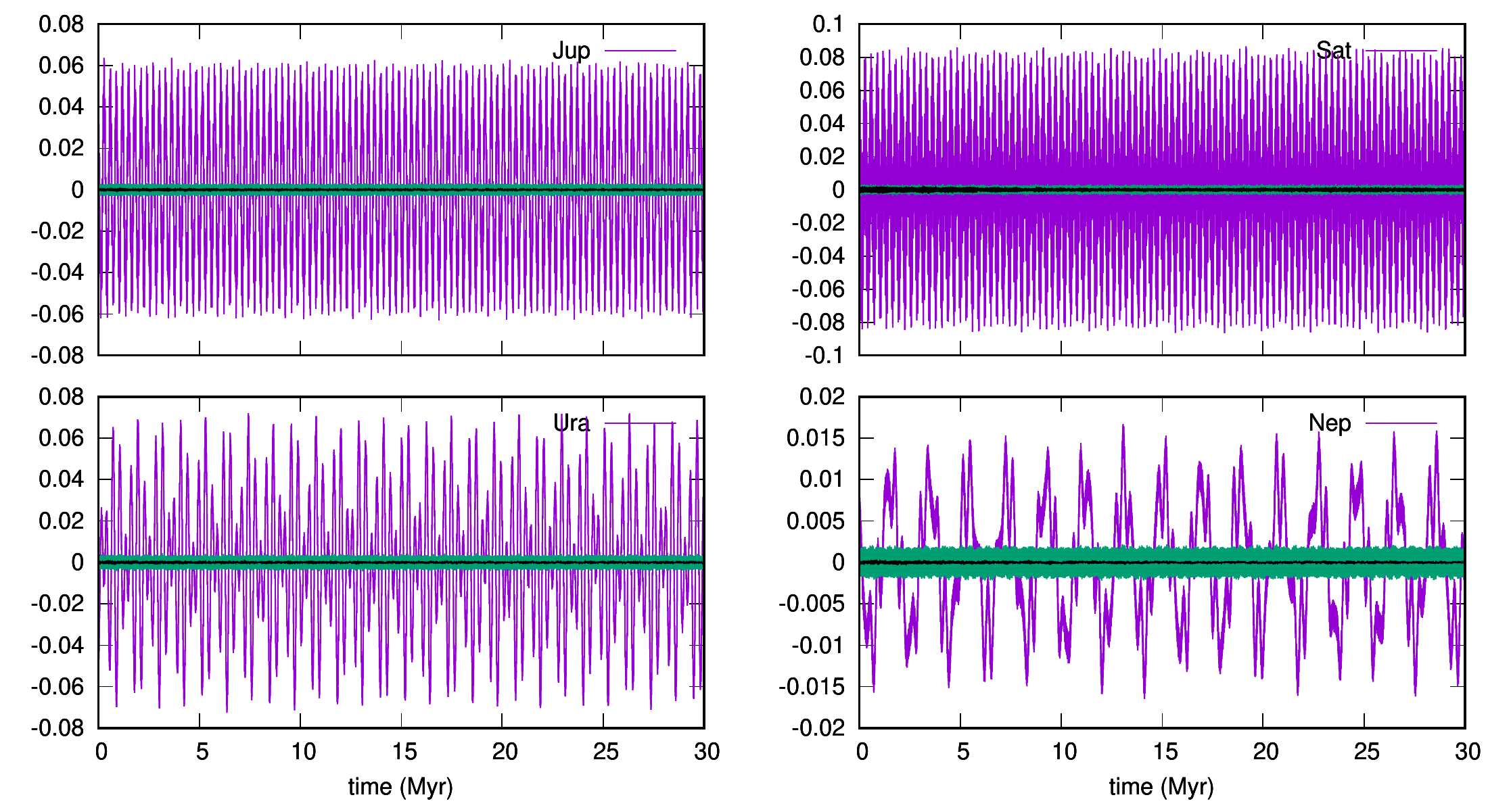}
    \caption{Real part of the eccentricity variable $\cX$ (Eq. \ref{eq:vXY}). The purple curve is the full LaX13b solution. The green curve is the residual after removing 
    the contribution of the secular model QPSO from LaX13b. The black curve is the same, after the removal of the main short-period terms from the residuals.}
    \label{fig:resX}
    \end{figure*}

\begin{figure*}[ht!]
    \centering
    \includegraphics[width=17cm]{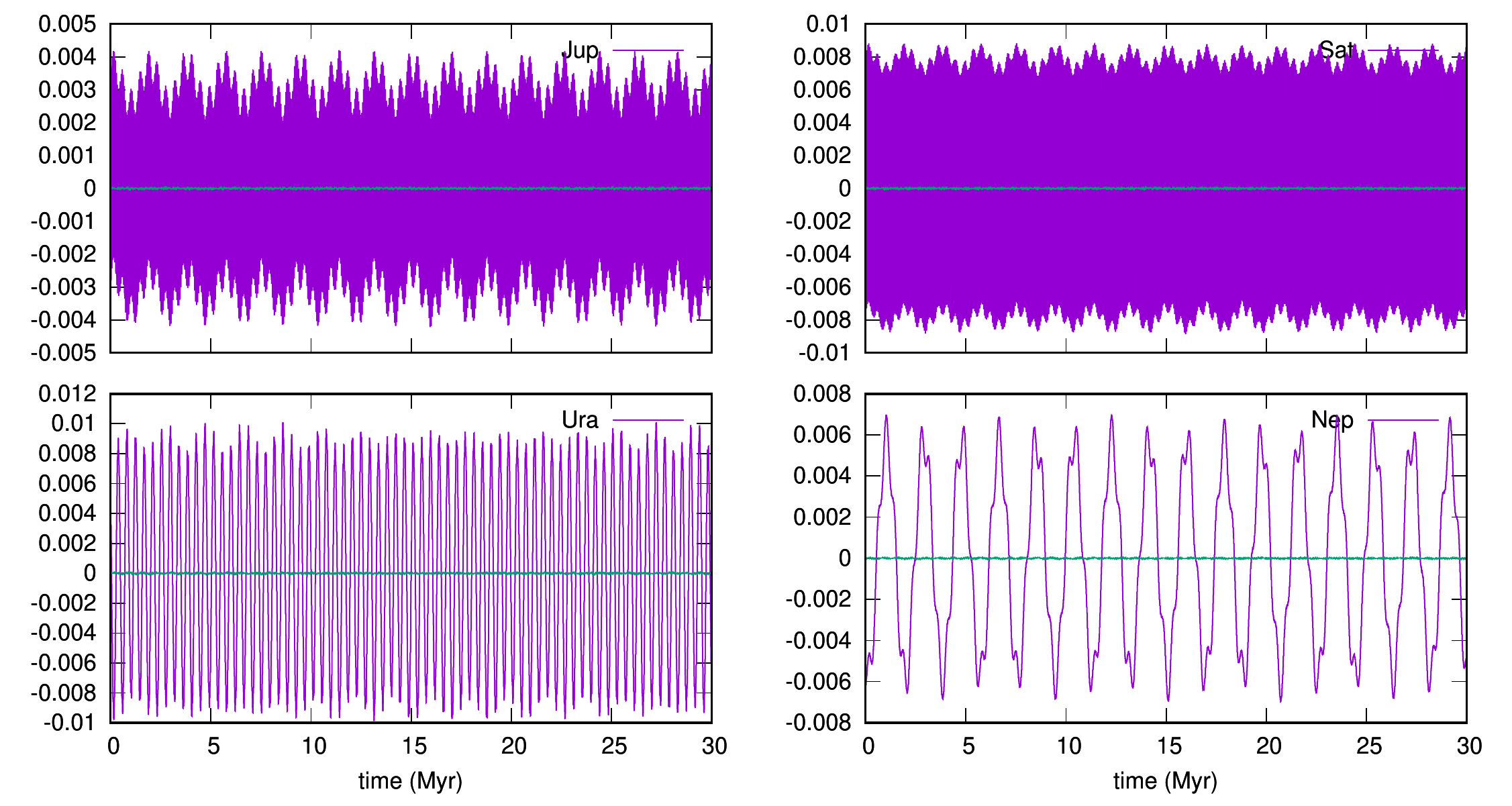}
    \caption{Real part of the inclination variable $\cY$ (Eq. \ref{eq:vXY}). The purple curve is the full LaX13b solution. The green curve is the residual after removing   the contribution of the secular model QPSO from LaX13b.}
    \label{fig:resY}
    \end{figure*}

In order to test this model, we have compared it to the full solution LaX13b (Figs. \ref{fig:resX} and \ref{fig:resY}). Only the real parts of $\cX$ and $\cY$ are 
represented, as the imaginary parts lead to very similar plots. The full solution from LaX13b is plotted in purple, while the 
residuals, once the analytical model QPSO removed, are plotted in green. For the inclination variables ($\cY_k$), these residuals are very small 
and appear as a straight line in the plots (Fig. \ref{fig:resY}). This is not the case for the eccentricity variables ($\cX_k$) where a significant 
band of residuals appears in green (Fig. \ref{fig:resX}). This is mostly due to the short-period terms that are present in the full LaX13b solution. When the largest of these terms are removed, the residuals become much smaller, as shown by the black curves (see also 
Table \ref{table:QPSO_diffs}). 

\subsection{Secular initial conditions}
To complete the QPSO solution, or to integrate the secular equations by Gauss's averaging method (Sec. \ref{sect:gauss_method}), one needs the values of the secular semi-major axes $\hat{a}_k$, or more precisely those of the secular canonical variables $\hat{\Lambda}_k \propto \!\!\sqrt{\hat{a}_k}$, which are constant in the secular system. In this work, the secular semi-major axes are defined as the square of the secular averages $\langle \!\!\sqrt{a_k} \rangle$ along the \LaX{} solution (Table \ref{table:CI}), i.e. $\hat{a}_k = \langle \!\!\sqrt{a_k} \rangle^2$. The mean mean motions $N_k$ (secular averages of the mean motions) and the values of the mean longitudes at the origin $\lambda_{0k}$ are derived as well from the LaX13b solution and provided in Table \ref{table:CI}. 
The remaining initial conditions of the secular system, corresponding to the complex variables $(\cX_k, \cY_k)$, are obtained by a least-square fit of a polynomial of degree 5 in time to the full LaX13b solution over the first few thousand years, after removal of the short-period component by Fourier filtering (Table \ref{table:CIXY}).  

\subsection{The La2004 invariant reference frame}
\label{sec:inv_plane}
All the solutions given here are established in the invariant reference frame, which would be, ideally, the $(x,y,z)$ reference frame whose $z$-axis is aligned with the total angular momentum of the system, and with the $x$-axis pointing towards the equinox J2000. Nevertheless, this convention is not very practical, as this invariant reference frame would change for any small variation of planetary masses or number of objects in the system. This is why in the numerical long-term integrations of \citep{Laskar2009,Laskar2011}, we have adopted a fixed reference frame, which is the conventional invariant reference frame of the orbital solution La2004 of \citep{Laskar2004}, that has been widely used in the paleoclimate community. We will call this reference frame the La2004 invariant reference frame. It is thus a fixed reference frame which is derived from the ICRF equatorial reference frame \citep{Ma1998} by the fixed transformation 
\begin{equation}
{\bf u}_{inv} = R_x(\theta_1) R_z(\theta_3) {\bf u}_{ICRF} \ ,
\end{equation} 
where $R_x$ and $R_z$ are rotation matrices defined as 
\begin{equation}
\begin{split}
R_x(\alpha) &=\Rotx{\alpha} \ ,\\
R_z(\alpha) &=\Rotz{\alpha} \ ,  
\end{split}
\end{equation} 
and 
\begin{equation}
\begin{split}
\theta_1  &=-0.4015807829125271 \ ,\\
\theta_3  &=-0.06724103544220839 \ .  
\end{split}
\end{equation}

\begin{table*}[ht!]
\centering
\begin{tabular}{R R R R R R R R R R}
\hline\hline
\rule{0pt}{1.1em} 	
 n &\multicolumn{1}{c}{$\operatorname{Re}(\widetilde{\cX}_n)$} &\multicolumn{1}{c}{$\operatorname{Im}(\widetilde{\cX}_n)$} & g_5 &g_6 &g_7 &g_8 &s_6 &s_7 &s_8 \\
\hline
\multicolumn{10}{c}{Jupiter } \\	
\hline	
   1 &  3.9355222649836610\mathrm{e-}2 &  2.0091205352205602\mathrm{e-}2 &    1 &   0 &   0 &   \phantom{-0}0 &   0 &   0 &   0 \\
   2 &  8.7066820380533507\mathrm{e-}3 & -1.3074566807550854\mathrm{e-}2 &    0 &   1 &   0 &   0 &   0 &   0 &   0 \\
   3 & -8.2479311026874128\mathrm{e-}4 &  1.6219245178173808\mathrm{e-}3 &    0 &   0 &   1 &   0 &   0 &   0 &   0 \\
   4 &  4.3644617155629635\mathrm{e-}4 &  3.6982470511425620\mathrm{e-}4 &   -1 &   2 &   0 &   0 &   0 &   0 &   0 \\
   5 & -1.6066790855383061\mathrm{e-}4 & -1.0789988494431164\mathrm{e-}4 &    1 &   1 &  -1 &   0 &   0 &   0 &   0 \\
   6 & -1.6498638598462650\mathrm{e-}4 & -1.0869857779342080\mathrm{e-}4 &   -1 &   1 &   1 &   0 &   0 &   0 &   0 \\
   7 &  1.0937458300137176\mathrm{e-}4 &  5.6433525785796263\mathrm{e-}5 &   -1 &   2 &   0 &   0 &   1 &  -1 &   0 \\
   8 &  8.5006502723719633\mathrm{e-}5 &  7.1060850357874042\mathrm{e-}5 &    1 &   0 &   0 &   0 &  -1 &   1 &   0 \\
   9 & -2.7299373985450095\mathrm{e-}5 &  6.4538424302377966\mathrm{e-}5 &    2 &   0 &  -1 &   0 &   0 &   0 &   0 \\
  10 &  3.2040993080287076\mathrm{e-}5 & -3.7642412972847125\mathrm{e-}5 &    0 &   2 &  -1 &   0 &   0 &   0 &   0 \\
  11 &  1.9199352898954454\mathrm{e-}5 &  5.4871593139050202\mathrm{e-}5 &    0 &   0 &   0 &   1 &   0 &   0 &   0 \\
  12 & -2.1420886245050815\mathrm{e-}5 &  2.0070283084590094\mathrm{e-}5 &   -2 &   3 &   0 &   0 &   0 &   0 &   0 \\
  13 &  9.7884950272160402\mathrm{e-}6 & -1.1528203620115796\mathrm{e-}5 &   -2 &   2 &   1 &   0 &   0 &   0 &   0 \\
  14 & -6.4760245885223307\mathrm{e-}6 &  9.7064946705620636\mathrm{e-}6 &    0 &  -1 &   0 &   0 &   2 &   0 &   0 \\
\hline	
\multicolumn{10}{c}{Saturn } \\	
\hline	
   1 & -2.6735064579495354\mathrm{e-}2 &  4.0147278432709689\mathrm{e-}2 &    0 &   1 &   0 &   0 &   0 &   0 &   0 \\
   2 &  2.9365236482935035\mathrm{e-}2 &  1.4991233172235453\mathrm{e-}2 &    1 &   0 &   0 &   0 &   0 &   0 &   0 \\
   3 & -1.4720163989740585\mathrm{e-}3 & -1.2472787791106869\mathrm{e-}3 &   -1 &   2 &   0 &   0 &   0 &   0 &   0 \\
   4 & -6.8748119559295436\mathrm{e-}4 &  1.3516157882362077\mathrm{e-}3 &    0 &   0 &   1 &   0 &   0 &   0 &   0 \\
   5 &  4.9631439815354689\mathrm{e-}4 &  3.3334599996898677\mathrm{e-}4 &    1 &   1 &  -1 &   0 &   0 &   0 &   0 \\
   6 &  5.0410851738721387\mathrm{e-}4 &  3.3210459157965804\mathrm{e-}4 &   -1 &   1 &   1 &   0 &   0 &   0 &   0 \\
   7 & -3.3825396912499196\mathrm{e-}4 & -1.7459211633013424\mathrm{e-}4 &   -1 &   2 &   0 &   0 &   1 &  -1 &   0 \\
   8 & -2.5925687680288125\mathrm{e-}4 & -2.1668870819116260\mathrm{e-}4 &    1 &   0 &   0 &   0 &  -1 &   1 &   0 \\
   9 & -1.0852577202837296\mathrm{e-}4 &  1.2750655350023438\mathrm{e-}4 &    0 &   2 &  -1 &   0 &   0 &   0 &   0 \\
  10 & -4.6777512194793370\mathrm{e-}5 &  1.2559406460037114\mathrm{e-}4 &    2 &  -1 &   0 &   0 &   0 &   0 &   0 \\
  11 &  6.9601743332890669\mathrm{e-}5 & -6.5337803141444139\mathrm{e-}5 &   -2 &   3 &   0 &   0 &   0 &   0 &   0 \\
  12 &  1.9059552532015572\mathrm{e-}5 &  5.4460189585479054\mathrm{e-}5 &    0 &   0 &   0 &   1 &   0 &   0 &   0 \\
  13 & -3.3260941738184332\mathrm{e-}5 &  3.9045961061362962\mathrm{e-}5 &   -2 &   2 &   1 &   0 &   0 &   0 &   0 \\
  14 & -1.8327645535629047\mathrm{e-}5 &  4.4381655444927067\mathrm{e-}5 &    2 &   0 &  -1 &   0 &   0 &   0 &   0 \\
  15 &  1.6832831093654628\mathrm{e-}5 & -2.5231972103884829\mathrm{e-}5 &   -2 &   3 &   0 &   0 &   1 &  -1 &   0 \\
  16 &  1.9485028357121861\mathrm{e-}5 & -1.8629237185287101\mathrm{e-}5 &    0 &   1 &   0 &   0 &  -1 &   1 &   0 \\
  17 &  1.3531984986636972\mathrm{e-}5 & -2.0332362024269671\mathrm{e-}5 &    0 &  -1 &   0 &   0 &   2 &   0 &   0 \\
\hline	
\multicolumn{10}{c}{Uranus } \\	
\hline	
   1 & -3.3516665527223363\mathrm{e-}2 & -1.7110181361426769\mathrm{e-}2 &    1 &   0 &   0 &   0 &   0 &   0 &   0 \\
   2 & -1.3168012223376812\mathrm{e-}2 &  2.5888592859517973\mathrm{e-}2 &    0 &   0 &   1 &   0 &   0 &   0 &   0 \\
   3 &  8.5812650851483144\mathrm{e-}4 & -1.2886280579699632\mathrm{e-}3 &    0 &   1 &   0 &   0 &   0 &   0 &   0 \\
   4 &  5.5001023804251395\mathrm{e-}4 &  1.5768772504826957\mathrm{e-}3 &    0 &   0 &   0 &   1 &   0 &   0 &   0 \\
   5 & -3.7488766418176093\mathrm{e-}4 & -2.5628502554155236\mathrm{e-}4 &   -1 &   0 &   2 &   0 &   0 &   0 &   0 \\
   6 &  1.8595341017043558\mathrm{e-}4 & -4.0628416200819552\mathrm{e-}4 &    2 &   0 &  -1 &   0 &   0 &   0 &   0 \\
   7 &  6.1419633897654503\mathrm{e-}5 & -3.8157567361894473\mathrm{e-}5 &    0 &   0 &   0 &   1 &   0 &  -1 &   1 \\
   8 & -4.8677965279021335\mathrm{e-}5 & -3.2236597182724679\mathrm{e-}5 &   -1 &   1 &   1 &   0 &   0 &   0 &   0 \\
   9 & -1.7354342210696565\mathrm{e-}5 &  4.5904026894682532\mathrm{e-}5 &    0 &   1 &   0 &   0 &   1 &  -1 &   0 \\
  10 &  2.2036143287116906\mathrm{e-}5 &  3.4065981233013441\mathrm{e-}5 &    0 &   0 &   1 &   0 &   0 &   1 &  -1 \\
  11 &  3.2445476543283863\mathrm{e-}5 &  2.7491971542182629\mathrm{e-}5 &   -1 &   2 &   0 &   0 &   0 &   0 &   0 \\
  12 &  1.4570603907287206\mathrm{e-}5 & -3.9111620103262798\mathrm{e-}5 &    2 &  -1 &   0 &   0 &   0 &   0 &   0 \\
  13 & -3.4710330596857471\mathrm{e-}5 & -1.2868365278212245\mathrm{e-}5 &    1 &  -1 &   1 &   0 &   0 &   0 &   0 \\
\hline	
\multicolumn{10}{c}{Neptune } \\	
\hline	
   1 &  3.0124292308931207\mathrm{e-}3 &  8.6106841210746767\mathrm{e-}3 &    0 &   0 &   0 &   1 &   0 &   0 &   0 \\
   2 &  1.6799504499698586\mathrm{e-}3 & -3.3014578256236110\mathrm{e-}3 &    0 &   0 &   1 &   0 &   0 &   0 &   0 \\
   3 &  1.6855558947835656\mathrm{e-}3 &  8.6062600108725124\mathrm{e-}4 &    1 &   0 &   0 &   0 &   0 &   0 &   0 \\
   4 &  9.2455797923682499\mathrm{e-}5 &  1.2548047080946322\mathrm{e-}4 &    0 &   0 &   1 &   0 &   0 &   1 &  -1 \\
   5 &  5.7121148764694937\mathrm{e-}5 & -8.5774015603758624\mathrm{e-}5 &    0 &   1 &   0 &   0 &   0 &   0 &   0 \\
   6 & -1.7043313379296477\mathrm{e-}5 &  5.5214648783483626\mathrm{e-}5 &    2 &   0 &  -1 &   0 &   0 &   0 &   0 \\
   7 & -5.0730275098891472\mathrm{e-}5 &  1.7656865611016711\mathrm{e-}5 &   -1 &   0 &   1 &   1 &   0 &   0 &   0 \\
   8 & -4.6217102122902557\mathrm{e-}5 &  8.6886583776811327\mathrm{e-}6 &    1 &   0 &  -1 &   1 &   0 &   0 &   0 \\
\hline	
\end{tabular}
\caption{Quasi-periodic decomposition (Eq. \ref{eq:qpXY}) of the  secular solution for the outer planets for the eccentricity variable $\cX$.
The first column is the index $n$ of the terms ranked by decreasing amplitude. Columns 2 and 3 are the real and imaginary parts of the complex 
amplitude $\widetilde{\cX}_n$. The last 7 columns are the integer coefficients $\vec{k}_n$ of the secular fundamental frequencies 
$(g_5,g_6,g_7,g_8,s_6,s_7,s_8)$.}      
\label{table:tabX} 
\end{table*}

\begin{table*}[ht!]
\centering
\begin{tabular}{R R R R R R R R R R}
\hline\hline	
\rule{0pt}{1.1em} 	
 n &\multicolumn{1}{c}{$\operatorname{Re}(\widetilde{\cY}_n)$} &\multicolumn{1}{c}{$\operatorname{Im}(\widetilde{\cY}_n)$} & g_5 &g_6 &g_7 &g_8 &s_6 &s_7 &s_8 \\
\hline	
\multicolumn{10}{c}{Jupiter } \\	
\hline	
   1 &  1.7488173284935253\mathrm{e-}3 & -2.6183244274251252\mathrm{e-}3 &    0 &   0 &   0 &   0 &   1 &   0 &   0 \\
   2 &  5.4086264886788772\mathrm{e-}4 &  2.0008212164609817\mathrm{e-}4 &    0 &   0 &   0 &   0 &   0 &   0 &   1 \\
   3 & -3.4991238273175804\mathrm{e-}4 &  3.2977145829984634\mathrm{e-}4 &    0 &   0 &   0 &   0 &   0 &   1 &   0 \\
   4 & -1.8308342659255464\mathrm{e-}5 & -1.5291916542971615\mathrm{e-}5 &    1 &  -1 &   0 &   0 &   0 &   1 &   0 \\
   5 &  1.4043531453627062\mathrm{e-}5 &  1.1934392862719320\mathrm{e-}5 &   -1 &   1 &   0 &   0 &   1 &   0 &   0 \\
   6 &  7.8810622671801405\mathrm{e-}6 & -1.1871211261358110\mathrm{e-}5 &    0 &   2 &   0 &   0 &  -1 &   0 &   0 \\
   7 &  5.9643858674840883\mathrm{e-}6 &  1.0197242568214183\mathrm{e-}5 &    1 &   0 &  -1 &   0 &   0 &   1 &   0 \\
   8 &  6.7104749336174370\mathrm{e-}6 &  7.0850042988014965\mathrm{e-}6 &   -1 &   0 &   1 &   0 &   0 &   1 &   0 \\
   9 &  7.5756401983704222\mathrm{e-}6 &  3.8603677665594590\mathrm{e-}6 &    1 &   1 &   0 &   0 &  -1 &   0 &   0 \\
  10 &  5.9778122148403010\mathrm{e-}6 &  3.0620950992005088\mathrm{e-}6 &    1 &  -1 &   0 &   0 &   1 &   0 &   0 \\
  11 & -2.1572046295338697\mathrm{e-}6 &  5.8221465688096679\mathrm{e-}6 &    2 &   0 &   0 &   0 &  -1 &   0 &   0 \\
  12 &  5.5060848892241336\mathrm{e-}6 &  1.0748962916018817\mathrm{e-}7 &    0 &   0 &   1 &  -1 &   0 &   1 &   0 \\
  13 &  3.6055121619276866\mathrm{e-}6 &  2.3835869329764091\mathrm{e-}6 &   -1 &   0 &   1 &   0 &   1 &   0 &   0 \\
  14 &  3.0826004389954454\mathrm{e-}6 &  2.0622614907196112\mathrm{e-}6 &    1 &   0 &  -1 &   0 &   1 &   0 &   0 \\
\hline	
\multicolumn{10}{c}{Saturn } \\	
\hline	
   1 & -4.3551504624732961\mathrm{e-}3 &  6.5205184658223064\mathrm{e-}3 &    0 &   0 &   0 &   0 &   1 &   0 &   0 \\
   2 &  5.2100147631152049\mathrm{e-}4 &  1.9273799677446370\mathrm{e-}4 &    0 &   0 &   0 &   0 &   0 &   0 &   1 \\
   3 & -2.8379662031061457\mathrm{e-}4 &  2.6741614761702752\mathrm{e-}4 &    0 &   0 &   0 &   0 &   0 &   1 &   0 \\
   4 &  4.6758845331323253\mathrm{e-}5 &  3.9087781691034812\mathrm{e-}5 &    1 &  -1 &   0 &   0 &   0 &   1 &   0 \\
   5 & -1.9580115731859662\mathrm{e-}5 &  2.9490631827283331\mathrm{e-}5 &    0 &   2 &   0 &   0 &  -1 &   0 &   0 \\
   6 & -1.7652537335071599\mathrm{e-}5 & -8.9921682226792706\mathrm{e-}6 &    1 &   1 &   0 &   0 &  -1 &   0 &   0 \\
   7 &  5.4933935882853390\mathrm{e-}6 & -1.4791745653763672\mathrm{e-}5 &    2 &   0 &   0 &   0 &  -1 &   0 &   0 \\
   8 & -9.8967096662986720\mathrm{e-}6 & -5.0668618867237504\mathrm{e-}6 &    1 &  -1 &   0 &   0 &   1 &   0 &   0 \\
   9 &  5.5991180901835754\mathrm{e-}6 &  9.2383183832156914\mathrm{e-}6 &    1 &   0 &  -1 &   0 &   0 &   1 &   0 \\
  10 & -8.3001644772275482\mathrm{e-}6 & -5.4822443203989674\mathrm{e-}6 &   -1 &   0 &   1 &   0 &   1 &   0 &   0 \\
  11 & -8.4231240803385387\mathrm{e-}6 & -5.6221713943930228\mathrm{e-}6 &    1 &   0 &  -1 &   0 &   1 &   0 &   0 \\
  12 &  4.3080242579548885\mathrm{e-}6 & -4.9727282168325052\mathrm{e-}6 &    0 &  -1 &   1 &   0 &   0 &   1 &   0 \\
  13 &  4.6767916324411297\mathrm{e-}6 &  4.9364726673458742\mathrm{e-}6 &   -1 &   0 &   1 &   0 &   0 &   1 &   0 \\
  14 &  5.3492390903931119\mathrm{e-}6 &  1.1516431592670172\mathrm{e-}7 &    0 &   0 &   1 &  -1 &   0 &   1 &   0 \\
\hline	
\multicolumn{10}{c}{Uranus } \\	
\hline	
   1 &  6.4509456587545097\mathrm{e-}3 & -6.0800433007360578\mathrm{e-}3 &    0 &   0 &   0 &   0 &   0 &   1 &   0 \\
   2 & -5.2144272869098863\mathrm{e-}4 & -1.9450453182598740\mathrm{e-}4 &    0 &   0 &   0 &   0 &   0 &   0 &   1 \\
   3 &  1.9588947834103037\mathrm{e-}4 & -2.9327923348973088\mathrm{e-}4 &    0 &   0 &   0 &   0 &   1 &   0 &   0 \\
   4 & -2.3212752397668531\mathrm{e-}4 & -1.9717826296471695\mathrm{e-}4 &   -1 &   1 &   0 &   0 &   1 &   0 &   0 \\
   5 & -1.2149356527579389\mathrm{e-}4 & -1.5156742923710744\mathrm{e-}4 &    1 &   0 &  -1 &   0 &   0 &   1 &   0 \\
   6 & -1.2811598024162592\mathrm{e-}4 & -1.3534877221356973\mathrm{e-}4 &   -1 &   0 &   1 &   0 &   0 &   1 &   0 \\
   7 & -5.2966451675750289\mathrm{e-}5 &  2.5077699446788398\mathrm{e-}5 &    0 &   0 &  -1 &   1 &   0 &   0 &   1 \\
   8 &  1.6984719605072953\mathrm{e-}5 & -1.9936933726180298\mathrm{e-}5 &    0 &   1 &  -1 &   0 &   1 &   0 &   0 \\
   9 & -2.4457738819518524\mathrm{e-}6 &  1.8969234909116482\mathrm{e-}5 &    2 &   0 &   0 &   0 &   0 &  -1 &   0 \\
  10 &  1.2046673620575406\mathrm{e-}5 &  1.5300754359230831\mathrm{e-}6 &    1 &   0 &   1 &   0 &   0 &  -1 &   0 \\
  11 & -7.8758536093341333\mathrm{e-}6 & -6.6184659114505215\mathrm{e-}6 &    1 &  -1 &   0 &   0 &   0 &   1 &   0 \\
\hline	
\multicolumn{10}{c}{Neptune } \\	
\hline	
   1 & -5.4490991487018952\mathrm{e-}3 & -2.0144507098382057\mathrm{e-}3 &    0 &   0 &   0 &   0 &   0 &   0 &   1 \\
   2 & -7.7303138004431132\mathrm{e-}4 &  7.2861855867743536\mathrm{e-}4 &    0 &   0 &   0 &   0 &   0 &   1 &   0 \\
   3 &  4.3096387136203190\mathrm{e-}5 &  3.6612167171941073\mathrm{e-}5 &   -1 &   1 &   0 &   0 &   1 &   0 &   0 \\
   4 & -5.7346979760257004\mathrm{e-}5 & -4.5712045540159025\mathrm{e-}7 &    0 &   0 &   1 &  -1 &   0 &   1 &   0 \\
   5 &  2.0971470945766545\mathrm{e-}5 & -3.1399605917613957\mathrm{e-}5 &    0 &   0 &   0 &   0 &   1 &   0 &   0 \\
   6 &  1.8895546258392125\mathrm{e-}5 &  1.9903787862297811\mathrm{e-}5 &   -1 &   0 &   1 &   0 &   0 &   1 &   0 \\
   7 & -6.3733664965820864\mathrm{e-}6 &  1.6355452734395612\mathrm{e-}5 &    1 &   0 &  -1 &   0 &   0 &   0 &   1 \\
   8 & -1.0680967225253630\mathrm{e-}5 &  1.2532923143655674\mathrm{e-}5 &    0 &   1 &  -1 &   0 &   1 &   0 &   0 \\
   9 &  7.6134109660753571\mathrm{e-}6 & -3.6727669507265366\mathrm{e-}6 &    0 &   0 &  -1 &   1 &   0 &   0 &   1 \\
  10 &  3.0786794242300075\mathrm{e-}7 & -6.0431372258343262\mathrm{e-}6 &    1 &   0 &   0 &  -1 &   0 &   1 &   0 \\
  11 & -5.1451015466983096\mathrm{e-}6 & -6.1085028343284465\mathrm{e-}7 &    1 &   0 &   1 &   0 &   0 &  -1 &   0 \\
\hline	
\end{tabular}
\caption{Quasi-periodic decomposition (Eq. \ref{eq:qpXY}) of the  secular solution for the outer planets for the eccentricity variable $\cY$.
The first column is the index $n$ of the terms ranked by decreasing amplitude. Columns 2 and 3 are the real and imaginary parts of the complex 
amplitude $\widetilde{\cY}_n$. The last 7 columns are the integer coefficients $\vec{k}_n$ of the secular fundamental frequencies $(g_5,g_6,g_7,g_8,s_6,s_7,s_8)$.}      
\label{table:tabY} 
\end{table*}

\clearpage


\end{appendix}

\end{document}